%% file: paper.tex
\documentclass[aps,superscriptaddress,nofootinbib,preprintnumbers]{revtex4}

\usepackage{amsmath}
\usepackage{graphicx,tabularx}
\usepackage{feynarts2}

\def\ie{{\it i.e.\;}}
\def\eg{{\it e.g.\;}}

\def\mgl{m_{\tilde{g}}}
\def\mev{~{\ensuremath\rm MeV}}
\def\gev{~{\ensuremath\rm GeV}}

\def\fb{~{\ensuremath\rm fb}}

\def\D0{D\O}  

\newcommand{\slashed}[1]{\ensuremath{{#1}{\!}{\!}{\!}{\!}{\:}/}}

\begin{document}

\date{\today}
\title{Charged-Higgs Collider Signals with or without Flavor}

\preprint{DO-TH 2006/14}
\preprint{LMU-ASC 22/05}
\preprint{MPP-2006-226}

\author{Stefan Dittmaier}
%\email{stefan.dittmaier@mppmu.mpg.de}
\affiliation{Max-Planck-Institut f\"ur Physik (Werner-Heisenberg-Institut), 
M\"unchen, Germany}

\author{Gudrun Hiller}
%\email{ghiller@physik.uni-dortmund.de}
\affiliation{Institut f\"ur Physik, Universit\"at Dortmund, Germany}

\author{Tilman Plehn}
%\email{tilman.plehn@cern.ch}
\affiliation{SUPA, School of Physics, University of Edinburgh, Scotland}

\author{Michael Spannowsky}
%\email{msp@theorie.physik.uni-muenchen.de}
\affiliation{ASC, Department f\"ur Physik,
             Ludwig-Maximilians-Universit\"at M\"unchen, Germany}

\begin{abstract}
  A charged Higgs boson is a clear signal for an
  extended Higgs sector, as for example predicted by supersymmetry.
  Squark mixing can
  significantly change the pattern of charged-Higgs production and
  most notably circumvent the chiral suppression for single Higgs
  production. We evaluate the LHC discovery potential 
  in the light of flavor physics, in the
  single-Higgs production channel and in association with a hard
  jet for small and moderate values of $\tan \beta$.  
  Thoroughly examining current flavor constraints we find that non-minimal 
  flavor structures can have a sizeable
  impact, but tend to predict moderate production rates.
  Nevertheless, charged-Higgs searches will
  probe flavor structures not accessible to
  rare kaon, bottom, or charm experiments, and can
  invalidate the assumption of minimal flavor violation.
\end{abstract}

\maketitle

%%%%%%%%%%%%%%%%%%%%%%%%%%%%%%%%%%%%%%%%%%%%%%%%%%%%%%%%%%%%%%%%%%%%%%
\section{Introduction}

Understanding the nature of electroweak symmetry breaking is the most
important endeavor in high-energy physics
over the coming years. With the LHC close to delivering
data and probing a new energy range in particle physics, we expect to be
close to solving this one remaining puzzle in the Standard Model.

The Standard Model chooses a particularly simple approach to electroweak
symmetry breaking: all masses are created by one Higgs doublet
acquiring a vacuum expectation value. This one doublet and its
conjugate give mass to up-type and down-type fermions.  For
example, supersymmetry does not allow 
for this simple mechanism. We
need two Higgs doublets to give mass to all fermions, if we want the Higgs 
fields to respect supersymmetry  and if we want to avoid
anomalies arising from fermionic supersymmetric Higgsinos. Such an
extended model with each Higgs doublet coupling exclusively to up-type
or down-type fermions is generally referred to as a
two-Higgs-doublet model of type~II~\cite{hhg}. Taking into account
electroweak precision data~\cite{lewwg}, a typical two-Higgs-doublet
model will predict a light Higgs scalar and a set of additional heavy
Higgs modes. In the most prominent two-Higgs-doublet model (the MSSM
Higgs sector) there is no doubt that we will see the light scalar
Higgs in the usual Standard Model search channels~\cite{wbf_susy}.
Unfortunately, to positively identify an extended Higgs sector it
might not be sufficient to simply study this light
Higgs~\cite{duehrssen}. An additional heavy charged Higgs is the most
distinct signature of a second Higgs doublet. In contrast to 
a heavy neutral scalar, it does not get faked by states
that are not linked to the Higgs sector.\bigskip

Over the years, many charged-Higgs search strategies at the LHC have
been proposed and studied. For a pure MSSM-type two-Higgs-doublet
model the entire leading-order parameter space is described by the
charged-Higgs mass
$m_{H^+}$ and $\tan\beta$, where $\tan\beta$ is the ratio of
the two vacuum expectation values. Almost all of the LHC search strategies
make use of a particularity in the type-II two-Higgs-doublet model: the
heavy-quark Yukawas $y_q$ to the heavy Higgs states are governed by
$y_b \tan\beta$ and by $y_t/\tan\beta$.  The most promising strategy
for finding a charged Higgs at the LHC will therefore include coupling
it to incoming or outgoing bottoms.

The most promising charged-Higgs production channel is in association
with a top quark~\cite{prod_top,prod_top_nlo,prod_tp1,prod_tp2}.  The rate
can consistently be computed in a 5-flavor or in a 4-flavor scheme,
\ie with or without using bottom parton densities~\cite{bottom_pdf}.
Because of the complexity of the top-associated final state, a charged-Higgs decay to
hadronic $\tau^+ \nu$~\cite{dec_tau_ph,dec_tau_ex} is easier to
extract from the background than the (likely undetectable) decay to
$t\bar{b}$~\cite{dec_top_ph,dec_top_ex}.  Recently, it has been shown
that the search for a light charged Higgs in anomalous top
decays $t \to H^+ b \to (\tau^+ \nu) b$ can be merged nicely with the search for a charged
Higgs produced with a top quark $\bar{b}g \to \bar{t} H^+ \to \bar{t}
(\tau^+ \nu)$~\cite{low_mass,prod_tp2}.\bigskip

Unfortunately, all strategies described above fail for
small $\tan\beta$. The bottom-induced search channels only cover
$\tan\beta \gtrsim 20$, leaving a hole $\tan\beta = 2 \cdots 20$
in the parameter space.
In the MSSM in this region we might only see a light SM-like Higgs, 
unless we are lucky enough to produce light Higgses in pairs coming from a
resonant heavy neutral Higgs~\cite{higgs_pairs}. There are several
ideas how to cover this region searching for a charged Higgs, such as, \eg,
the production in association with a $W$~\cite{prod_w} or pair
production. The latter occurs at tree level with incoming bottom
quarks, $b\bar{b} \to H^+ H^-$, it can also be loop mediated, $gg \to H^+H^-$, 
or for low and intermediate
$\tan\beta$ we can search for $q\bar{q} \to H^+H^-$~\cite{prod_pair}. Unfortunately, none
of these strategies are too promising, because the rates without
$\tan\beta$ enhancement are small.\bigskip

Looking beyond bottom-mediated production channels reveals an opportunity
linked to charged-Higgs searches: while it is well known how to absorb the
leading supersymmetric loops into an effective bottom 
Yukawa coupling~\cite{bottom_yuk,prod_tp2}, the production via light-flavor
quarks can be heavily affected by the 
flavor structure of the model embedding the two Higgs doublets. 
Within the Standard Model flavor symmetry breaking is
governed solely by the Yukawa interactions.
This simple, highly predictive 
mechanism is successful in explaining a multitude of flavor-changing 
quark transitions. Applying this concept to extensions of the 
Standard Model leads to the notion of minimal flavor violation (MFV)~\cite{mfv}:
in an MFV model there are no other sources of flavor
violation other than the Yukawas,  the spurions of flavor symmetry breaking. 
For the case of the MSSM with unbroken $R$ parity, the MFV condition is 
automatically satisfied for supersymmetric gauge couplings ($D$ terms)
and for scalar couplings derived from the superpotential ($F$ terms).
However, general soft SUSY breaking introduces new sources of flavor violation. In MFV (i) all soft scalar squark masses need to be diagonal in flavor space and (ii) all triscalar $A$-terms describing the
squark--squark--Higgs couplings have to be proportional
to the Yukawas.
Corrections consistent with the Standard Model flavor symmetry 
are induced by higher powers in the Yukawas \cite{mfv,Hiller:2002um,Altmannshofer:2007cs}.
This set of MFV assumptions automatically
passes a large fraction of experimental constraints.\bigskip

Such an MFV assumption is not necessary. 
While some flavor-non-diagonal MSSM couplings are tightly constrained,
others can be of order one (see \eg \cite{Hall:1985dx,fcnc-susy,vacuum}).
In general, constraints from flavor-changing-neutral-current (FCNC) 
$K$- and $B$-physics data having external down-type 
quarks are stronger on flavor violation among down-squarks, 
because down-squark effects can occur
via strongly interacting gluino loops, as opposed to up-squark effects mediated
by the weak interaction.
With the exception of the recent $D^0 \bar D^0$-mixing
measurements, which mostly constrain flavor mixing between
first- and second-generation squarks \cite{Nir:2007ac},
currently there are only upper bounds on charm or top FCNCs.
Some of the most stringent limits on the flavor structure including the third
generation
come from $B$- and $B_s$-meson measurements and involve the
$b \to s$ and $b \to d$ quark transitions in meson mixing and decays.
Particularly constraining are the radiative
$B \to X_s \gamma$ and $B \to \rho \gamma$, 
semileptonic $B \to X_s \ell^+ \ell^-$ and $B \to \pi \ell^+ \ell^-$  decays
and the $B_d{-}\overline{B}_d$
mass differences~\cite{b_s_gamma_ex,b_s_gamma_th,rhogamma,bsll,susy-Zpenguin,B2pill,data,bs_mix,Bertolini:1990if,mixing-th}.

Even with the strong current constraints from flavor physics taken into 
account, 
the MSSM beyond MFV has regions of parameter space where
the couplings of a charged Higgs to light quarks are substantially modified
by SUSY loops. 
For small $\tan\beta$ charged-Higgs searches at the LHC are
a sensitive probe of supersymmetric flavor physics, in a similar way to
rare decays at $B$ factories: they will never guarantee
charged-Higgs discovery, but their experimental verification
would shed light on otherwise poorly constrained aspects of the MSSM
flavor sector, linked to the physics of supersymmetry breaking.
\bigskip

The paper is organized as follows: in Section~\ref{sec:singleH} we study the single-charged-Higgs 
production $q \bar q^\prime \to H^\pm$ 
in the MSSM, assuming MFV and allowing for 
general flavor violation. A brief discussion 
of flavor violation in supersymmetric models 
is included in this section. 
We improve on earlier work~\cite{single_higgs} by a more 
general treatment of squark mixing and by taking into account
FCNC 
constraints.
In Section~\ref{sec:flavor} we discuss current constraints on soft-breaking parameters from flavor-physics data and theory.
In Section~\ref{sec:hardjet} we calculate charged-Higgs production rates in association with a hard jet, within and beyond MFV. A brief background study for the LHC environment is included. We summarize in Section~\ref{sec:outlook} and
provide details about flavored quarks and squarks in the appendix.

%%%%%%%%%%%%%%%%%%%%%%%%%%%%%%%%%%%%%%%%%%%%%%%%%%%%%%%%%%%%%%%%%%%%%%
\section{Single-Charged-Higgs Production} 
\label{sec:singleH}

\begin{figure}[t]
 \begin{center}
   \includegraphics[width=3.5cm]{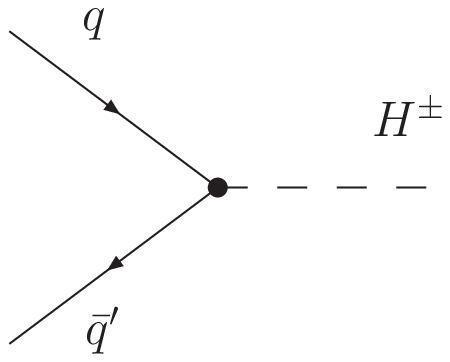} \hspace*{5mm}
   \includegraphics[width=3.5cm]{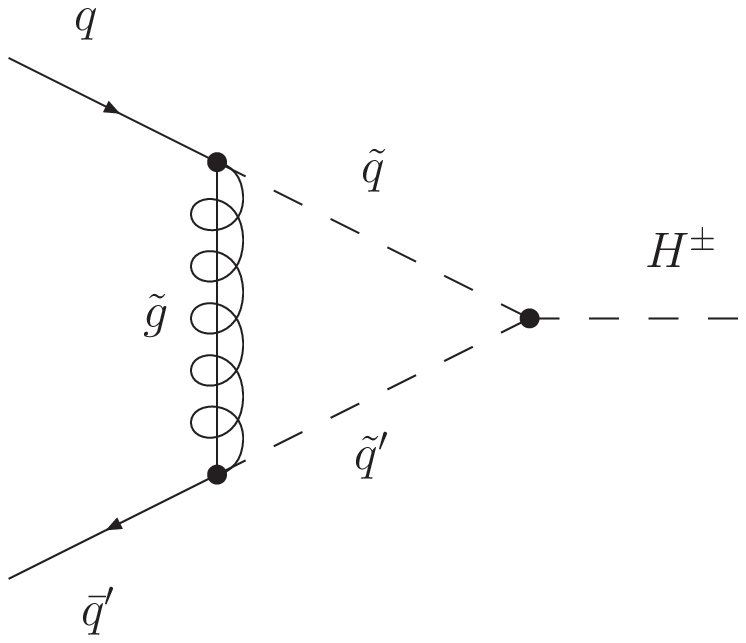}\hspace*{5mm}
   \includegraphics[width=3.5cm]{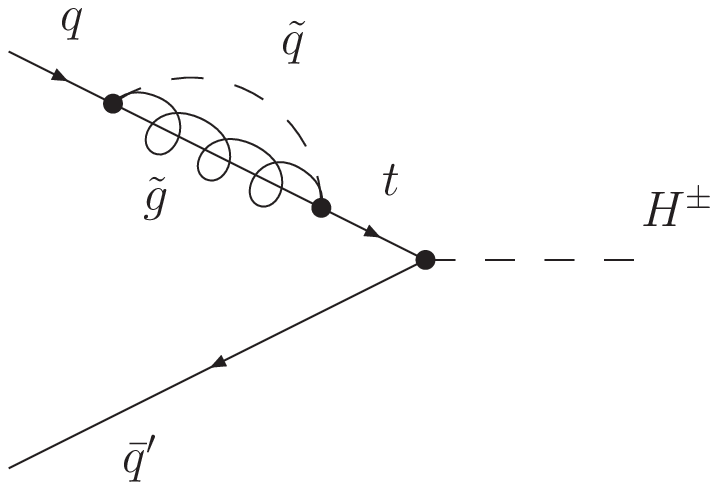} \hspace*{5mm}
   \includegraphics[width=4.0cm]{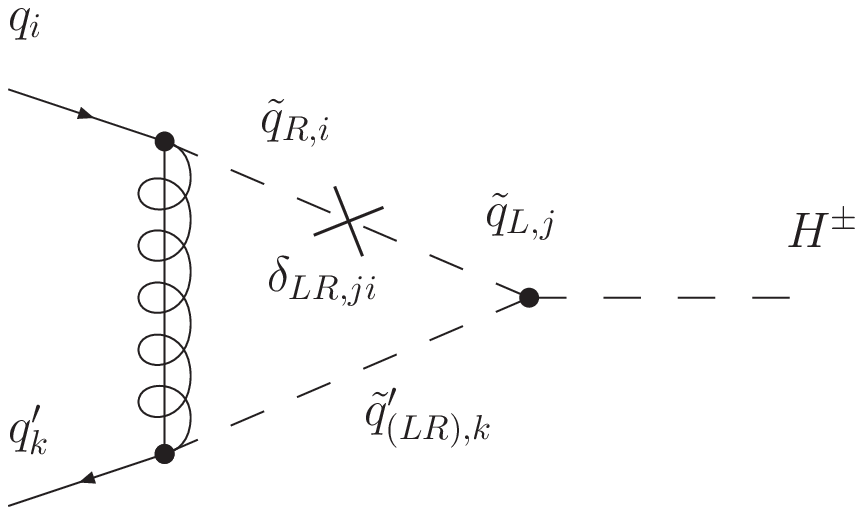}
 \end{center}
 \caption{Feynman diagrams contributing to $q \bar q^\prime \to
   H^\pm $ in the MSSM at tree level and at one-loop level. The last 
   diagram is shown only to illustrate the contributions arising in SUSY 
models beyond MFV. Instead of the
   mass insertion approximation, we use the complete squark-mass matrix for 
   the numerical analyses throughout the paper.}
 \label{fig:feyn_single}
\end{figure}

We start by considering single-charged-Higgs production from quark--antiquark
scattering at the LHC. To leading order this process can be described
by a general type-II two-Higgs-doublet model. In
Fig.~\ref{fig:feyn_single} we show the Drell-Yan-like diagram for $q
\bar q^\prime \to H^\pm$. In the quark mass basis the corresponding
coupling is given by
\begin{equation}  
  \mathcal{L}_{H^\pm qq'} = \sqrt{2}
   V_{ij} \; \bar u_i 
     \left(  \frac{m_{d_j}}{v} P_R \tan \beta 
           + \frac{m_{u_i}}{v} P_L \cot \beta 
     \right) 
   d_j H^+ \; + \; \text{h.c.}
\label{eq:Hqq}
\end{equation}
with the quark fields $u,d$, their masses $m_{u,d}$ and the
CKM matrix elements $V_{ij}$ ($i,j=1,..,3$).
The Yukawas are given in terms
of $v= 2 m_W/g = 246$~GeV, $g=e/\sin \theta_w$. Here $\tan
\beta=v_u/v_d = \left< H_{u}^0 \right>/\left< H_{d}^0 \right>$ denotes the
ratio of vacuum expectation values of the two Higgs doublets.  The
physical charged Higgs scalar in terms of interaction eigenstates is
$H^+=\sin \beta (H_d^-)^* + \cos \beta H_u^+$. The chiral projectors
are defined as $P_{L,R} = (1 \mp \gamma_5)/2$.

Following Eq.(\ref{eq:Hqq}) the amplitude for single-Higgs production
in the type-II two-Higgs-doublet model is proportional to the quark
Yukawa, \ie it is small unless third-generation quarks are
involved.  This chiral suppression is generic and with proper
assumptions survives radiative corrections, like the SUSY-QCD
corrections shown in Fig.~\ref{fig:feyn_single}.  Every
gauge-invariant operator linking quark--antiquark--Higgs fields 
involves a chirality flip,
hence vanishes with $m_q \to 0$ as long as the theory has a chiral
limit.  The renormalizable operators contributing up to dimension 4
are (modulo hermitian conjugates)~\cite{operators}
\begin{equation}  
  \overline Q H^C_u U, \qquad 
  \overline Q H_d^C D, \qquad \qquad \qquad
  \overline Q H_d U,   \qquad 
  \overline Q H_u D,
\label{eq:4Dops}
\end{equation}
where $H^C=i \tau_2 H^*$, and $Q$ and $U,D$ are the $SU(2)$ weak-interaction
eigenstate doublets and singlets, respectively.  
In general, capital letters describe interaction
eigenstates, while small letters denote fermionic
mass eigenstates.

While the first two operators in Eq.(\ref{eq:4Dops}) are the usual
tree-level Yukawa interactions, the second two operators involve the
`wrong' Higgs fields, and do not occur in the plain type-II
two-Higgs-doublet model. Such `wrong' Higgs operators are 
induced by SUSY
breaking. They are proportional to a soft SUSY-breaking parameter
like the gluino mass or an $A$-term and couple the Higgs to a 
squark loop~\cite{bottom_yuk}. Since after spontaneous symmetry breaking
all operators in Eq.(\ref{eq:4Dops}) contribute to the fermion masses, 
the lowest-order relation
between the quark masses and the Yukawas is broken.
This effect becomes numerically important for large $\tan \beta$. 
Since we are 
only interested in small and moderate $\tan \beta$, we 
can safely neglect this effect.
Wrong Higgs couplings increase also with increasing $\mu$ term
\cite{bottom_yuk,mfv}.
As far as the chiral limit of the MSSM is concerned, it is not spoiled 
as long as the soft-breaking
$A^{u,d}$ terms are proportional to the respective quark Yukawa
$Y^{u,d}$. Sometimes, this proportionality is made
explicit by rescaling the $A^{u,d}$ terms and splitting off the Yukawa
matrix as a prefactor.\bigskip

For single-Higgs production we are
limited to the four operators in Eq.(\ref{eq:4Dops}), including necessarily 
some kind of chirality flip. We can build an extended set of operators 
to contain fermions of the same chirality by
simply adding an external gauge field.
We will entertain this possibility  in Section~\ref{sec:hardjet}.

\subsection{Tree-Level Single-Higgs Production}

Because the top quark is too heavy for the gluon to split into a
collinear $t\bar t$ pair at the LHC, the large flavor-diagonal CKM element
$V_{tb}$ does not play any role in single-charged-Higgs production. Instead, in a two-Higgs-doublet model all
interactions in Eq.(\ref{eq:Hqq}) suffer suppression either from
light-flavor quark masses or from quark mixing, parameterized by
CKM entries such as $V_{cb}\simeq
0.04$~\cite{single_higgs}.
Modulo differences in parton densities, from Eq.(\ref{eq:Hqq}) we expect
the largest production rates from 
bottom--charm fusion or strange--charm fusion, since
$m_s V_{cs}$ and $m_b V_{cb}$ are of similar size. Using the
 $\overline{\rm MS}$ quark masses given in Table~(\ref{tab:masses}) 
at typical Higgs-mass scales we find explicitly that the charm--bottom channel
is favored.
Hence, for large enough values of $\tan \beta $ the biggest contribution to
the single-charged-Higgs production cross section  will always be
proportional to $|m_b V_{cb} \tan \beta|^2$.

For example, for $\tan\beta = 7$ and a charged-Higgs mass of $m_{H^\pm}
= 188\gev$ we find LHC cross sections for $H^+$ production of
$\sigma_{cs} = 10.1\fb$ and $\sigma_{cb} = 25.3\fb$. If we neglect the
theoretically poorly defined strange-quark Yukawa, the cross
section decreases to $\sigma_{cs} = 0.56\fb$. Neglecting the
charm Yukawa does not visibly shift $\sigma_{cb}$. The more
we then increase $\tan\beta$, the more we will be dominated by the enhanced
bottom Yukawa in $\bar{b}{-}c$ scattering, in spite of its
strong CKM suppression.\bigskip

The charged Higgs can best be found in $H \to \tau \nu$ decays. In
general, charged-Higgs decays are very similar to $W$ decays, with a
bias towards heavy fermions, because of the Yukawa instead of the
generation-universal gauge couplings. The irreducible background to
our searches is single-$W$ production, mediated by
\begin{equation}
  \mathcal{L}_{W^\pm qq'}= - V_{ij} \frac{g}{\sqrt{2}} \; 
   \bar u_i \, \gamma^\mu P_L \, d_j \, W^+_\mu \; + \; \text{h.c.}.
\label{eq:Wqq}
\end{equation}
This coupling is much bigger than the couplings in Eq.(\ref{eq:Hqq}):
$g/\sqrt{2} \sim {\mathcal{O}}(0.5) \gg Y^{u,d}$. Hence, the
$W^+$ production cross section of $90 \cdot 10^6\fb$
will be a serious challenge to our $H^+$ search in the
two-Higgs-doublet model.
Applying a phase-space cut on the transverse mass $m_T$ of the W~boson
between the Jacobian peaks from $W^\pm$ and $H^\pm$
production, which appear at $m_T=m_W$ and $m_T=m_{H^+}$, respectively,
reduces the W~production cross section by a factor of $10^2$ to
$10^3$. This drastic background reduction is still not enough for a
significant signal/background ratio on the basis of integrated cross
sections. In practice, one thus has to investigate whether a
shoulder from the Higgs Jacobian peak can be resolved in the W~transverse
mass spectrum. The corresponding significance and
further background suppression in this spectrum
can only be seriously investigated by including detector effects such
as efficiencies and momentum smearing, a task that we have to leave
with experimental experts.

\subsection{Loop-Induced Single-Higgs Production in the Flavored MSSM}

Not assuming MFV has serious impact on the production rate
for $q \bar q^\prime \to H^\pm$. Squark loops will weaken the CKM
suppression at the charged-Higgs--bottom vertex through flavor mixing.
The squark mixing matrix collects $D$ and $F$ terms, and soft terms 
from the SUSY-breaking Lagrangian defined in Eq.(\ref{eq:lsoft-superCKM}), 
the latter being susceptible to flavor violation beyond MFV.

The hermitian $6 \times 6$ squark mass matrices ${\mathcal M}_q^2$ for 
up- and down-type squarks 
are composed out of the left- and right-handed blocks $M^2_{q \, AB}$
$(A,B=L,R)$. Each block is a $3 \times 3$ matrix in generation space:
\begin{equation}
{\mathcal M}_q^2 =\left( \begin{array}{cc}
M^2_{q \, LL} & M^2_{q \, LR} \\ 
M_{q \, LR}^{2 \, \dagger} & M^2_{q \,RR}
\end{array} \right), \qquad q=u,d.
\label{eq:mass_matrix}
\end{equation}
The explicit expressions for the $M^2_{q \, AB}$ are given in 
Eq.(\ref{eq:muentries}) and (\ref{eq:muentries-d}). 
Following the quark notation, 
doublet squarks are labeled as $L$, as opposed to 
$SU(2)$ singlets, which are marked as $R$. Squark 
mass matrices are given in the
basis defined by diagonal quark Yukawas (super-CKM basis).\bigskip

$R$-parity-conserving effects beyond MFV are confined to the soft-breaking
Lagrangian. Hence, if we assume MFV the off-diagonal elements 
of the sub-blocks vanish
in the super-CKM basis---modulo tiny effects from renormalization-group 
running~\cite{susyflavorRGE}, that is,  
$M^2_{q \, LL \, ij} \propto m_0^2 \delta _{ij}$,
$M^2_{q \, RR \, ij} \propto m_{0q}^{\prime 2}\delta _{ij}$ and
$M^2_{q \, LR \, ij} \propto m_{q_i} A_0\delta _{ij}$. The SUSY-breaking mass parameters are the  
generation-universal SUSY-breaking scalar masses $m_0^2, m_{0q}^{\prime 2}$ 
and the trilinear term $A_0$. \bigskip

To trace back and discuss the sources of new-physics flavor violation, 
it is useful to define the dimensionless mass insertions~\cite{Hall:1985dx,fcnc-susy} 
\begin{equation}
 \delta_{AB,ij}^q \equiv \frac{M^2_{q \, AB \, ij}}
                              {\tilde m^2}.
 \label{eq:deltas}
\end{equation}
The denominator is the geometric mean $\tilde{m}^2 = m_{A \, ii} m_{B \,jj}$
of the squared scalar masses of 
$\tilde q_{A i}$ and $\tilde q_{B j}$.
Following the previous discussion, the off-diagonal entries of
$\delta_{AB \, ij}^q, i \neq j$ are significant only in non-MFV models 
and
can be complex, inducing CP violation. We 
confine ourselves to real $\delta_{AB}^q$.
Note that in our numerical calculations we 
diagonalize the squark mass matrices and do not
employ a perturbative expansion in the $\delta_{AB}^q$, which would avoid the 
calculation of the squark unitary transformations~\cite{Hall:1985dx}.
We only use the intuitive mass-insertion approximation for 
illustration and order-of-magnitude estimates, see also the appendix of 
Ref.~\cite{Colangelo:1998pm} for formulae.
\bigskip

For our analysis of charged-Higgs production involving squark loops the
three-scalar couplings of
squarks and Higgses are relevant. They stem from three different sources:
\begin{equation}
  \mathcal{L}_{H^\pm \tilde q \tilde q^\prime} = 
     D\text{-term} + F\text{-term} +A\text{-term}.
\end{equation}
The $D$ term couples the charged Higgs to two doublet
squarks, \ie the combination $LL$:
\begin{equation}
   \mathcal{L}_{H^\pm \tilde q \tilde q^\prime}|_D 
     = - \frac{V_{ij} \, g \, m_W}{\sqrt{2}} \sin (2 \beta) \; 
       \tilde u_{L i}^* \tilde d_{L j} H^+ \; 
     + \; \text{h.c.}.
\label{eq:hsqsq}
\end{equation}
This $D$-term contribution is proportional to $\sin(2\beta)$, \ie it
is suppressed by $1/\tan \beta$ for large $\tan \beta$.  Most
importantly, it does not break chirality.

While $D$ terms are gauge couplings, $F$ terms arise from the superpotential.
$F$-term couplings of squarks to Higgses are Yukawa induced and 
involve all four possible combinations of $L$ and $R$ squarks:
\begin{alignat}{5}
 \mathcal{L}_{H^\pm \tilde q \tilde q^\prime}|_F 
   =  \frac{gV_{ij}}{\sqrt{2}m_W} H^+
    & \left[ \; \tilde u_{L,i}^{\ast } \tilde d_{L,j}
              \left( m_{d,j}^2 \tan \beta +m_{u,i}^2 \cot \beta \right)  
      \right.  \notag\\
    & \left. + \tilde u_{R,i}^{\ast} \tilde d_{R,j} m_{u,i} m_{d,j}
              \left( \cot \beta +\tan \beta \right) 
             + \mu m_{d,j} \tilde u_{L,i}^{\ast } \tilde d_{R,j} 
             + \mu  m_{u,i} \tilde u_{R,i}^{\ast } \tilde d_{L,j}
      \right].
\end{alignat}

$A$-terms and soft masses are general soft SUSY-breaking
parameters.
$A$-terms occur with a chirality-flipping squark combination.
We keep the soft terms $A^{u,d}$ with all flavor indices $i,j,k$ 
and without a Yukawa prefactor:
\begin{equation}
 \mathcal{L}_{H^\pm \tilde q \tilde q^\prime}|_A 
    =  \tilde d_{Li} V_{ki}   A^u_{kj} \tilde u_{Rj}^* \cos \beta \, H^+ 
     + \tilde u_{Li} V^*_{ik} A^d_{kj} \tilde d_{Rj}^* \sin \beta \, H^- \;
     + \; \text{h.c.}.
\end{equation}
Both $D$- and $F$-term contributions to the charged-Higgs--squark coupling
are driven by the respective CKM element, as a result of being MFV. This
is different for the $A$-terms induced by SUSY breaking.
We note that our MSSM Lagrangian is defined at the
weak scale, so all parameters are evaluated 
at the scale of the charged-Higgs mass.
\bigskip

We address the
question of how large the $H^\pm$ production cross sections in the MSSM
can be with general flavor after taking into account experimental and 
theoretical constraints.  The dominant one-loop 
corrections are due to the 
gluino vertex and self-energy diagrams shown in Fig.~\ref{fig:feyn_single}
at ${\mathcal{O}}(\alpha_s)$, having the largest gauge couplings.
Beyond MFV, the loop diagrams do not have to include a 
quark mass to yield a chiral operator.
Instead, we can, for example, combine a gaugino mass with a 
left-right mixing $\delta_{LR}$ among the squarks. This combination can 
lift the supersymmetric charged-Higgs
production rate above the two-Higgs-doublet model 
prediction, despite its loop suppression.

We are mainly interested in mixing in the
up-squark sector, because here bigger beyond-Standard-Model 
effects are possible.
As it turns out, the leading contribution to 
charged-Higgs production involves 
$\tilde t_L{-}\tilde u(c)_R$ mixing rather than 
$\tilde t_R{-}\tilde u(c)_L$: while the latter can have particularly 
big impact on rare $K$ and $B$ decays through a modified FCNC $Z$-boson vertex~\cite{susy-Zpenguin,Colangelo:1998pm}, the former 
escapes these constraints, as we will explain in Section~\ref{sec:flavor}.
Contributions not involving a third-generation squark are  
negligible.\bigskip

We first give order-of-magnitude estimates for $H^\pm$~production from
the gluino loop versus the tree-level strange--charm 
${\mathcal{A}_{cs}}$ and bottom--charm ${\mathcal{A}_{cb}}$ 
amplitude discussed in the previous section:
\begin{alignat}{3}
 \frac{\mathcal{A}_{\rm gluino-loop}}{\mathcal{A}_{cs}} \,
    \propto & \frac{\alpha_s}{4 \pi} \;
             \frac{\mgl}{m_c} 
             \delta^u_{LR,3i}, \notag \\
 \frac{\mathcal{A}_{\rm gluino-loop}}{\mathcal{A}_{cb}}  
    \propto & \frac{\alpha_s}{4 \pi} \;
             \frac{\mgl}{V_{cb} m_b} \,
             \frac{1}{\tan^2 \beta} \; \delta^u_{LR,3i},
 \qquad \qquad i=1,2.
\label{eq:ratio}
\end{alignat}
For these ratios we approximate the diagonal CKM elements $V_{tb}, V_{cs} \simeq 1$.
Both ratios in Eq.(\ref{eq:ratio}) exhibit an enhancement of the gluino 
loop that can be as large as ${\mathcal{O}}(10)$ for suitable SUSY
masses and $\tan \beta$. Depending on the initial state, up ($i=1$) or
charm ($i=2$) quarks can induce such a genuine MSSM contribution.

With this estimate in mind we then calculate $H^+$ production from quark--antiquark fusion
including the dominant
squark--gluino loops.
Generally, the amplitude $ \mathcal{A}^{ij}$ for 
$u_{i}\bar{d}_{j}\rightarrow H^{+}$ production can be
written with quark $u_q$ and antiquark $v_q$ spinors as
\begin{alignat}{5}
 \mathcal{A}^{ij}        = \sum_\sigma
  \mathcal{F}^{ij,\sigma} \mathcal{M}^{ij,\sigma} \qquad \text{with} \qquad
 \mathcal{M}^{ij,\sigma} & = \bar{v}_{d_{j}}~P_{\sigma}~u_{u_{i}},   \qquad
 \mathcal{F}^{ij,\sigma} & =\mathcal{F}_0^{ij,\sigma}+\mathcal{F}_S^{ij,\sigma}+\mathcal{F}_V^{ij,\sigma}, ~~ \sigma=L,R.
\end{alignat}
We obtain for the
tree-level contribution $\mathcal{F}_0$ and to leading order in the mass 
insertion expansion for the one-loop self-energy $\mathcal{F}_S$ and 
vertex $\mathcal{F}_V$ contributions  
\begin{alignat}{2}
\mathcal{F}^{ij,R}_0 &=
 \frac{e V_{ij}^*}{\sqrt{2} m_W \sin \theta_w} m_{u_{i}} \cot\beta, \notag\\
\mathcal{F}^{ij,L}_0 &=
 \frac{e V_{ij}^*}{\sqrt{2} m_W \sin \theta_w} m_{d_{j}} \tan\beta, \notag\\
\mathcal{F}_S^{ij,R} & =
 \frac{ \sqrt{2}eV_{3j}^\ast}{m_W \sin \theta_w}
                       \frac{\alpha_s}{4\pi} C_F \frac{m_{\tilde{g}}}{\tan \beta }\ \delta_{LR,3i}^u \;
                       \tilde{m}^2 \;
                       \mathcal{I}_{12}(m_{\tilde{g}},m_{\tilde{q}}), \notag\\
\mathcal{F}_V^{ij,R} & =
 \frac{ \sqrt{2} eV_{3j}^\ast}{m_W \sin \theta_w}
                       \frac{\alpha_s}{4\pi} C_F \left( \frac{m_t^2}{\tan\beta} - m_W^2 \sin(2\beta) 
                       \right) m_{\tilde{g}} \delta_{LR,3i}^u \; \tilde{m}^2 \;
                       \mathcal{I}_{13}(m_{\tilde{g}},m_{\tilde{q}}),
\label{eq:susyloop}
\end{alignat}
where we define
\begin{alignat}{2}
\mathcal{I}_{lm}(m_{\tilde{g}},m_{\tilde{q}}) = \int \frac{d^4 q}{i\pi^2}
\frac{1}{(q^2-m^2_{\tilde{g}})^l (q^2-m_{\tilde{q}}^2)^m}, \qquad
l+m>2.
\end{alignat}
Here, $m_{\tilde{q}}$ denotes  a generic squark mass scale in the loops.
Note that the functions $\mathcal{I}_{lm}$ scale as $M_{SUSY}^{4-2 l -2 m}$
for $M_{SUSY}\sim m_{\tilde{g}} \sim m_{\tilde{q}}$.
The left-chiral contributions $\mathcal{F}_{S,V}^{ij,L}$
vanish if all quarks but the top quark are massless. 
For bottom--up fusion Eqs.~(\ref{eq:susyloop}) show explicitly 
that the gluino loops with $\delta_{LR,3i}^u$ are proportional to 
$V_{tb} m_{\tilde g}$, hence avoid the CKM and quark-mass suppression 
present in the non-SUSY amplitudes.
We note the cancellation of $F$-term ($\propto m_t^2$)
and $D$-term ($\propto m_W^2$) contributions in the vertex correction 
$\mathcal{F}_V^{ij,R}$. Therefore, the self-energies give the dominant 
MSSM contribution with parametric dependence as in Eq.(\ref{eq:ratio}).
Our analytical formulae 
are in agreement with Ref.~\cite{single_higgs}, where
only stop--scharm mixing in $A$-terms has been considered.
\bigskip

As already stressed, we do not use the mass-insertion series
in our numerical analysis presented in the next 
section, but diagonalize the full squark mass matrix.
We also investigate effects of
$LL$ and $RR$ squark mixing with stops.
Specifically we use the program {\sl FeynArts}~\cite{feynarts} for
the generation of graphs and amplitudes, the package
{\sl FormCalc/LoopTools}~\cite{formcalclooptools} for their
evaluation, and the program {\sl HadCalc}~\cite{hadcalc} for the 
convolution with the CTEQ6~\cite{Pumplin:2002vw} parton distribution functions.
Parts of the calculations have been checked with in-house routines.

\subsection{Supersymmetric Parameter Space beyond MFV}
\label{sec:single-higgs-results}

\begin{figure}[t]
 \begin{center}
   \includegraphics[width=7.5cm]{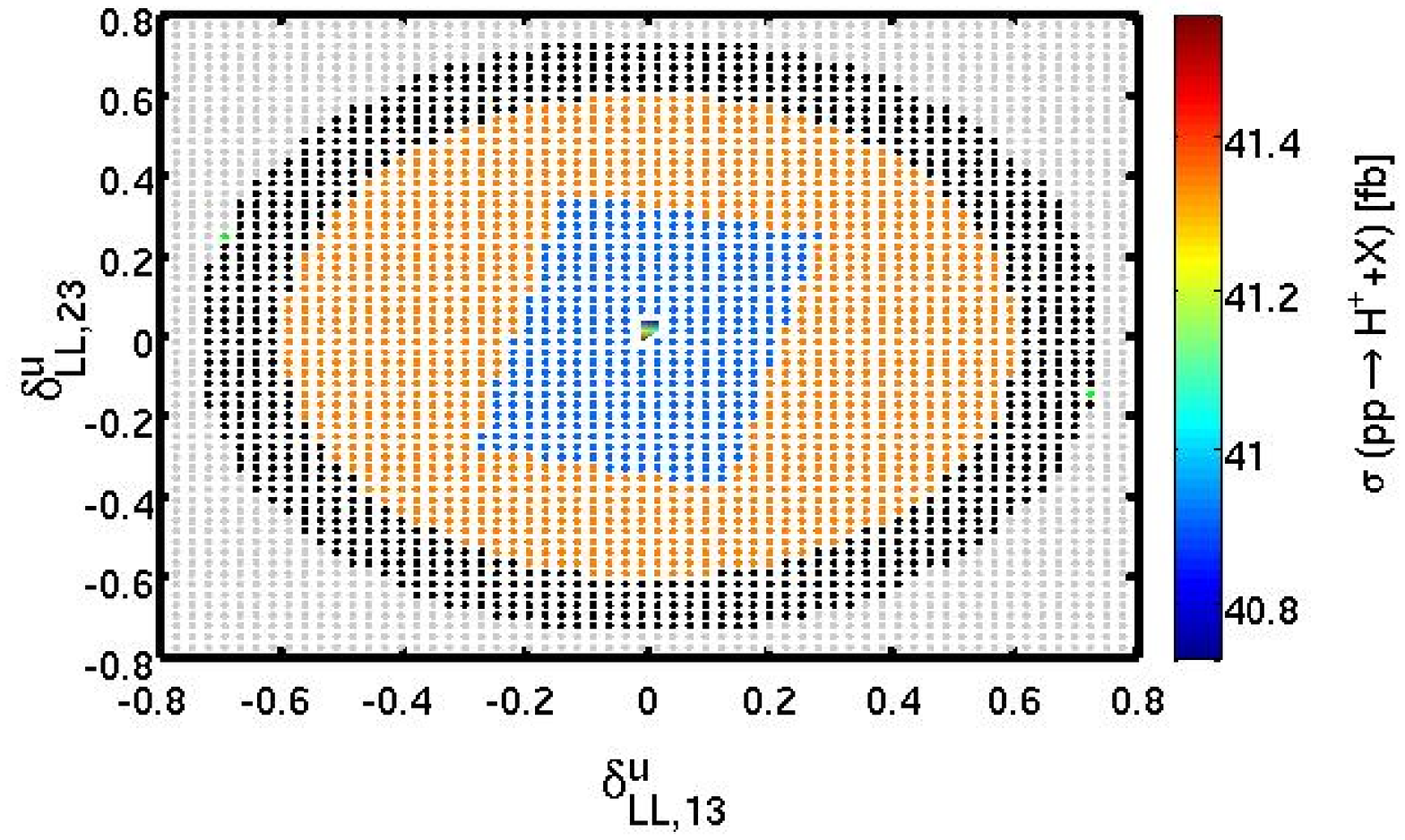} \hspace*{0mm}
   \includegraphics[width=7.5cm]{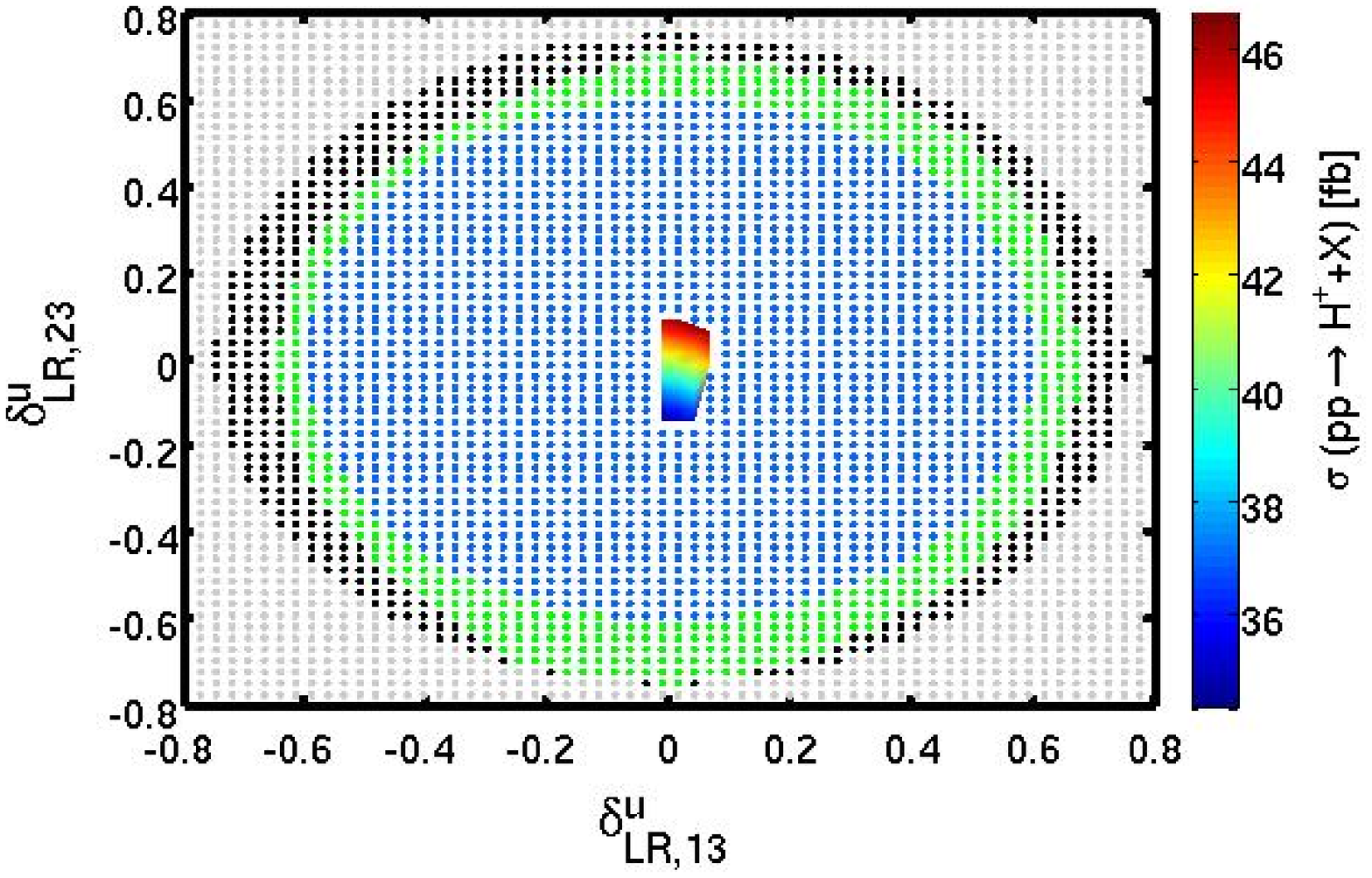} \hspace*{0mm}
   \includegraphics[width=7.5cm]{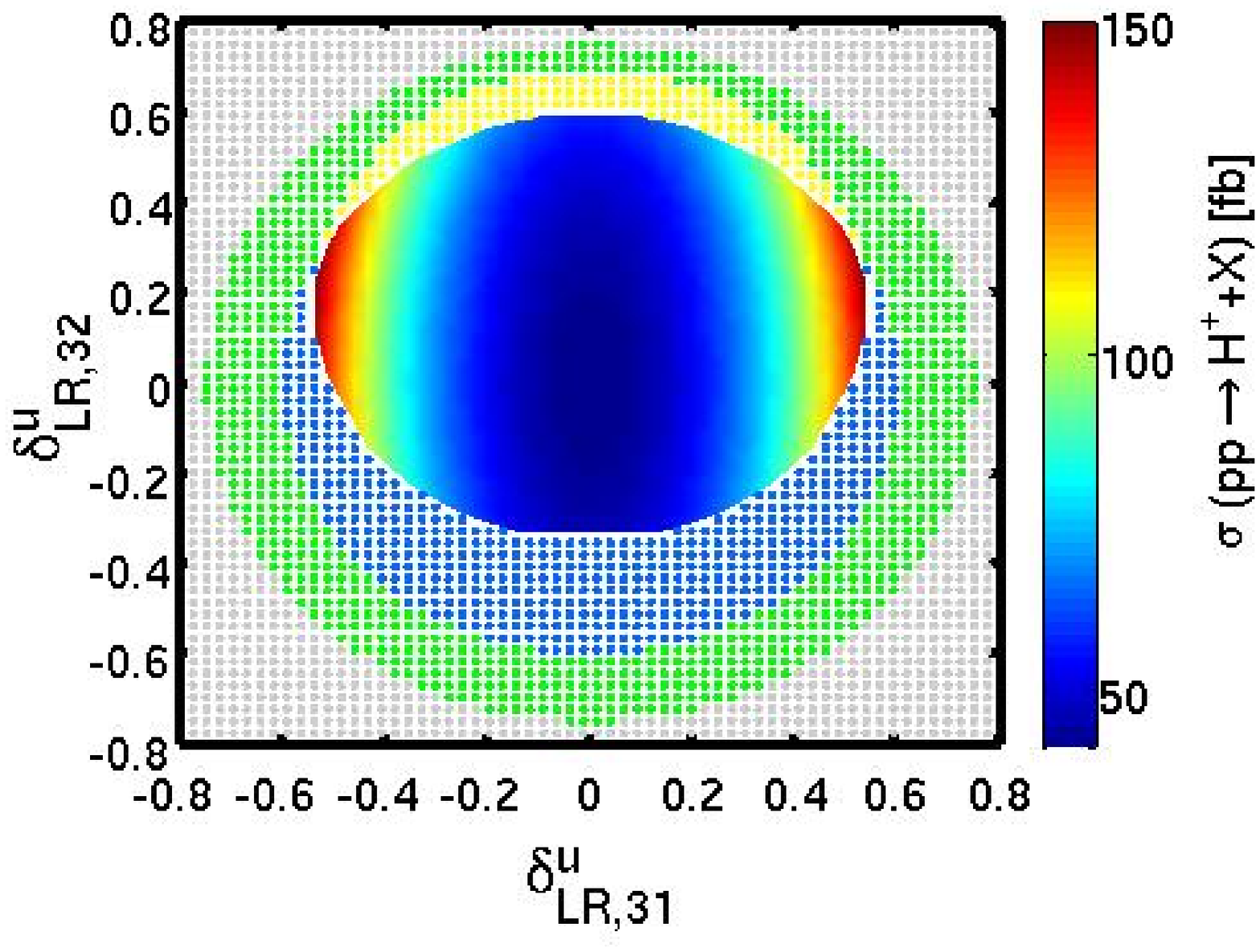} \hspace*{0mm}
   \includegraphics[width=7.5cm]{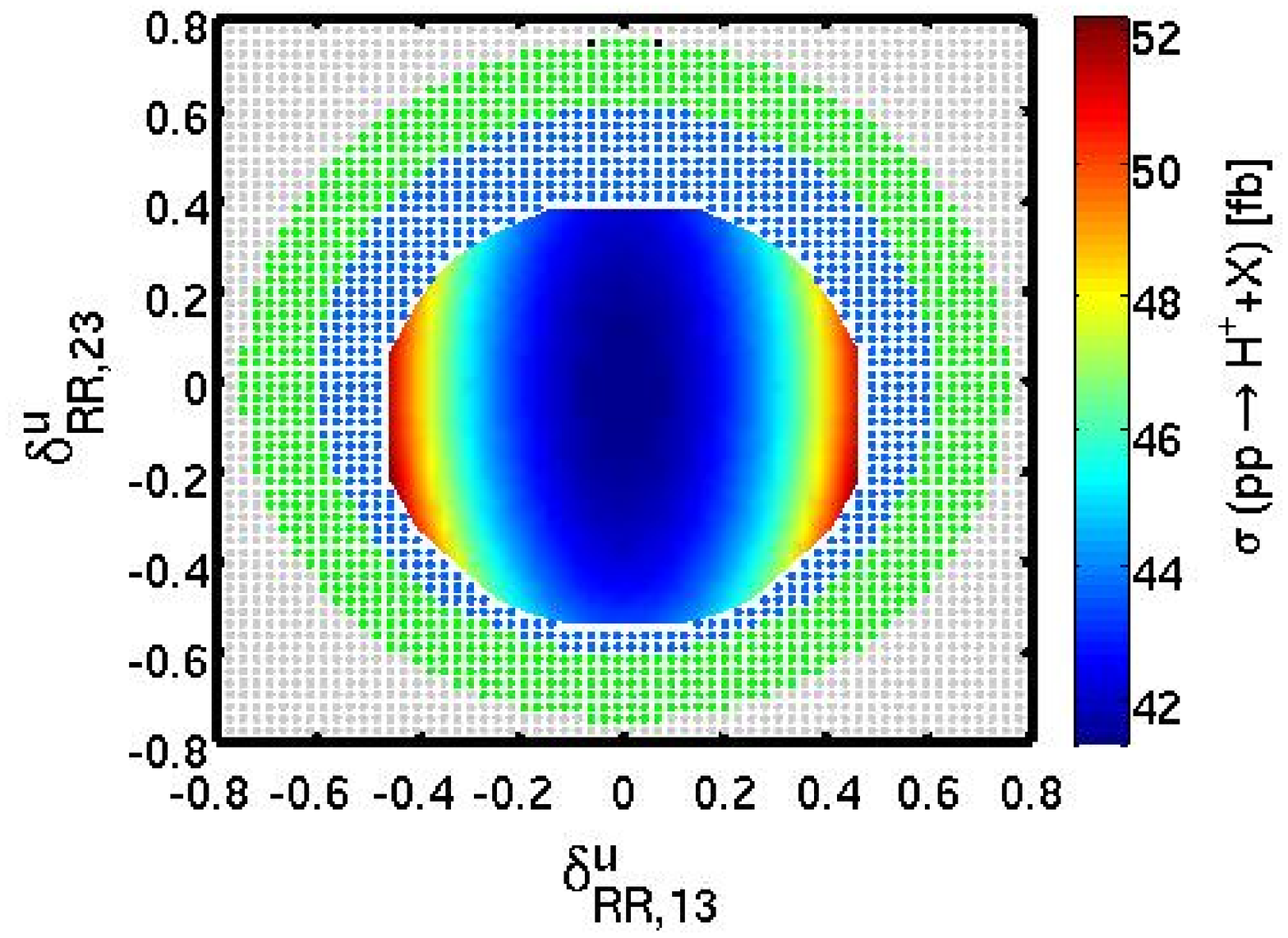}
 \end{center}
 \caption{Single-charged-Higgs production cross sections at the LHC. In the
 rainbow-colored area we include beyond-MFV parameters around the
 lower-mass parameter point (\ref{eq:paras1}). Two
 $\delta_{AB,ij}^u$ are varied in each panel, all others are set to
 zero. The area outside the rainbow is ruled out experimentally.}  
\label{fig:single_scan1}
\end{figure}

To test the effects of flavor structures on the single-Higgs
cross section we start with a generic MFV
SUSY parameter point which
does not violate any current bounds. We then allow for flavor
violation beyond MFV, as illustrated by $\delta_{AB}^q$ as defined in
Eq.(\ref{eq:deltas}). Because of current experimental and theoretical 
constraints discussed in detail in Section~\ref{sec:flavor},
the up-squark parameters $\delta^u_{LR, 3i}$ and  $\delta^u_{RR, 3i}$
involving 1-3 and 2-3 mixing 
are the least constrained and therefore expected to 
cause the biggest effects. 
%The mass insertions $\delta^u_{LL}$ and  $\delta^u_{LR,i3}$ are stronger
%bounded by flavor physics, and are less promising for Higgs searches. 
An insertion of $\delta^q_{LR, ji}$ is illustrated in the last Feynman 
diagram of Fig.~\ref{fig:feyn_single}.
Specifically, we are dealing here with gluino--squark loop contributions
to $u \bar b \to H^+$ and $c \bar b \to H^+$ processes,
which are not CKM suppressed by means of the genuine SUSY flavor breaking
parameters $\delta^u_{3i}$.  
Incoming first- and second-generation quarks have larger luminosities, 
but supersymmetric loop effects are suppressed 
by small squark-mixing 
couplings such as $\delta^{u,d}_{LR \, 11}$ and $\delta^{u,d}_{LR \, 22}$.
\bigskip

Our starting (lower-mass) parameter point is given by:
\begin{alignat}{6}
 \tan\beta    &= 7, &       m_A &= 170\gev,\qquad  &     \mu &=-300 \gev,          \notag\\
 m_{\tilde U_{LL,RR} \, ii}= m_{\tilde D_{LL,RR} \, ii} &=600\gev,   
                    &       M_2 &= 700\gev,        &    \mgl &= 500\gev,           \notag\\
 A^{u,c}      &= 0, & A^{d,s,b} &= 0,              &     A^t &= 1460\gev, 
\label{eq:paras1}
\end{alignat}
where $m_A$ denotes the mass of the CP-odd Higgs leading 
to $m_{H^+}= 188\gev$.
$M_2$ is the SUSY-breaking wino mass.
The diagonal soft-breaking entries in the squark mass matrices 
defined in Eq.(\ref{eq:lsoft-superCKM}) is chosen
universal.
All parameters are given at a scale of order $m_{H^+}$.
The large value of $A^t$
(corresponding to $\delta^u_{LR,33}$) increases the light Higgs mass
to $119.9 \gev$ at two loops~\cite{higgs_mass}. For this parameter
choice the tree-level $H^{+}$ production cross section at the LHC in the
two-Higgs-doublet model is $41.2 \fb$. \bigskip

The production cross sections as a function of the dominant
beyond-MFV mass insertions in the up-sector are shown in
Fig.~\ref{fig:single_scan1}. Beyond-MFV effects can enhance the
single-Higgs rate to values above $100\fb$. The size of the production cross
section is encoded in the rainbow scale in all panels of
Fig.~\ref{fig:single_scan1}, while the parameter choices outside this
area are ruled out. We will discuss the constraints
in more detail in Section~\ref{sec:flavor}. 
The different experimental
constraints impacting the (lower-mass) parameter point shown in
Fig.~\ref{fig:single_scan1} include:
\begin{itemize} 
\item[--] Tevatron searches for mass-degenerate first- and second-generation 
squarks put constraints on their masses \cite{tev_limits}. The 
\D0 analysis has been performed within minimal supergravity, but
assuming similar decay chains the mass bounds hold in a general MSSM context.
In our analysis we require $m_{\tilde q} > 200$~GeV. This rules out
the yellow points.
\item[--] Squark searches and radiative and semileptonic $B$-decay limits
rule out the green points. 
\item[--] Black points are forbidden by the squark-mass limits, $B$ mixing, and radiative and semileptonic $B$ decays. 
\item[--] Blue points indicate a
violation of the radiative and semileptonic $B$ decay bounds only.
\item[--] Orange points correspond to a violation of the $B$ mixing and radiative and semileptonic decay limits.
\item[--] Red points are ruled out by $B$ mixing alone. 
\item[--] Grey points on the outside of the panels indicate a negative squark
mass square after diagonalizing the squark mass matrix. 
\end{itemize}
\bigskip

In Fig.~\ref{fig:single_scan1} we see that the limits on radiative
  and semileptonic decays followed by the 
Tevatron limit on light-flavor 
squark masses define two distinct boundaries of
forbidden parameter space.
After taking into account all limits, the off-diagonal
entry $\delta^u_{LR,31}$ has the strongest impact on the rate. 
It yields a maximal single-Higgs rate for
$|\delta^u_{LR,31}| \sim 0.6$ (third panel). 
The effect of $\delta^u_{LR,32}$ is
similar to $\delta^u_{LR,31}$, except that the process now requires an
incoming $c_R$. The latter is disfavored with respect to
incoming $u_R$ by smaller parton luminosity.
Another MFV pattern that leads to 
an enhanced production rate is $|\delta^u_{RR,13}| \sim 0.5$ 
%in association with $\delta^u_{RR,23}$ 
(fourth panel).
This contribution requires a further $LR$ switch through the squarks,
which could be an $A$- or $F$-term squark--$H^\pm$ coupling.
Since $A^u_{33}$ is typically large (see Eq.(\ref{eq:A33-bound})), 
the relevant combination $\delta^u_{RR,13} \delta^u_{LR,33}$
is numerically sizeable, as is the $F$-term contribution 
$\propto m_t \mu \delta^u_{RR,13}$. 

We recall that
for the numerical analysis we do not use mass
insertions. Otherwise, values of $\delta^u_{AB,ij}$ close to unity
would not give numerically reliable predictions.
Current experimental limits, for
example from squark searches generally imply $\delta^u<1$, but not
necessarily $\delta^u \ll 1$. 
\bigskip

\begin{figure}[t]
 \begin{center}
   \includegraphics[width=8.5cm,height=5.0cm]{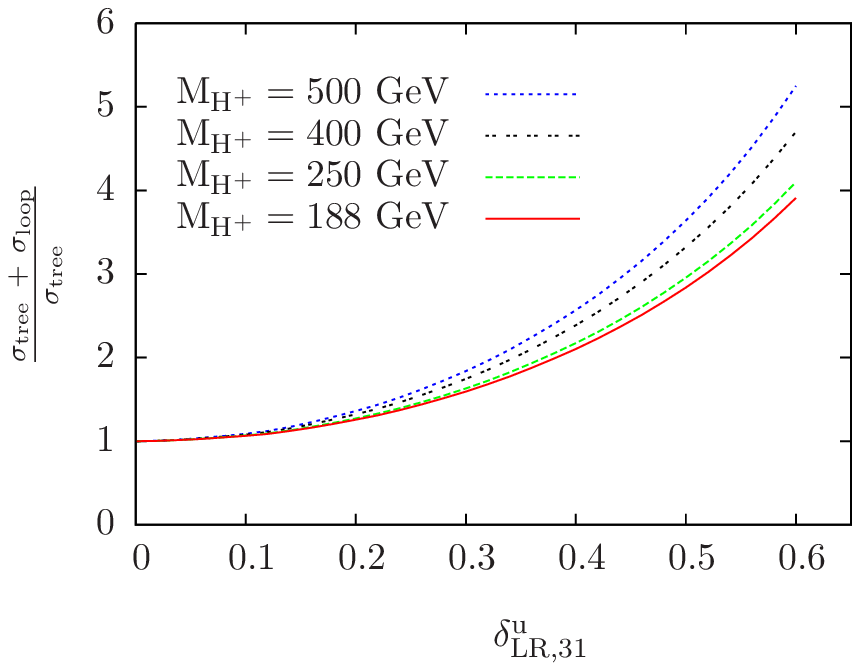} 
 \end{center}
 \caption{Ratio of single-charged-Higgs cross sections including
 supersymmetric beyond-MFV loops vs. in the two-Higgs-doublet
 model. All supersymmetric parameters are given in
 Eq.(\ref{eq:paras1}). All beyond-MFV parameters except for
 $\delta^u_{LR,31}$ are zero.\label{fig:single_kfac}}
\end{figure}

In Fig.~\ref{fig:single_kfac} we show the ratio of the cross section
including beyond-MFV diagrams over the (tree-level)
two-Higgs-doublet-model cross section. At tree level we include all
Standard-Model Yukawas. For the different curves we vary the charged-Higgs mass
between 188 and 500~GeV and find little impact on the relative size
of the contributions. All supersymmetric parameters correspond to the
lower-mass parameter choice (\ref{eq:paras1}). To show the typical
size of the observed effect, we vary the dominant beyond-MFV
parameter $\delta^u_{LR,31}$ within its allowed range, with all other
beyond-MFV parameters zero.  While beyond-MFV diagrams
are formally of higher order, namely supersymmetric 
one-loop corrections, we can already read off Eq.(\ref{eq:ratio})
that they lead to larger effects. This is indeed confirmed by 
Fig.~\ref{fig:single_kfac}. Supersymmetric corrections by factors of
$\mathcal{O}(5)$ are not a reason to worry about the stability of
perturbation theory. Instead, they reflect an additional source of
fermionic mass insertions, which can be large compared to
five light-flavor Yukawas, as discussed at the beginning of this
section.\bigskip

The impact of the experimental squark bound depends crucially on the squark
masses we choose. For an illustration we consider the eigenvalues $m^2_i$ of a ($2\times 2$) mass matrix 
with off-diagonal mixing $\delta$ and a diagonal sfermion mass $m_0$:
\begin{equation}
M^2 = m_0^2 
\left( \begin{array}{ll} 1  &  \delta \\ \delta  &  \Delta \end{array} \right),
\qquad
m^2_i=m_0^2\left(
\frac{1+\Delta}{2} \pm \sqrt{\frac{(1-\Delta)^2}{4} + \delta^2} \right).
\end{equation}
We also allow for non-degenerate diagonal entries 
$\Delta$ not too far from one (as possible in models beyond MFV).
Both $\delta$ and $1-\Delta$ increase the mass splitting.
From an experimental limit $m_i > m_{\rm bound}$ we obtain a bound on $\delta$ as a function of $m_0$:
\begin{equation}
\label{eq:nochmehrdelta}
\delta < \sqrt{(1-r^2)(\Delta-r^2)},  
\qquad
r=\frac{m_{\rm bound}}{m_0} < 1,\Delta
\end{equation}
or simply $\delta < 1-r^2$ for degenerate diagonal 
matrix elements.
For  $\Delta <1 (\Delta >1)$, the constraint on the mixing $\delta$ improves (eases) with respect to the $\Delta=1$ case.
Clearly, for increasing values of the squark mass scale $m_0$ the bound on the
off-diagonal mixing from  
direct search limits weakens and the flavor constraints having a different
decoupling behaviour are of most 
importance.
\bigskip

\begin{figure}[t]
 \begin{center}
   \includegraphics[width=7.5cm,height=5cm]{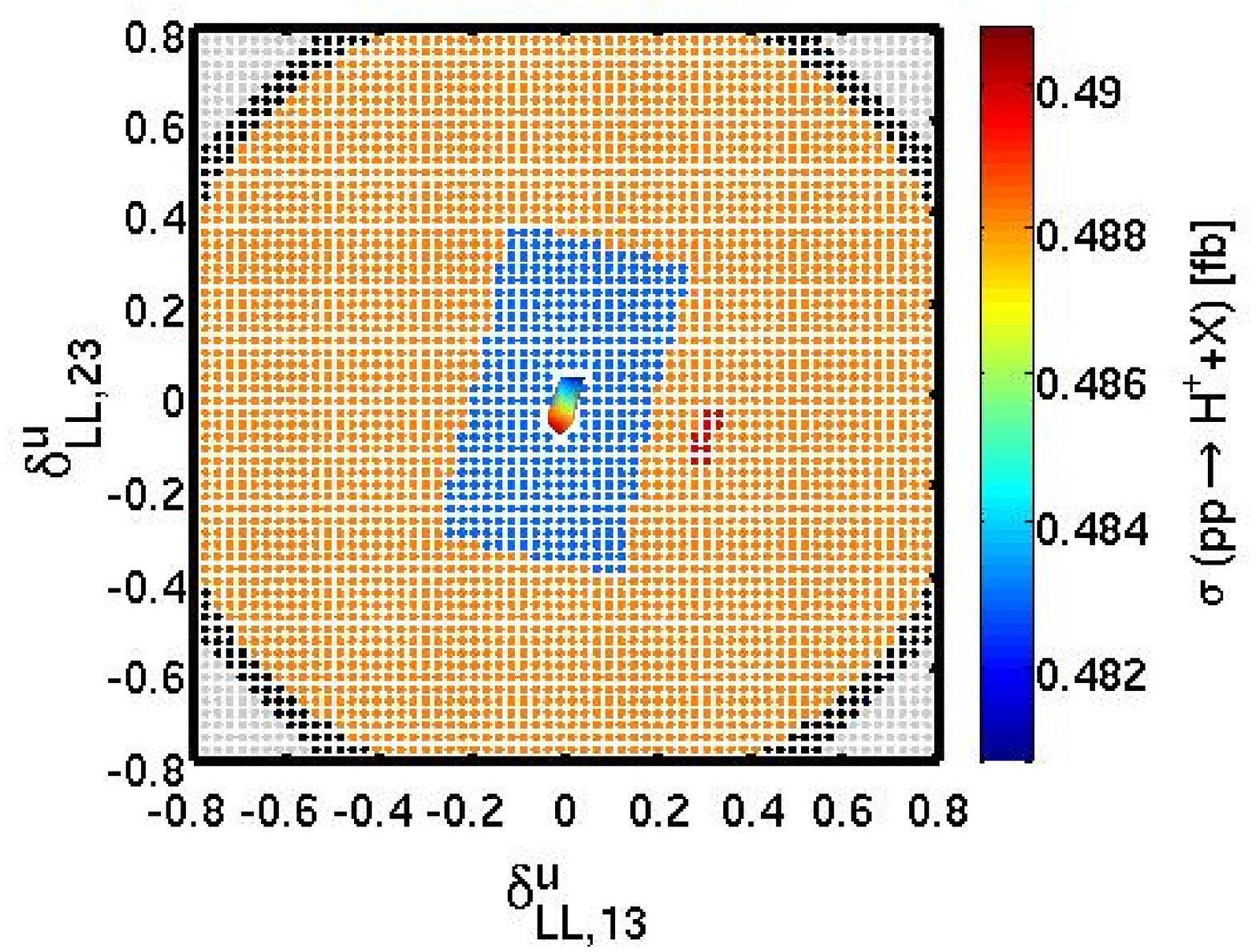} \hspace*{0mm}
   \includegraphics[width=7.5cm,height=5cm]{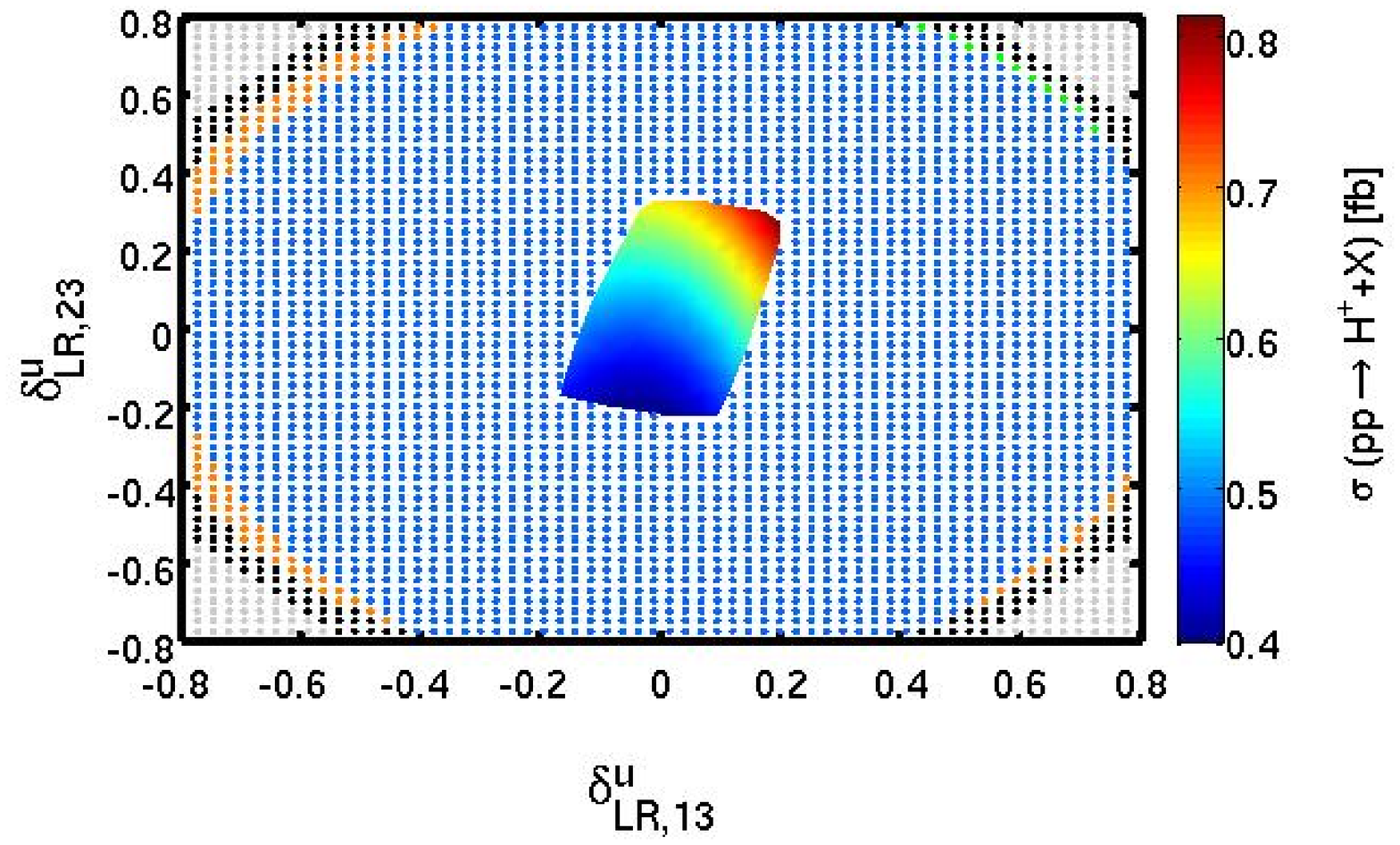} \hspace*{0mm}
   \includegraphics[width=7.5cm,height=5cm]{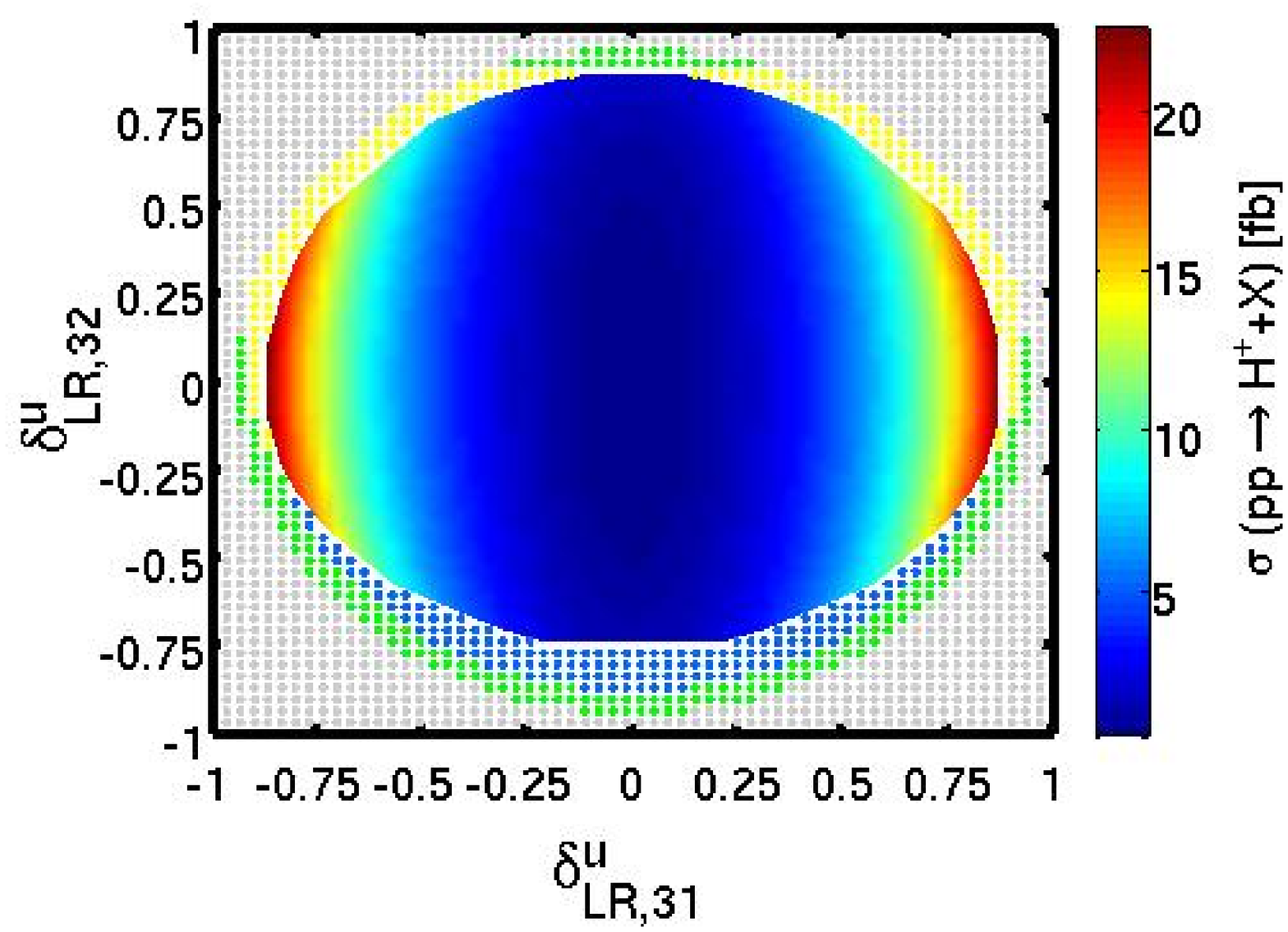} \hspace*{0mm}
   \includegraphics[width=7.5cm,height=5cm]{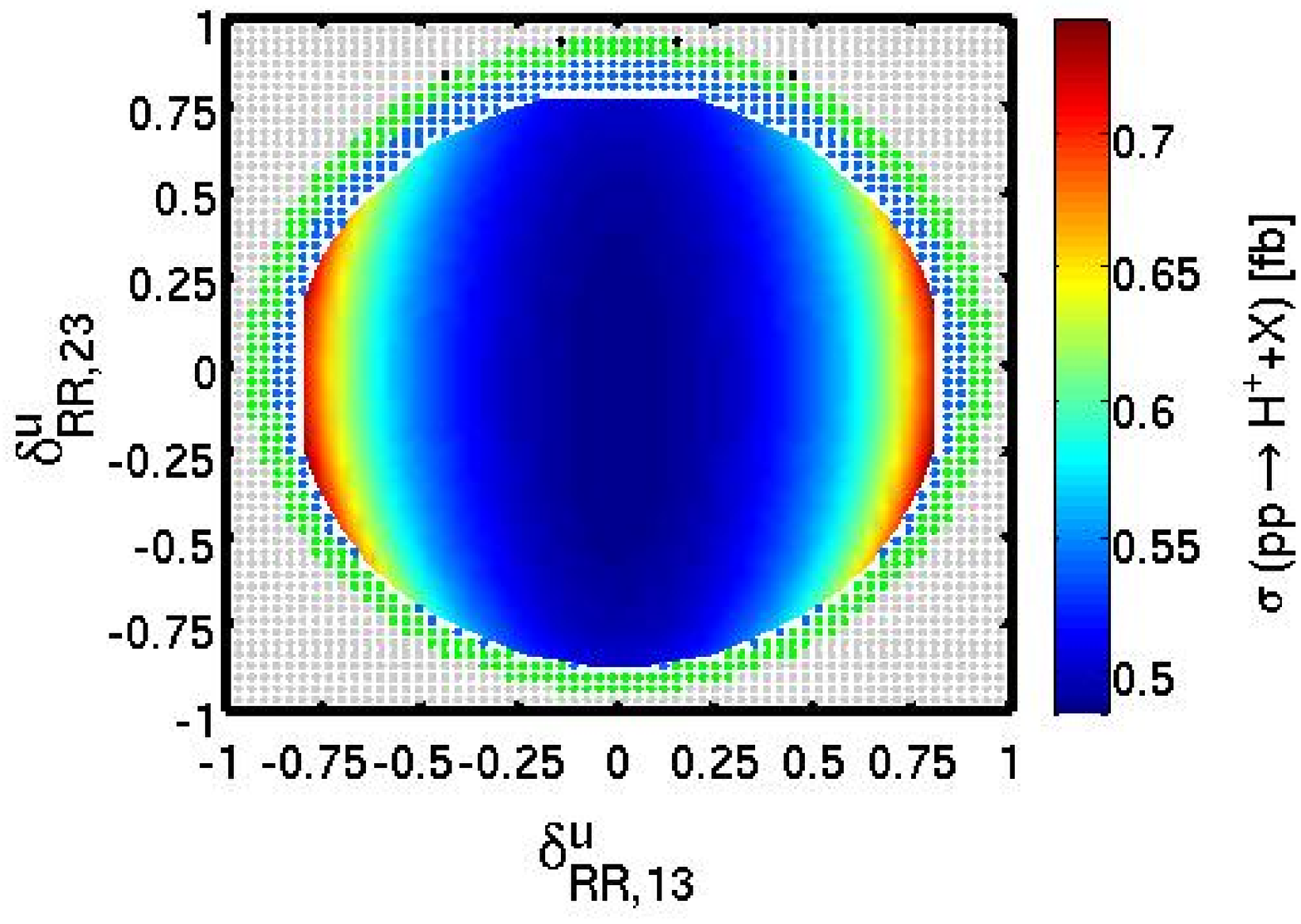}
 \end{center}
 \caption{Single-charged-Higgs production cross sections at the LHC. In the
 rainbow-coded area we include beyond-MFV parameters around the
 higher-mass parameter point (\ref{eq:paras2}). Two
 $\delta_{AB,ij}^u$ are varied in each panel, all others are set to
 zero. The area outside is ruled out.}  \label{fig:single_scan2}
\end{figure}

We can make this explicit by 
slightly increasing the soft-breaking
squark masses and $m_A$, which gives us another (higher-mass) parameter point:
\begin{alignat}{5}
 \tan\beta    &= 5, &       m_A &= 500\gev,\qquad  &     \mu &=-200 \gev,          \notag\\
 m_{\tilde U_{LL,RR} \, ii}= m_{\tilde D_{LL,RR} \, ii} &=800\gev,    
                    &       M_2 &= 500\gev,        &    \mgl &= 500\gev,           \notag\\
 A^{u,c}      &= 0, & A^{d,s,b} &= 0,              &     A^t &= 1260\gev.
\label{eq:paras2}
\end{alignat}
The charged-Higgs mass is now $m_{H^+}=507\gev$.  The tree-level cross
section of $0.48\fb$ in the two-Higgs-doublet model is suppressed by
this heavy final-state mass.  
The color coding for the different
constraints in Fig.~\ref{fig:single_scan2} is the same as in
Fig.~\ref{fig:single_scan1}. 
The basic features of the higher-mass
parameter point and the previously discussed lower-mass parameter
point are similar. The effects of the squark-mixing parameters
$\delta^u_{LR,3i}$ and $\delta^u_{RR,i3}$ can be seen in
Fig.~\ref{fig:single_scan2}: for non-zero values of $\delta^u_{LR, 31}$ the
production rate can be enhanced by about a factor of 40.  As
before, rare $B$ decays strongly limit the parameter space,
complemented by similarly strong limits from the direct searches at
the Tevatron.  The main difference compared to the low-mass point is
the size of the allowed region.  
Instead of a typical value of
$\delta^u \lesssim 0.5$ for 600~GeV squark masses with heavier squarks 
we can have bigger mixing $\delta^u \lesssim 0.8$. 
Note that the shift in the charged-Higgs 
production including flavor structures beyond MFV from the parameters of
Eq.(\ref{eq:paras1}) to Eq.(\ref{eq:paras2}) is mostly due to the heavier
Higgs mass.

%%%%%%%%%%%%%%%%%%%%%%%%%%%%%%%%%%%%%%%%%%%%%%%%%%%%%%%%%%%%%%%%%%%%%%
\section{Flavor Physics Constraints}
\label{sec:flavor}

The vast number of past and ongoing 
flavor-physics measurements has serious impact
on flavor physics at the LHC. From the previous
section and the rough estimate in Eq.(\ref{eq:ratio}) it is obvious that 
without any constraints on squark mixing the charged-Higgs
production rates could be arbitrarily large. However, flavor physics
strongly constrains the structure of the general squark matrices 
in Eq.(\ref{eq:mass_matrix}).  The important parameters are the $LR$ entries 
induced by the $A$-terms $A^{u,d}$
and the corresponding ($LL,RR$)-type mass matrices 
$m_{\tilde U_L, \tilde D_L, \tilde U_R, \tilde D_R}^2$, which occur
at tree level in the SUSY-breaking Lagrangian in the super-CKM basis, which we 
write out in Eq.(\ref{eq:lsoft-superCKM}).\bigskip

We summarize the theoretical and experimental constraints acting on the 
relevant flavored SUSY parameters:
\begin{itemize}
\item[--] $A_{ii}^{u,d}$: diagonal $A$-term entries contribute
  to up- and down-quark masses at one loop:
\begin{equation}
\delta m_{q_i} \propto 
\frac{\alpha_s}{4 \pi} m_{\tilde g} \delta^q_{LR \, ii} , ~~~~~~ (q=u,d; ~~~ i=1,2,3) \; .
\label{eq:massrenorm}
\end{equation}
  For the exact dimensionless loop functions see \eg Ref.~\cite{Pierce:1996zz}.
  We require perturbativity of SUSY-QCD corrections 
  $\delta m_q \lesssim m_q$. This effectively limits the set of
  $A_{ii}^{u,d} = \delta^{u,d}_{LR \, ii} M_{SUSY}^2/v_{u,d}$ to 
  large values of $A_{33}$ only.
  
\item[--] $A_{33}^{u,d}$: loop corrections 
  lift the lighter MSSM Higgs mass from $m_Z$ 
  to above the LEP2 limits. 
  For fixed stop (and for very large $\tan\beta$ also sbottom) masses  
  this translates into an upper bound~\cite{higgs_mass}
  \begin{equation}
\label{eq:A33-bound}
    | A^{u}_{33}|   \lesssim {\mathcal{O}}(3) \; y_t \; M_{\rm SUSY}.
  \end{equation}

\item[--] $A_{13}^{u,d},A_{23}^{u,d}$, $A_{31}^{u,d},A_{32}^{u,d}$: 
general vacuum
  stability constraints limit the inter-generational $A$-terms~\cite{vacuum}:
  \begin{alignat}{5}
    |A_{i3}^{d}|,|A_{3i}^{d}| &\leq 
        \frac{m_{b}}{v_{d}}\sqrt{2 \tilde m(d)^2+ \tilde m(\ell)^2 } \simeq \sqrt{3}  \;  y_b  \; M_{\rm SUSY}, \notag\\
    |A_{i3}^{u}|,|A_{3i}^{u}| &\leq 
        \frac{m_{t}}{v_{u}}\sqrt{2 \tilde m(u)^2+2 \tilde m(\ell)^2 } \simeq \sqrt{3}  \; y_t  \;  M_{\rm SUSY} ,
    \qquad \qquad i=1,2.
\label{eq:vac-stab}
  \end{alignat}
  The masses $\tilde m(u), \tilde m(d), \tilde m(\ell)$ are the mean squark 
and slepton masses defined for Eq.(\ref{eq:deltas}). Because of the
  smaller Yukawas the down
  sector is subject to much stronger constraints than the up sector.
We do not explicitly show analogous bounds for $LR$ mixing among the first and 
second generations, which are strongly suppressed by the
strange and charm Yukawas.

\item[--] $A_{23}^{u,d}, (m_{\tilde U_{L}, \tilde D_{L,R}}^2)_{23}$:
  mixing between the second and third generation in the up and
  in the down sector is constrained by ($b \to s$)-type measurements, like 
  $B \to X_s \gamma$~\cite{b_s_gamma_ex,b_s_gamma_th,data}
and  $B \to X_s \ell^+ \ell^-$~\cite{b_s_gamma_th,bsll,susy-Zpenguin,data} 
at the $B$~factories and the
  $\overline{B}_s{-}B_s$ mixing mass difference $\Delta m_s$ from the
  Tevatron~\cite{bs_mix,Bertolini:1990if,mixing-th}.
Using CDF data only the latter implies  the 90\% C.L. range
\begin{equation}
%\frac{\Delta m_s}{\Delta m_s^{\rm SM}} = 1.00 \pm 0.27
0.56 < \frac{\Delta m_s}{\Delta m_s^{\rm SM}} < 1.44,
\end{equation}
dominated by theory uncertainty.
  To include the constraints from 
  $B \to X_s \gamma$ decays we demand 
   $2.94 \cdot 10^{-4}< {\rm BR}(B \to X_s \gamma) < 4.14 \cdot 10^{-4}$~\cite{b_s_gamma_ex,b_s_gamma_th}.
For  ${\rm BR}(B \to X_s \ell^+ \ell^-)$ we use the data averaged over
electrons and muons for dilepton masses above $0.2$ GeV, leaving us with
$2.8 \cdot 10^{-6}< {\rm BR}(B \to X_s \ell^+ \ell^-) < 6.2 \cdot 10^{-6}$~\cite{data}.
    
\item[--] $A_{13}^{u,d}, (m_{\tilde U_{L}, \tilde D_{L,R}}^2)_{13}$:
  similarly, mixing between the first and third generation in the up and
  the down sector is constrained by $b \to d$ transitions:
  $B \to \rho \gamma$~\cite{rhogamma},
 $B \to \pi \ell^+ \ell^-$ decays~\cite{B2pill} 
and $\Delta m_d$ in $\overline{B}_d{-}B_d$ mixing at 90\% C.L.~\cite{data,Bertolini:1990if,mixing-th}:
\begin{equation}
%\frac{\Delta m_d}{\Delta m_d^{\rm SM}} = 1.00 \pm  0.54 
0.46 < \frac{\Delta m_d}{\Delta m_d^{\rm SM}} < 1.54.
\end{equation}
The first signal of $b \to d \gamma$ transitions has recently been
seen by BaBar and Belle in $B \to (\rho,\omega) \gamma$ decays~\cite{rhogamma}. At 90\% C.L. we 
use $0.63 \cdot 10^{-6} <{\rm BR}(B^0 \to \rho^0 \gamma) < 1.24 \cdot 10^{-6}$.
For semileptonic decays there exists only an upper bound from 
BaBar
${\rm BR}(B \to \pi \ell^+ \ell^-) <9.1 \cdot 10^{-8}$ at 90\% C.L.~\cite{B2pill}.

\item[--] $m_{\tilde{U}_{L}}^{2}$ and $m_{\tilde{D}_{L}}^{2}$: because
  SUSY breaking respects the $SU(2)$ gauge symmetry, the doublet
  soft-breaking masses are identical. Using the definitions (\ref{eq:su2})
  in the super-CKM basis this means
  \begin{equation}
\label{eq:weakisospin}
     m_{\tilde{U}_{L}}^{2}=V\cdot m_{\tilde{D}_{L}}^{2}\cdot V^{\dagger }.
  \end{equation}
Hence,  
universal $m_{\tilde{D}_{L} \, ij}^{2}= m_0^2 \delta_{ij}$
implies 
$m_{\tilde{U}_{L} \, ij}^{2}= m_0^2 \delta_{ij}$, and vice versa.

\item[--] 
  Inter-generational mixing involving the third generation also
  affects the lightest Higgs mass and the $\rho$
  parameter~\cite{schwen,top_fcnc}.  However, the constraints from
  rare decays and direct squark searches are generally
  stronger~\cite{top_fcnc}.

\end{itemize}

Let us summarize the generic features of the above  
constraints: the bounds on down-squark matrices 
$A^d$ and $m^2_{\tilde D_{L ,R}}$ are in general stronger than those for 
up-squark matrices $A^u$ and $m^2_{\tilde U_{L ,R}}$. This is
due to theoretical arguments such as
Eq.(\ref{eq:vac-stab}) and existing data on 
kaon and $B$~FCNCs, which involve down-squark mixing
via strongly coupling gluino loops. Particularly strong bounds follow
from radiative FCNC decays on the chirality-flipping coupling $A^d$ due to
an $m_{\tilde g}/m_b$ enhancement.
Hence, 
we can limit our analysis to up-squark mixing between
different generations
while neglecting down-squark mixing, as long as it
is not required by Eq.(\ref{eq:weakisospin}).
Furthermore, mixing between
first- and second-generation squarks is tightly constrained by 
$K$-physics, \eg, \cite{fcnc-susy,Colangelo:1998pm} and by the 
recent measurements of
$D^0 \bar D^0$ mixing~\cite{Nir:2007ac}. 
We therefore investigate effects on charged-Higgs production from
mixing involving the
third-generation up-type squarks, parameterized by $\delta^u_{i3}$, ($i=1,2$).
Since we do not consider in this work CP violation in the MSSM Lagrangian 
electric dipole moments do not pose constraints on the (real) soft terms.
\bigskip

Among the up-squark parameters, $A^u_{i3}$ 
and $m^2_{\tilde U_{L} \, i3}$ are constrained by data on
$b \to s$ and $b \to d$ transitions, as well as 
by the weak isospin relation (\ref{eq:weakisospin}).
Note that we strictly use the convention
$A_{ij} = A_{L_i R_j} \neq A_{ji}$.
On the other hand, 
$A^u_{3i}$ and $m^2_{\tilde U_{R} \, i3}$ are only very loosely bounded by 
flavor physics, 
the $LR$ chirality flip by Eq.(\ref{eq:vac-stab}).
The reason is that these entries involve right-handed squarks $\tilde u_R$
and $\tilde c_R$; those enter FCNC processes with external 
down quarks only via higgsino vertices proportional to the small up and charm 
Yukawa. To circumvent this Yukawa suppression, we could combine
$\tilde t {-} \tilde u_L (\tilde c_L)$ mixing with a subsequent 
generational-diagonal left-right mixing $\tilde u_R {-}\tilde u_L$ 
$(\tilde c_R {-}\tilde c_L)$. However, generation-diagonal mixing is
strongly constrained by the quark masses (\ref{eq:massrenorm}).

Further constraints on flavor mixing could arise from
$B$-meson decays into $\tau \nu$ final states, which also 
receive contributions from a charged-Higgs exchange. 
$B$-factory experiments determine the  $B_u^- \to \tau \bar \nu$ 
branching ratio to be in agreement with the 
Standard Model, within substantial theoretical and experimental 
uncertainties~\cite{Btaunu}.
Since for our moderate values of $\tan \beta$ 
the $H^\pm$-mediated amplitude  
cannot compete with the tree-level $W$~exchange, 
$B_u^- \to \tau \bar \nu$ data do not put additional constraints on the 
up squarks.

We have seen that  $\delta^u_{LR \, 3 i}$ and $\delta^u_{RR \, i3}$ ($i=1,2$)
are currently the least constrained flavored SUSY couplings. 
Kaon, charm, and $B$-physics experiments are largely
insensitive to the mixing of 
$\tilde u_R$ or $\tilde c_R$ with stops. The latter has impact
on FCNC top decays, see also~\cite{top_fcnc}.
In this work we investigate 
the potential impact of these relevant 
$\delta^u_{3 i}$ on charged-Higgs collider searches.

We implement the constraints on the supersymmetric
flavor sector into our code, and apply
them at 90\% C.L.. Since we are interested in big effects only,
we neglect flavor-diagonal SUSY contributions to the FCNCs. Recall
that we are not in the large-$\tan \beta$ region, where these corrections
can be sizeable.
Thus, we get complicated constraints in the higher-dimensional 
parameter space of the various $\delta$s, which depend on
squark and gaugino masses and wino-higgsino mixing. Note that
all FCNC constraints vanish for mass-degenerate
squarks because of the super-GIM mechanism and reappearance of
flavor symmetry, respectively.

%%%%%%%%%%%%%%%%%%%%%%%%%%%%%%%%%%%%%%%%%%%%%%%%%%%%%%%%%%%%%%%%%%%%%%
\section{Charged-Higgs Production with a hard Jet \label{sec:hardjet}}

\begin{table}
 \begin{center}
 \begin{tabular}{|cc|ll|lll|lll|}
 \hline
 $m_{H^+}$ & $\tan\beta$ & $\sigma_{\rm 2HDM}$ & $\sigma_{\rm 2HDM}^{(m_s=0)}$
 & $\sigma_{\rm MFV}$ & $\sigma_{\rm MFV}^{(m_s=0)}$ & $\sigma_{\rm
MFV}^{(m_q=0)}$
 & $\sigma_{\rm SUSY}$ & $\sigma_{\rm SUSY}^{(m_s=0)}$ & $\sigma_{\rm
SUSY}^{(m_q=0)}$ \\[1mm]
 \hline
 188\gev & 3 & $2.5 \cdot 10^{-1}$ \quad & $1.9 \cdot 10^{-1}$ &
 $2.6 \cdot 10^{-1}$ \quad & $2.0 \cdot 10^{-1}$ &
 $6.7 \cdot 10^{-4}$ \quad & $14.3 \cdot 10^{0}$ &
 $14.2 \cdot 10^{0}$ \quad & $13.9 \cdot 10^{0}$ \\
 188\gev & 7 & $9.9 \cdot 10^{-1}$ & $6.0 \cdot 10^{-1}$ &
 $1.1 \cdot 10^{0}$ & $6.5 \cdot 10^{-1}$ &
 $1.5 \cdot 10^{-4}$ & $4.6 \cdot 10^{0}$ &
 $4.4 \cdot 10^{0} $ & $3.0 \cdot 10^{0}$ \\[2mm]
 400\gev & 3 & $4.0 \cdot 10^{-2}$ & $3.0 \cdot 10^{-2}$ &
 $4.2 \cdot 10^{-2}$ & $3.2 \cdot 10^{-2}$ &
 $4.2 \cdot 10^{-4}$ & $2.4 \cdot 10^{0}$ &
 $2.4 \cdot 10^{0}$ & $2.3 \cdot 10^{0}$ \\
 400\gev & 7 & $1.6 \cdot 10^{-1}$ & $1.0 \cdot 10^{-1}$ &
 $1.7 \cdot 10^{-1}$ & $1.1 \cdot 10^{-1}$ &
 $9.1 \cdot 10^{-5}$ & $7.9 \cdot 10^{-1}$ &
 $7.3 \cdot 10^{-1}$ & $5.4 \cdot 10^{-1}$ \\[2mm]
 500\gev & 3 & $2.0 \cdot 10^{-2}$ \quad & $1.44 \cdot 10^{-2} $ &
 $2.1 \cdot 10^{-2}$ & $1.5 \cdot 10^{-2}$ &
 $3.5 \cdot 10^{-4}$ & $1.3 \cdot 10^{0} $ &
 $1.3 \cdot 10^{0} $ & $1.2 \cdot 10^{0} $ \\
 500\gev & 5 & $4.2 \cdot 10^{-2}$ & $2.7 \cdot 10^{-2}$ &
 $4.4 \cdot 10^{-2}$ & $2.9 \cdot 10^{-2}$ &
 $1.4 \cdot 10^{-4}$ & $5.5 \cdot 10^{-1}$ &
 $5.4 \cdot 10^{-1}$ & $5.0 \cdot 10^{-1}$ \\
 500\gev & 7 & $7.9 \cdot 10^{-2}$ & $5.1 \cdot 10^{-2}$ &
 $8.4 \cdot 10^{-2}$ & $5.4 \cdot 10^{-2}$ &
 $7.6 \cdot 10^{-5}$ & $4.0 \cdot 10^{-1}$ &
 $3.7 \cdot 10^{-1}$ & $2.8 \cdot 10^{-1}$ \\
 \hline
 \end{tabular}
 \end{center}
 \caption{Cross sections (in fb) for
 the associated production of a charged Higgs with a hard jet:
 $p_{T,j}>100\gev$.
 The label 2HDM denotes a two-Higgs-doublet of type~II, while 
 MFV and SUSY refer to the complete set of supersymmetric
 diagrams, assuming MFV and beyond.
 The SUSY parameters are given in Eq.(\ref{eq:paras1}). Beyond MFV 
 we choose $\delta^u_{LR,31} = 0.5$.
 The label $(m_s=0)$ means a zero
 strange Yukawa, $(m_q=0)$ indicates that
 all quark (except top) Yukawas are neglected. In this case only $D$-term
 couplings contribute within MFV.  \label{tab:hj}}
\end{table}

The generic chiral suppression that characterizes single-Higgs production and
limits the cross section at tree level can be removed by adding an external
gluon to the operator basis.
Such operators can be of the
form $i\,\overline Q \gamma_\mu Q \, H_u
\overset{\leftrightarrow}{D^\mu} H_u^C$, leading 
to higher-dimensional $q \bar q'Hg$ operators after electroweak
symmetry breaking. For a detailed discussion of the operator basis see 
\eg Ref.~\cite{operators}. It is of course by no means guaranteed that
all possible operators are actually induced at the one-loop level in the MSSM. 
Some operators can be forbidden by symmetry.\bigskip

To probe such operators at the LHC, we study charged-Higgs searches in
association with a hard jet. Simple diagrams for this process can be
derived from all single-Higgs production diagrams just radiating an
additional gluon. The infrared divergences that occur for soft jets or
jets that are collinear to the incoming partons are excluded by requiring
a hard jet with transverse momentum $p_{T,j}>100\gev$.

Similar to single-Higgs production we are interested in supersymmetric
loop corrections in and beyond MFV. Such diagrams are suppressed by
$\alpha_s$, which means that when comparing them to tree-level rates
in the two-Higgs-doublet model we should consistently compute the
next-to-leading-order corrections to the tree-level. On the other
hand, we know from single-Higgs production that the flavor effects we
are interested in can be much larger than we expect next-to-leading-order 
QCD effects to be. Therefore, we ignore all gluonic
next-to-leading order corrections to charged-Higgs production with a
hard jet and limit our analysis to tree-level rates in the
two-Higgs-doublet model and additional supersymmetric one-loop
corrections with and without the MFV assumption.
Final-state top quarks introducing a top Yukawa
we do not consider, because they lead to a completely different
signature. \bigskip

Unlike the amplitude for single-Higgs production, the amplitude for
Higgs production with a hard jet does not vanish in the limit of zero
quark masses, even in a two-Higgs-doublet model. There,
contributions to non-chiral operators arise at two loops, when the
charged Higgs couples to neutral Higgses and gauge bosons and not
directly the fermions. However, such non-supersymmetric two-loop
contributions have to be compared with the tree-level processes: modulo
parton-density effects the bottom Yukawa competes with
the weak coupling multiplied with two loop factors
$(g^2/(16\pi^2))^2 \sim 10^{-5}$, so we can safely 
neglect the two-loop non-chiral contributions as well.\bigskip

In the first two columns of Table~\ref{tab:hj} we list the 
hadronic tree-level cross sections
for charged-Higgs-plus-jet production 
for a non-supersymmetric two-Higgs-doublet type-II model. 
Sharing this feature with the single-Higgs production discussed 
previously, the bottom Yukawa in the absence of a final-state
top appears with a
CKM suppression, leading to effective couplings of the order
$m_b V_{cb} \sim m_s V_{cs}$.
Parton densities will lightly enhance the strange-quark 
contribution compared to incoming bottoms. 
This numerical behavior is what we see in
Table~\ref{tab:hj}: at tree level the strange 
and the bottom Yukawas
contribute at a comparable rate.\bigskip

In the following analysis of charged-Higgs-plus-jet production in supersymmetry
we also include Higgs decays. As
long as the Higgs mass is small, $m_{H^+} \lesssim 200\gev$, the Higgs decay into
a hadronic $\tau$ lepton is the most
promising~\cite{dec_tau_ph,dec_tau_ex}. For the lower-mass 
parameter point in Eq.(\ref{eq:paras1}) with its charged-Higgs mass of
188~GeV,  we find 
${\rm BR}(H^- \to \tau \bar\nu) = 71\%$, with a 
taggable hadronic $\tau$ branching ratio of around roughly two thirds~\cite{data}. 
The dominant background to this signature is clearly $W$+jet production,
again with the $W$ decaying to a hadronic~$\tau$.
For $p_{T,j} >100 \gev$ the corresponding cross section is about 1 nb.

\subsection{MFV Loops and Decoupling}

\begin{figure}[t]
\centerline{\footnotesize \input{udhg_sqcd.tex} }
\vspace*{-2em}
\caption{SUSY QCD diagrams for  
$u \bar d\to g H^\pm$ with MFV and massless quarks.}
\label{fig:feyn_higgsjet}
\end{figure}
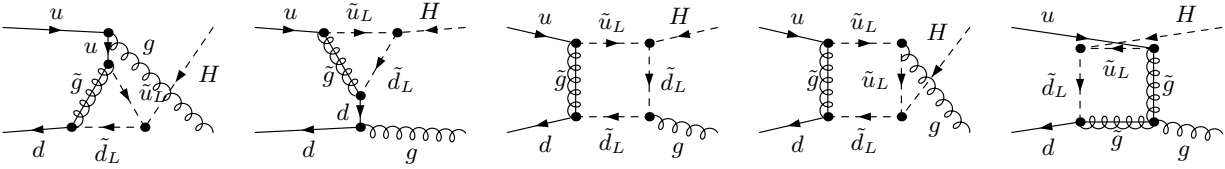

The difference between the two-Higgs-doublet model and
higher-dimensional operators realized by supersymmetric one-loop
diagrams are additional Higgs couplings to squarks. We discuss those in
Section~\ref{sec:singleH}: assuming MFV, $F$-term and $A$-term
couplings of the Higgs to two squarks are proportional to the quark
masses, which means that supersymmetric one-loop amplitudes are
expected to be of the size of typical supersymmetric NLO
corrections. In contrast, the $D$-term couplings shown in
Eq.(\ref{eq:hsqsq}) are gauge couplings, which means they could be
considerably larger than light-flavor Yukawas. This formal enhancement is a
novel aspect of associated charged-Higgs production with a hard
jet. For single-Higgs production, such $D$-term couplings do not
contribute, because they are $LL$ diagonal in the squarks
and do not introduce the necessary left-right mixing without
additional beyond-MFV contributions.

Since it circumvents the Yukawa suppression of the amplitude in the two-Higgs-doublet model,
one might expect the $D$-term contribution to charged-Higgs production with
a jet to be significant.
The corresponding gluino--squark diagrams are shown in 
Fig.~\ref{fig:feyn_higgsjet}. Chargino and neutralino loops are neglected 
due to their smaller gauge coupling. At the LHC, a mixed
quark--gluon initial state yields the largest cross section
for heavy-particle production, because it is a good compromise between
the high-$x$ valence quark parton densities and the large gluon
luminosity at lower $x$.\bigskip

The general amplitude for the partonic subprocess $u_i + \bar{d}_j
\rightarrow H^+ + g$ is given in terms of form factors as
\begin{equation}
  \mathcal{A}^{ij } = \sum_\sigma 
        \sum_{k=1}^6 \; \mathcal{F}_k^{ij,\sigma } 
                                             \mathcal{M}_k^{ij,\sigma} \; ,
~~~~\sigma=L,R
\end{equation}
with 12 standard matrix elements~\cite{denner}
\begin{alignat}{6}
 \mathcal{M}^{ij,\sigma}_1 &= \bar{v}_j(p_2) \slashed{\varepsilon} P_\sigma u_i(p_1)                      \; ,
&\mathcal{M}^{ij,\sigma}_4 &= \bar{v}_j(p_2) \slashed{k}_2 \slashed{\varepsilon} P _\sigma  u_i(p_1)      \; ,\notag\\
 \mathcal{M}^{ij,\sigma}_2 &= \bar{v}_j(p_2) \slashed{k}_2 P_\sigma u_i(p_1) \; (\varepsilon \cdot p_1)   \; ,\qquad \qquad
&\mathcal{M}^{ij,\sigma}_5 &= \bar{v}_j(p_2) P_\sigma u_i(p_1) \; (\varepsilon \cdot p_1)                 \; ,\notag\\
 \mathcal{M}^{ij,\sigma}_3 &= \bar{v}_j(p_2) \slashed{k}_2 P_\sigma u_i(p_1) \; ( \varepsilon \cdot p_2)  \; , 
&\mathcal{M}^{ij,\sigma}_6 &= \bar{v}_j(p_2) P_\sigma u_i(p_1) \; (\varepsilon \cdot p_2)                 \; .
\end{alignat}
The momenta are assigned as $u_i(p_1), \bar{d}_j(p_2), H^+(k_1),
g(k_2)$, the corresponding Mandelstam variables are 
$s=(p_1+p_2)^2$, $t=(p_1-k_1)^2$, $u=(p_1-k_2)^2$, and
$\varepsilon$ is the polarization vector of the gluon.
$SU(3)$ gauge invariance implies a Ward identity, which means the
amplitude has to vanish if we replace the external gluon polarization vector 
by the gluon momentum. This relates the different form factors
to each other:
\begin{equation}
  \mathcal{F}_1^{ij,\sigma} 
+ \mathcal{F}_2^{ij,\sigma} \left( p_1 \cdot k_2 \right) 
+ \mathcal{F}_3^{ij,\sigma} \left( p_2 \cdot k_2 \right) 
=0 ~, ~~
 \mathcal{F}_5^{ij,\sigma} \left( p_1 \cdot k_2 \right) 
+ \mathcal{F}_6^{ij,\sigma} \left( p_2 \cdot k_2 \right) 
=0 .
\label{eq:gaugeinvariance}
\end{equation}
\bigskip

Numerical results for hadronic charged-Higgs-plus-jet production in MFV 
are presented in the second set of rows in Table~\ref{tab:hj}. We  
show the cross sections for the lower-mass parameter point
(\ref{eq:paras1}), and vary $\tan \beta$ and $m_{H^+}$ as indicated. 
All supersymmetric loop diagrams share the usual loop-suppression
factors. This means that in MFV the additional
supersymmetric contributions are unlikely to numerically dominate over 
the tree-level rates in the two-Higgs-doublet model.

The purely $D$-term-induced contributions ($m_q=0$) are numerically
negligible, despite the fact that they avoid the chiral suppression.
The reason is that the loop amplitude suffers from an additional mass
suppression $1/M_{\rm SUSY}^4$ in the limit
$m_{H^+}^2, m_W^2, s, |t|,|u| \ll M_{\rm SUSY}^2$,
where $M_{\rm SUSY}$ denotes a common squark and gluino mass.
It is not easy to see this decoupling in the explicit analytical result
for the form factors, which is given in the appendix.
The decoupling can be understood by applying power-counting in $M_{\rm
SUSY}$ to the individual form factors in combination with
the gauge-invariance relation (\ref{eq:gaugeinvariance}).
Naive power-counting suggests a scaling $\propto 1/M_{\rm SUSY}^2$ for
the form factors, but including the Lorentz structure of the
loop integrals reveals that only $\mathcal{F}_1^{ij,\sigma}$
can receive contributions of this order, while the other form factors
scale $\propto 1/M_{\rm SUSY}^4$. Thus, Eq.(\ref{eq:gaugeinvariance})
shows that all contributions in $\mathcal{F}_1^{ij,\sigma}$
proportional to $1/M_{\rm SUSY}^2$ have to cancel. We have explicitly
verified this fact by performing
a large-mass expansion \cite{Smirnov:1996ng} of the SUSY-QCD diagrams
in the relevant SUSY masses, confirming that
the one-loop amplitude with $D$-term couplings scales like
\begin{equation}  
 \mathcal{A}^{\tilde{q} \tilde{g}}_{\rm D-term} \propto 
  \frac{g_s^3 g}{M_{\rm SUSY}^4} \sin (2\beta) \; .
\label{eq:hj_amp1}
\end{equation}
This means that the pure $D$-term contribution to the
charged-Higgs plus a hard jet cross section  
decouples as $\sigma \propto 1/M_{\rm SUSY}^8$, four powers of $M_{\rm SUSY}$
faster than the leading supersymmetric cross section (with finite
quark masses or not imposing MFV).
\bigskip

\begin{figure}[t]
 \begin{center}
   \includegraphics[width=7.0cm]{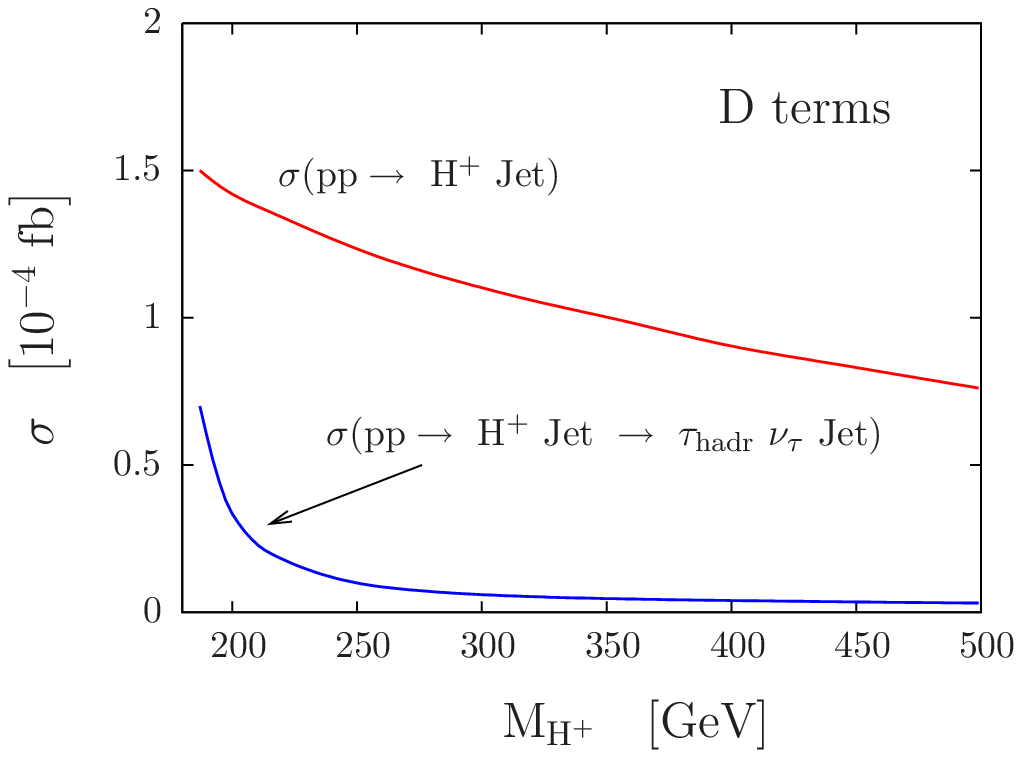} \hspace*{10mm}
   \includegraphics[width=7.0cm]{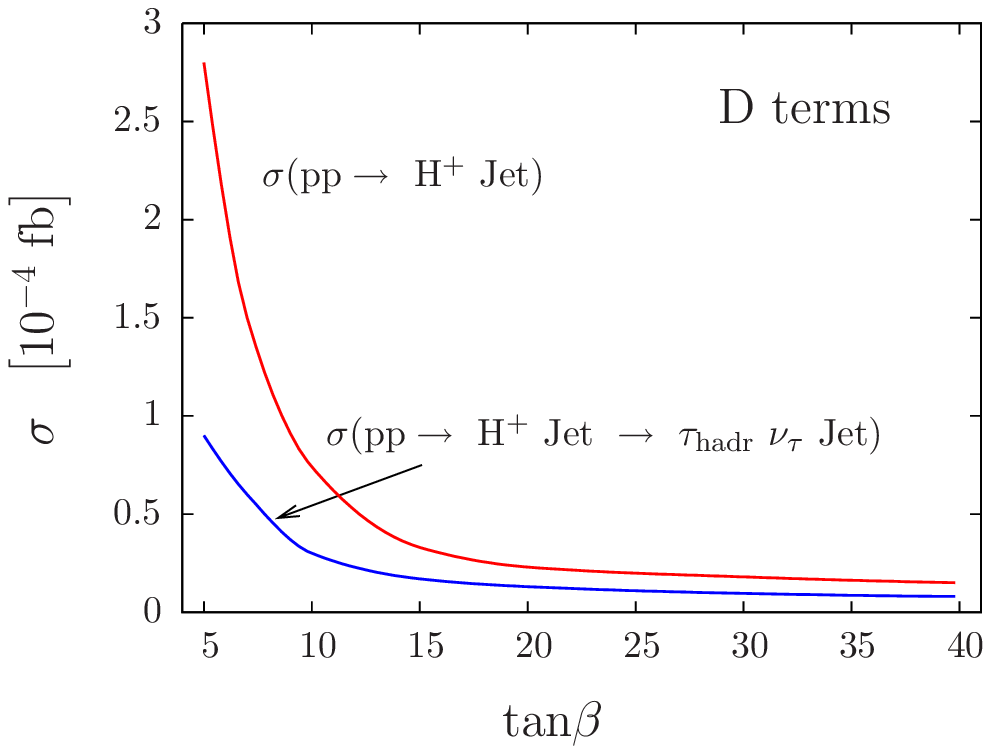} 
   \includegraphics[width=7.0cm]{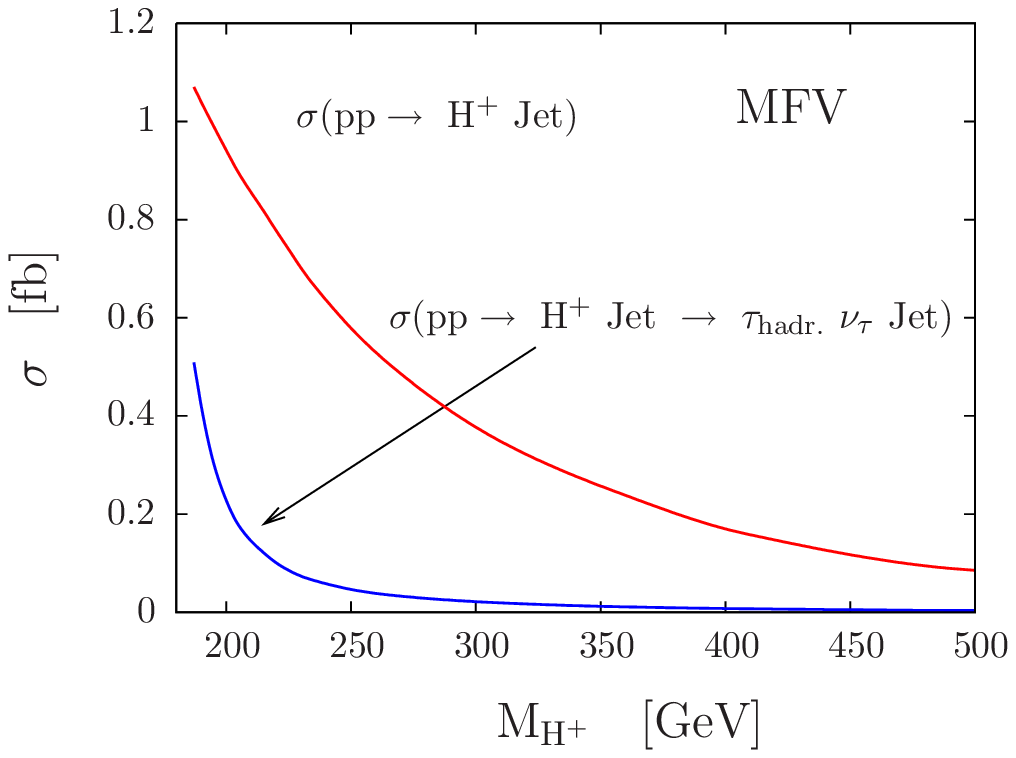} \hspace*{10mm}
   \includegraphics[width=7.0cm]{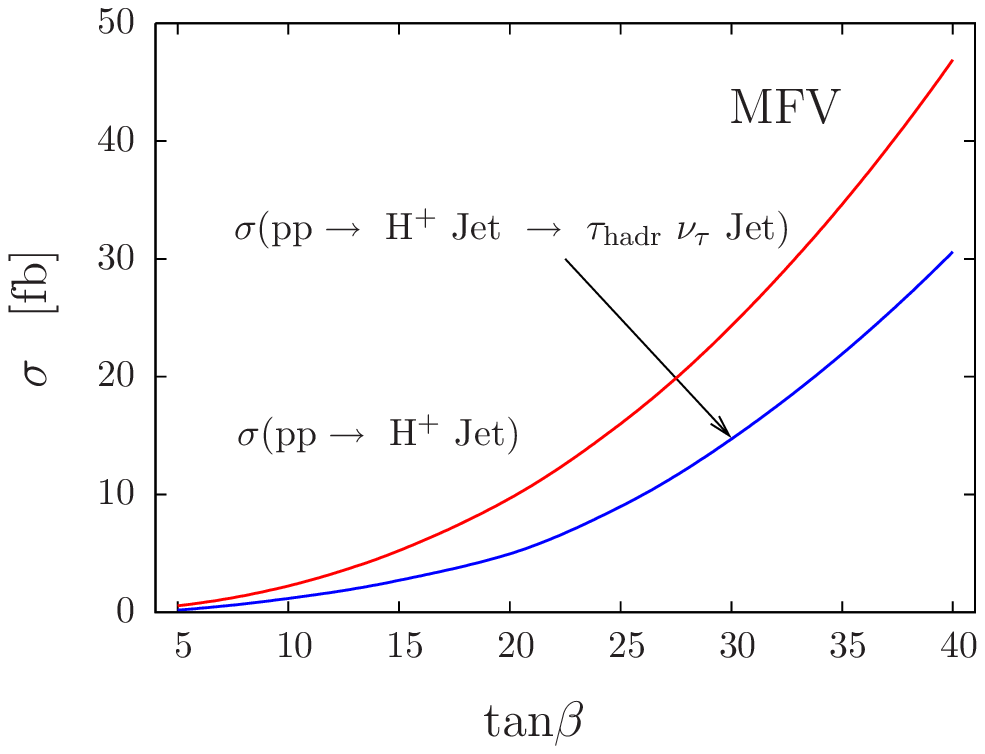}
 \end{center}
 \caption{Production rates for a
 charged Higgs with a hard jet including SUSY loops. Upper: 
 assuming MFV with $D$-term contributions only.
 Lower: assuming MFV, but including all couplings.
 The supersymmetric parameters 
 are listed in Eq.(\ref{eq:paras1}). We
 vary only the Higgs-sector parameters via 
 the charged-Higgs mass and $\tan\beta$.
 We include the 
 Higgs decay into a hadronic 
 $\tau$ plus $\nu_\tau$ (lower curves). 
 \label{fig:asso_mfv}}
\end{figure}

Comparing the different Yukawas, Table~\ref{tab:hj} also shows that
similarly to single-Higgs production and to the two-Higgs-doublet model 
the contribution of the strange
Yukawa is non-negligible.
To see the typical behavior of the MFV amplitudes we show
the LHC cross section of a charged Higgs boson with a hard jet
$(p_{T,j}>100\gev)$ as a function of $m_{H^\pm}$ and $\tan \beta$
in Fig.~\ref{fig:asso_mfv}, with and without the branching ratio to 
hadronic $\tau$'s. The upper panels show the contributions from $D$ terms 
only, while the lower panels include all supersymmetric MFV contributions.
We start from the lower-mass
parameter point (\ref{eq:paras1}).
As expected, the rates drop dramatically for heavier Higgs 
masses, even worse once we include the Higgs decay. The
$\tan\beta$ dependence still shows the original motivation to
consider such loop-induced processes, and in particular the $D$ terms: 
for those, the rates are largest for small values of
$\tan\beta$, where all other known searches fail. However, because of the
unexpectedly large mass suppression, Yukawa couplings are numerically 
dominant, as indicated by the different scales on the $y$ axes in 
Fig.~\ref{fig:asso_mfv}.
Possible large supersymmetric corrections in this process can only occur 
beyond MFV --- just
like for single-Higgs production.

\subsection{Beyond MFV}

\begin{figure}[t]
 \begin{center}
  \includegraphics[width=7.5cm,height=5.5cm]{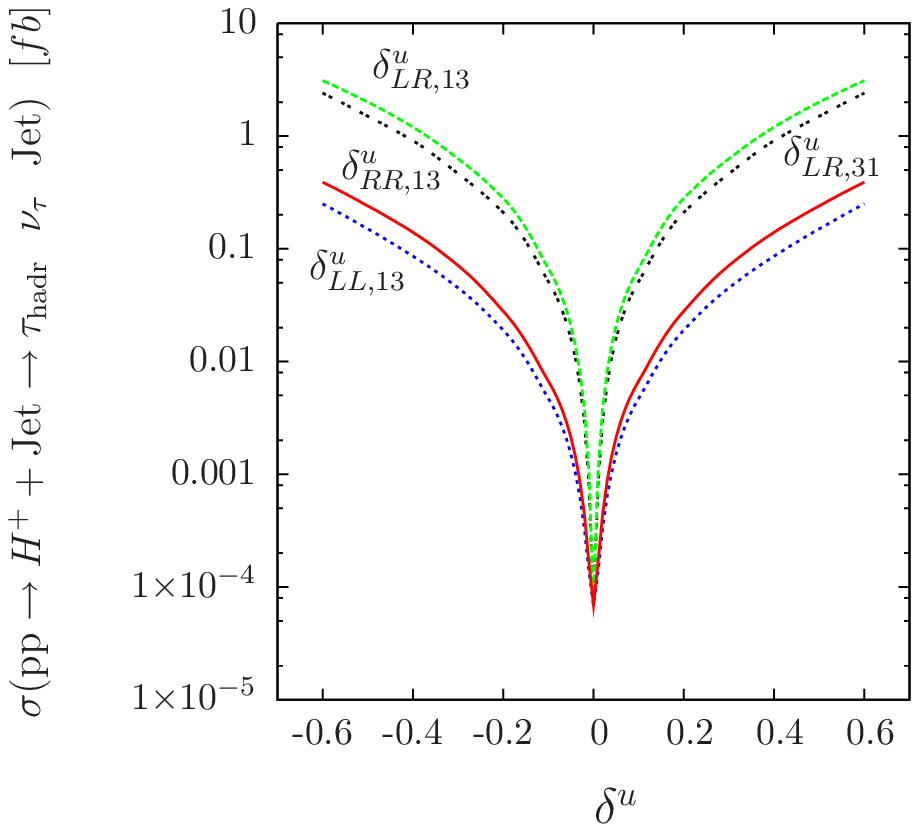} \hspace*{10mm}
  \includegraphics[width=8.0cm,height=6.0cm]{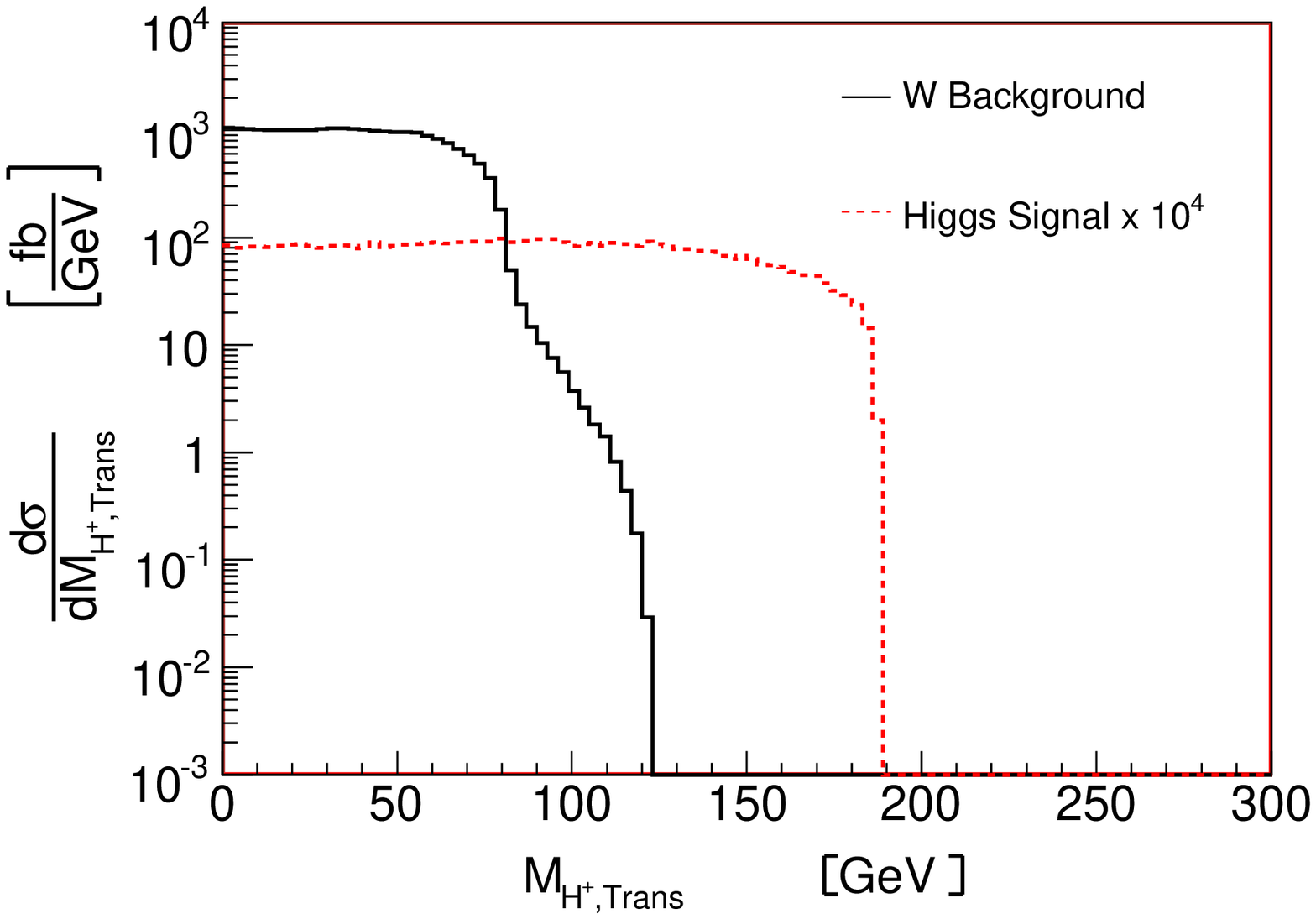}
 \end{center}
 \caption{Left: 
  Cross sections for
  charged-Higgs production with a hard jet including the decay into 
 a hadronic $\tau$ with beyond-MFV
 effects. 
 The flavor-diagonal parameters 
 correspond to the lower-mass scenario 
 (\ref{eq:paras1}). We vary different 
 $\delta^u_{AB,ij}$, one at a time, and
 assume $m_q=0$. 
 Right: transverse-mass distributions for
 charged-Higgs production with a jet including a decay assuming beyond-MFV
 ($\delta^u_{LR,31}=0.5$). We also show the $W$+jet
 background~\cite{madevent}.  \label{fig:asso_nmfv}}
\end{figure}

In contrast to single-Higgs production, the operator basis for Higgs-plus-jet production does not
get significantly extended when we introduce beyond-MFV
effects. However, just like for single-Higgs
production the effective vertices shown in
Fig.~\ref{fig:feyn_single} will get significantly enhanced once we
allow for sizeable $\delta^u_{AB,ij}$. Of course, to get a reliable account 
of the size of such effects we have to take into account the current limits 
on the flavor sector beyond MFV.\bigskip

In this section we consider squark mixing between the first and 
third generation. The corresponding cross sections from 
second- and third-generation mixing are very similar, but slightly reduced
due to the reduced charm parton density.
The largely unconstrained 
$\delta^u_{RR, 13}$ and $\delta^u_{LR, 31}$ can have a sizeable effect
on the charged-Higgs production rate. We already see this in the last
set of columns in Tab.~\ref{tab:hj}: independent of the Yukawas, flavor
effects beyond MFV can enhance the rate by a factor of five, compared
to the tree-level process or compared to the MFV case. The same effect
we see in the left panel of Fig.~\ref{fig:asso_nmfv}, where we show
the variation of the Higgs cross section times branching ratio to a
hadronic $\tau$ as a function of the $\delta^u$, each of them varied
independently. For example, $|\delta^u_{LR, 31}|>0.2$ outgrows the
tree-level results for the SUSY parameters listed in
Eq.(\ref{eq:paras1}).

The bounds on the four considered $\delta^u$ mixings are different:
The mass-matrix entries $\delta^u_{LL, 13}$ and $\delta^u_{LR,13}$ 
are quite constrained. Their impact shown in Fig.~\ref{fig:asso_nmfv}
would not be allowed by flavor-physics constraints if 
only one of the $\delta$'s was varied at a time.
We nevertheless show the curves, because there might be 
cancellations induced by correlations between different deltas
in the rare-decay observables. The four curves illustrate that the 
contribution of the different parameters beyond MFV are generically of 
similar size.
To indicate how we would attempt to reduce the $W$ background we also
show the distributions in the transverse mass 
\begin{equation}
  m^2_{T,H} = (|\vec{p}_{T, {\rm hadr}}| + |\vec{p}_{T, {\rm miss}}|)^2 
            - (\vec{p}_{T, {\rm hadr}} + \vec{p}_{T, {\rm miss}})^2  
\end{equation}
for the Higgs signal and for the $W$ background.
For sufficiently large Higgs masses and modulo detector-resolution
effect mostly on the missing transverse momentum vector, we could use
such a distribution to enhance the signal over the
background.\bigskip

\begin{figure}[t]
 \begin{center}
  \includegraphics[width=7.8cm,height=5.8cm]{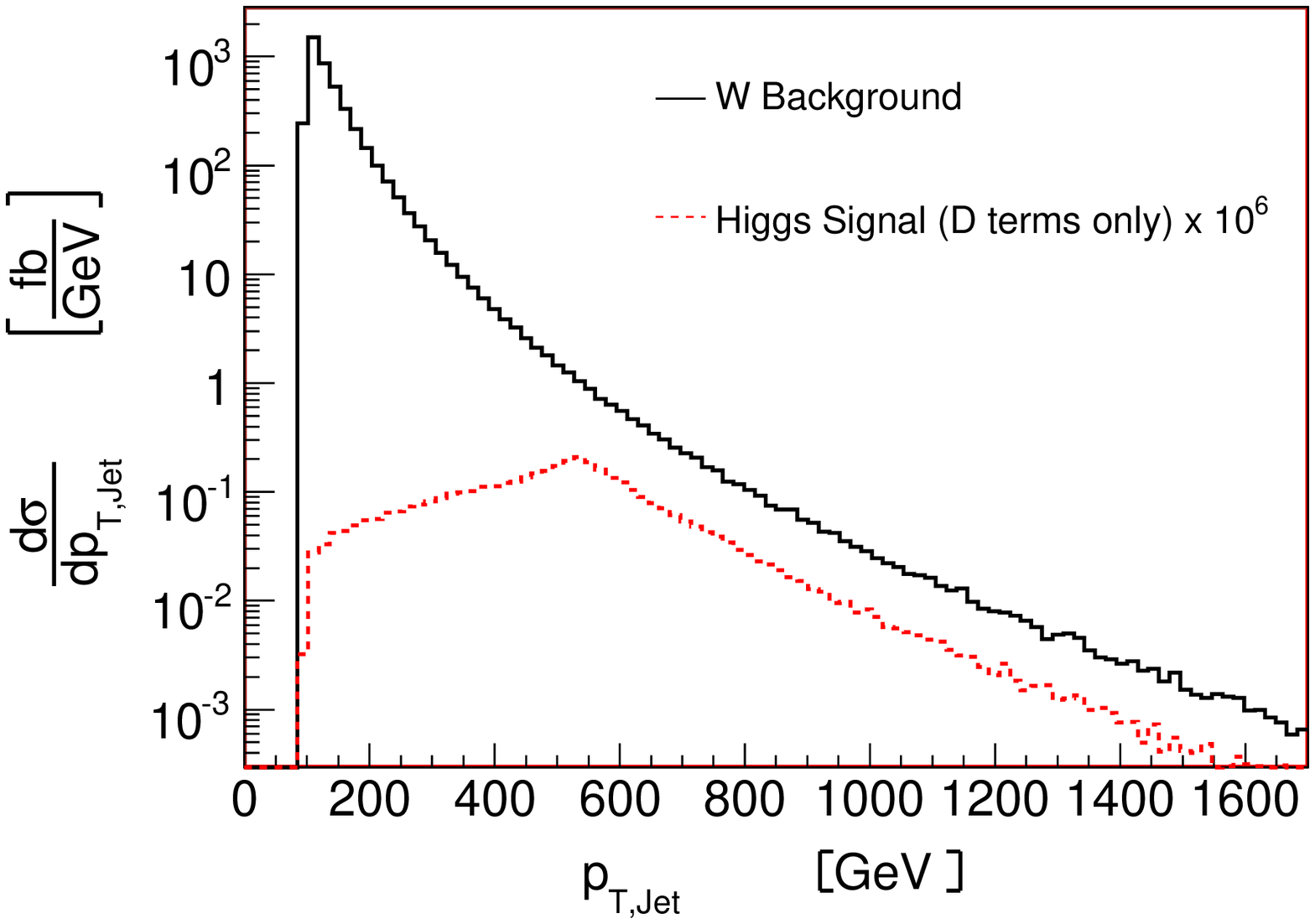} \hspace*{10mm}
  \includegraphics[width=7.8cm,height=5.8cm]{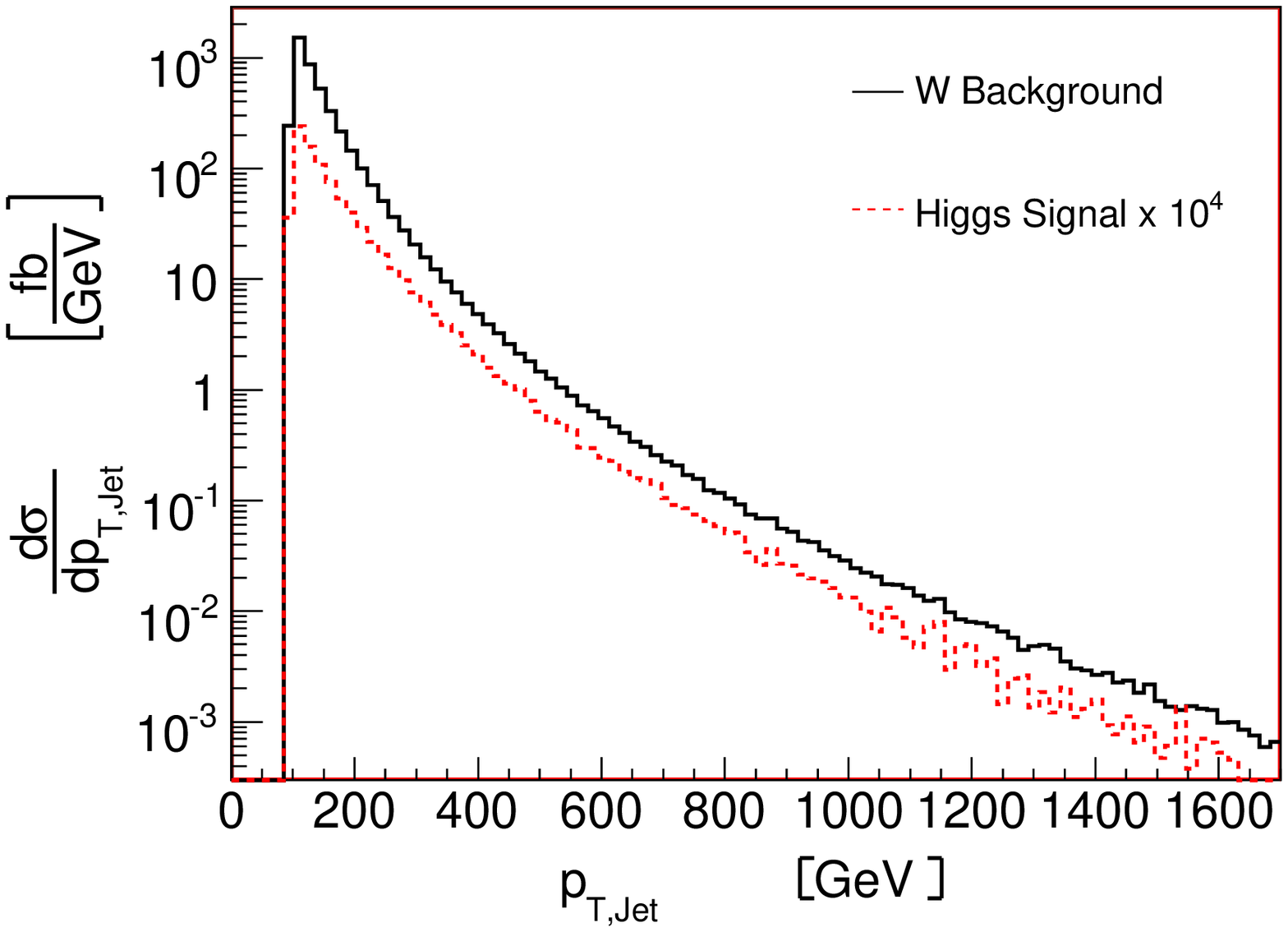} 
 \end{center}
 \caption{Transverse-momentum distributions for charged-Higgs
 production with a jet including the decay to a hadronic $\tau$.
 We also show the scaled background distributions from
 $W$+jet
 production~\cite{madevent}. The left panel shows MFV and $D$ terms only with
$m_q=0$, the right panel includes 
 beyond-MFV effects ($\delta^u_{LR,31}=0.5$). All other parameters 
 given in Eq.(\ref{eq:paras1}).
\label{fig:pt_nmfv}}
\end{figure}

An interesting side aspect of Higgs-plus-jet production via different
supersymmetric couplings can be seen in Fig.~\ref{fig:pt_nmfv}: in
the left panel we show the $p_{T,j}$ distribution (equivalent to
$p_{T,H}$) only taking into account $D$-term couplings in
squark--gluino boxes and vertices. 
For small transverse momenta the cross section
is finite, because the loops with $D$-term
couplings have no counterpart in single-Higgs production and
the $2 \to 2$ process is not an infrared-sensitive real-emission correction. 
Moreover, the heavy particles in the box define the
typical energy scale of the process and show a threshold behavior
around $p_T \sim 500\gev$. On the other hand, in the right panel we see that the $p_T$
distributions for the Higgs signal and the $W$ background look very
similar. Both are infrared divergent for small values $p_T$. 
This infrared (soft and collinear) divergences will of course be canceled 
by virtual corrections and factorization contributions to the single-Higgs
or single-$W$ processes. A proper description of the $p_T$ spectrum in
the small-$p_T$ domain would require soft-gluon resummation.

In Figs.~\ref{fig:asso_nmfv} and \ref{fig:pt_nmfv} we see how the
Standard--Model background to charged-Higgs production is still
overwhelming. On the other hand, the transverse--mass distribution
also shows the background cut off above $m_T = m_W$. While detector
effects will smear out this distribution, it might allow us to improve
the signal--background ratio to a level where other cuts become
useful. Probably, the transverse momentum of additional jets would be
one of those signatures.

%%%%%%%%%%%%%%%%%%%%%%%%%%%%%%%%%%%%%%%%%%%%%%%%%%%%%%%%%%%%%%%%%%%%%%
\section{Outlook \label{sec:outlook}}

According to the current state of the art, charged-Higgs searches at
the LHC have to rely on a $\tan \beta$ enhanced bottom Yukawa 
for a sufficiently large production cross section. We
studied two types of loop-induced production mechanisms which can
significantly increase the production cross section for small Higgs
masses and small values of $\tan\beta$: \bigskip

Single-charged Higgs production in $pp$ collisions in a general 
two-Higgs-doublet model is
suppressed by either light-generation quark Yukawas or by
small CKM mixing. 
For models with minimal flavor violation, this chiral  
suppression is generic and cannot be lifted by, \eg,
supersymmetric loops. If we allow for general squark mixing,
additional loop-induced contributions arise. Here, the
left-right chiral flip does not require a quark mass, but can proceed via
squark mixing. We find that such contributions
can enhance the single-charged-Higgs production cross section by
almost an order of magnitude, even after including all current bounds
on squark-flavor mixing.

Charged-Higgs production in association with a hard jet can be
induced by supersymmetric $D$ terms. These are proportional to the 
weak gauge coupling and therefore appear in the one-loop amplitudes
even in the limit of vanishing quark masses. 
We find, however, that although chirally not suppressed, the $D$-term
contribution is only a small fraction of the supersymmetric
amplitude, due to its faster decoupling with heavy superpartner masses.
Just like in the single-charged Higgs case, only
beyond-MFV contributions can enhance the associated
charged-Higgs rate significantly above
the two-Higgs-doublet model.

We find that the dominant source of genuine
supersymmetric flavor enhancement in the charged-Higgs production rate is the 
soft-breaking 
$A$-term for up-type squarks, specifically $A^u_{3i}$. It mixes the 
doublet-stop with light-generation singlets.
This term is essentially unconstrained by flavor 
physics data, which are
however sensitive to the chirality flipped term $A^u_{i3}$.
Based on theory prejudice 
these off-diagonal $A$-terms should be small~\cite{Nir:1993mx},
\begin{eqnarray}
\delta^q_{LR \, ij} \sim \frac{ m_{q_i} m_{q_j} }{\tilde m^2} ~~~~ \mbox{(alignment)}, \\
\delta^u_{LR \, 3j} \sim \frac{V_{jb}^* m_{u_j} }{\tilde m}, ~~~~
\delta^u_{LR \, i3} \sim \frac{V_{ti}^* m_{t} }{\tilde m}~~\mbox{(abelian flavor)}. \label{eq:U1}
\end{eqnarray}
{}From Eq.(\ref{eq:U1}) it follows further that squark mixing involving
a doublet stop $\delta^u_{LR \, 3j}$ is suppressed with respect to
a singlet stop $\delta^u_{LR \, i3}$ by a factor $m_{u_j}/m_t$.\bigskip

We stress that the effects involving mixing of 
$\tilde u_R$ or $\tilde c_R$ with stops are unvisible to
kaon, charm, and $B$~experiments. Hence, 
collider searches for enhanced 
charged-Higgs production cross sections probe a unique sector of flavor.
A discovery would signal besides a breakdown of the Standard Model 
a quite non-standard solution to the flavor puzzle, including a breakdown of 
the minimal-flavor-violation hypothesis (see also \cite{Grossman:2007bd} for 
MFV tests at the LHC).

At present, we cannot firmly claim that these
flavor-induced charged-Higgs production rates at small $\tan\beta$
rates lead to observable signals over the large $W$-production
backgrounds; we leave the conclusions to a detailed signal--background
analysis, which carefully has to include detector effects.

\subsection*{Acknowledgments}

GH is happy to thank Thorsten Feldmann for a stimulating discussion.
We are particularly grateful to Michael Rauch for his help with
FormCalc/HadCalc.
MS and TP are grateful to the Max-Planck-Institute for Physics for
their continuous hospitality. MS is grateful to Sebastian J\"ager for helpful
discussions.
The work of GH is supported in part 
by Bundesministerium f\"ur Bildung und Forschung, Berlin-Bonn.
This work is supported in part by the European Community's Marie-Curie
Research Training Network HEPTOOLS under contract MRTN-CT-2006-035505.

%%%%%%%%%%%%%%%%%%%%%%%%%%%%%%%%%%%%%%%%%%%%%%%%%%%%%%%%%%%%%%%%%%%%%%
\section*{Appendix}

\renewcommand{\theequation}{\Alph{section}-\arabic{equation}}
\setcounter{section}{1}
\setcounter{equation}{0}

In this appendix we give details about the super-CKM basis, the superpotential,
supersymmetry breaking, and scalar mass matrices. Moreover, we give all
numerical details in computing the $\overline{\rm MS}$ quark masses
as well as explicit analytical results for the $D$-term-induced form factors
for $H^+$-plus-jet production.

\subsection{Super-CKM Basis}

Following the SUSY conventions of Ref.~\cite{Rosiek:1995kg}
except for $Y^d_{\rm here}=-Y^d_{\rm Rosiek}$ and $A^u_{\rm here}=-A^u_{\rm Rosiek}$
the MSSM superpotential is given as
\begin{equation}  
  W=Q_i Y^u_{ij} U_j H_u - Q_i Y^d_{ij} D_j H_d + \mu H_u H_d,
  \label{eq:W}
\end{equation}
where we make flavor indices $i,j=1,2,3$ explicit.
The symbols $Q,U,D,H_d,H_u$ used for the superfields should
not be confused with the same symbols used for the quark and Higgs fields
in the main text.
The superfields $Q,U,D$ and $H_d, H_u$ transform under
$SU(3)_C \times SU(2)_L \times U(1)_Y$ as
\begin{equation}
 Q \equiv (3,2,\frac{1}{6}),           \qquad 
 U \equiv ( \bar 3, 1, - \frac{2}{3}), \qquad
 D \equiv ( \bar 3, 1, \frac{1}{3}),   \qquad
 H_d \equiv (1,2,-\frac{1}{2}),        \qquad
 H_u \equiv (1,2,\frac{1}{2}).
\end{equation}
In the superpotential (\ref{eq:W}) and in the soft-breaking Lagrangian, see
Eq.(\ref{eq:lsoft}) below, we suppress $SU(2)$ contractions for all doublets, 
\eg
$H_d H_u \equiv \epsilon_{ij} H_{di} H_{uj}$ with $\epsilon_{12}=+1$. 
(Note that $\epsilon_{12}^{\rm Rosiek}=-1$.)
The two Higgs doublets in the
supersymmetric Lagrangian are defined in terms of their components
$H_d = (H_d^0,H_d^-)^T$ and $H_u = (H_u^+,H_u^0)^T$.
The scalar (fermionic) parts of the superfields $Q$, $U$, and $D$ are denoted
as $\tilde Q (\Psi_Q)$, $\tilde U^* (\Psi_U^C)$, and 
$\tilde D^* (\Psi_D^C)$ below, respectively, where
$\Psi^C$  is the charge-conjugate of the fermion field $\Psi$.\bigskip

Translating the three generations of flavor or weak eigenstates 
$Q,U,D$ with $Q=(U_L,D_L)^T$ (denoted by capital letters)
into mass eigenstates $u_L,d_L,u_R,d_R$ and 
$\tilde u_L, \tilde d_L, \tilde u_R, \tilde d_R$ 
(denoted by small letters) defines 
the unitary transformations $V^{u,d},U^{u,d}$, 
\begin{align} \label{eq:CKM-trafo}
 u_L \equiv V^u \Psi_{U_L}, & \qquad
 d_L \equiv V^d \Psi_{D_L}, &
 u_R \equiv U^u \Psi_U,   & \qquad
 d_R \equiv U^d \Psi_D, \\
 \tilde u_L = V^u \tilde U_L, & \qquad
 \tilde d_L = V^d  \tilde D_L, &
 \tilde u_R = U^u \tilde U,   & \qquad
 \tilde d_R = U^d \tilde D, \nonumber
\end{align}
such that the fermion mass matrices are diagonal: 
\begin{equation}
V^u Y^{u *} U^{u \dagger} = \text{diag}(y_u,y_c,y_t) 
= \text{diag}\left(\frac{m_u}{v_u},\frac{m_c}{v_u},
\frac{m_t}{v_u}\right), 
\qquad 
V^d Y^{d *} U^{d \dagger} = \text{diag}(y_d,y_s,y_b) 
= \text{diag}\left(\frac{m_d}{v_d},\frac{m_s}{v_d},
\frac{m_b}{v_d}\right).
\end{equation}
Note that the one-to-one map between Yukawas and masses receives
corrections from non-holomorphic terms, relevant
for down-type fermions at large $\tan \beta$~\cite{bottom_yuk}.
The CKM matrix is given as $V \equiv V^u V^{d \dagger}$. It is 
parameterized according to the quark flavors it connects in the charged-current 
interaction, 
\begin{eqnarray}
V =\left( \begin{array}{lll}
V_{ud}  &  V_{us} & V_{ub}\\
V_{cd}  &  V_{cs} & V_{cb}\\
V_{td}  &  V_{ts} & V_{tb}
\end{array} \right).
\end{eqnarray}

Given all these
fields we can write the relevant flavored part of the soft-breaking Lagrangian
in (gauge eigenstate) component fields with flavor indices $i,j=1,..3$:
\begin{equation}
   \mathcal{L}_{\rm soft} 
 = -  \tilde U^*_i m_{\tilde U ij}^2 \tilde U_j
   - \tilde D^*_i m_{\tilde D ij}^2 \tilde D_j
   - \tilde Q^\dagger_i m_{\tilde Q ij}^2 \tilde Q_j 
 -  \left[ \tilde Q_i \bar A^u_{ij} \tilde U^*_j H_u 
         - \tilde Q_i \bar A^d_{ij} \tilde D^*_j H_d \; + \; \text{h.c.} 
    \right].  
\label{eq:lsoft}
\end{equation}
Diagonalizing the quark fields according to 
Eq.(\ref{eq:CKM-trafo}) leads to the quark mass basis.
Simultaneous rotation of the squarks leads to  
$\mathcal{L}_{\rm soft}$ in the super-CKM basis:
\begin{eqnarray}
   \mathcal{L}_{\rm soft}
 &=& -  \tilde u_{Ri}^* m_{\tilde U_R ij }^2 \tilde u_{Rj} 
   - \tilde d_{Ri}^* m_{\tilde D_R ij}^2 \tilde d_{Rj}
-  \tilde u_{Li}^* m_{\tilde U_L ij}^2 \tilde u_{Lj} 
   - \tilde d_{Li}^* m_{\tilde D_L ij}^2 \tilde d_{Lj}
\nonumber \\
& -&  \left[ \tilde u_{Li} A^u_{ij} \tilde u^*_{Rj} H_u^0
- \tilde d_{Li} V_{ki} A^u_{kj} \tilde u^*_{Rj} H_u^+  
- \tilde u_{Li} V_{ik}^* A^d_{kj} \tilde d^*_{Rj} H_d^-
+ \tilde d_{Li} A^d_{ij} \tilde d^*_{Rj} H_d^0 \; + \; \text{h.c.} 
    \right],  
\label{eq:lsoft-superCKM}
\end{eqnarray}
where for $q=u,d$ 
\begin{equation}
A^{q} = V^{*q} \bar A^q U^{q T}, ~~
 m_{\tilde U_R}^2 = U^u  m_{\tilde U}^2 U^{u \dagger}, ~~
 m_{\tilde D_R}^2 = U^d  m_{\tilde D}^2 U^{d \dagger}, ~~
 m_{\tilde U_L}^2 = V^u  m_{\tilde Q}^2 V^{u \dagger}, ~~
 m_{\tilde D_L}^2 = V^d  m_{\tilde Q}^2 V^{d \dagger}. 
\label{eq:su2}
\end{equation}

\subsection{Squark Masses}

The entries in the $6 \times 6$ squark mass matrices $\mathcal{M}_q^2$, 
($q=u,d$) in Eq.(\ref{eq:mass_matrix}) stem from
soft-breaking $A$-terms in ${\mathcal L_{\rm
soft}}$ given in Eq.(\ref{eq:lsoft-superCKM}) and from $D$ and $F$ terms. For up-squarks, with $Q_u=2/3$ and $T_3^u=1/2$, $i,j=1,..,3$ they read
\begin{alignat}{5}
 M_{u \, LL \, ij}^2 
  = &  m_{\tilde U_L \, ij}^2 + \left( m_{u_i}^2
     + (T_3^u-Q_u\sin^2 \theta_w ) m_Z^2 \cos 2\beta \right) \delta_{ij},  \notag\\
 M^2_{u \, RR \, ij}
  = &  m_{\tilde U_R \, ij}^2 + \left( m_{u_i}^2
     + Q_u\sin^2 \theta_w m_Z^2 \cos 2\beta \right) \delta_{ij} \notag, \\
  M^2_{u \, LR \, ij}
  = &  \left< H_u^0 \right> A_{ij}^u- m_{u_{i}} \mu \cot \beta \, \delta_{ij},
\label{eq:muentries}
\end{alignat}
while for down-squarks they read with $Q_d=-1/3$ and $T_3^d=-1/2$ 
\begin{alignat}{5}
 M_{d \, LL \, ij}^2 
  = &  m_{\tilde D_L \, ij}^2 + \left( m_{d_i}^2
     + (T_3^d-Q_d \sin^2 \theta_w ) m_Z^2 \cos 2\beta \right) \delta_{ij},  \notag\\
 M^2_{d \, RR \, ij}
  = &  m_{\tilde D_R \, ij}^2 + \left( m_{d_i}^2
     + Q_d\sin^2 \theta_w m_Z^2 \cos 2\beta \right) \delta_{ij} \notag, \\
  M^2_{d \, LR \, ij}
  = &  \left< H_d^0 \right> A_{ij}^d- m_{d_{i}} \mu \tan \beta \, \delta_{ij}.
\label{eq:muentries-d}
\end{alignat}
We recall that throughout this paper the SUSY-breaking parameters and the 
$\mu$-term are real quantities.

The full squark mass matrices $\mathcal{M}_q^2$ can be diagonalized 
with unitary transformations
$Z^q$ to obtain the squark mass eigenstates $\tilde q_{1i}, \tilde q_{2i}$:
\begin{equation}
 Z^q \mathcal{M}_q^2 Z^{q \dag} 
  = \text{diag} \left( m_{\widetilde{q}_{1i}}^{2},m_{\widetilde{q}_{2i}}^{2}
                \right), \qquad \qquad 
 Z^{u}
  \begin{pmatrix}
   \widetilde{u}_L \\ 
   \widetilde{c}_L \\ 
   \widetilde{t}_L \\ 
   \widetilde{u}_R \\ 
   \widetilde{c}_R \\ 
   \widetilde{t}_R
  \end{pmatrix}
 =
  \begin{pmatrix}
   \widetilde{u}_1 \\ 
   \widetilde{u}_2 \\ 
   \widetilde{u}_3 \\ 
   \widetilde{u}_4 \\ 
   \widetilde{u}_5 \\ 
   \widetilde{u}_6
  \end{pmatrix},
\qquad 
 Z^{d}
  \begin{pmatrix}
   \widetilde{d}_L \\ 
   \widetilde{s}_L \\ 
   \widetilde{b}_L \\ 
   \widetilde{d}_R \\ 
   \widetilde{s}_R \\ 
   \widetilde{b}_R
  \end{pmatrix}
 =
  \begin{pmatrix}
   \widetilde{d}_1 \\ 
   \widetilde{d}_2 \\ 
   \widetilde{d}_3 \\ 
   \widetilde{d}_4 \\ 
   \widetilde{d}_5 \\ 
   \widetilde{d}_6
  \end{pmatrix}.
\end{equation}

\subsection{Quark Masses}

We use the running quark masses at next-to-leading order
\begin{equation}
m(\mu)  = m(\mu_0) \; 
          \left[ \frac{\alpha_s(\mu)}{\alpha_s(\mu_0)}
          \right]^{\gamma_{m}^{(0)}/(2\beta_0)} \;
          \left[ 1 \, + \, \left( \frac{\gamma_m^{(1)}}{2\beta_0}
                                 -\frac{\beta_1 \gamma_m^{(0)}}{2\beta_0^2}
                           \right) 
                        \frac{\alpha_s(\mu)-\alpha_s(\mu_0)}{4\pi}\right].
\end{equation}
Here, $\beta_0 = 11-2/3~N_f$, $\beta_1= 102-38/3~N_f$,
$\gamma_m^{(0)}=6C_F$ and $\gamma_m^{(1)}= C_F\, \left(
3C_F+97-10/3N_f\right) $, where $N_{f}$ denotes the number of quarks
with $m_f \leq \mu$.\bigskip

For our numerical analysis we use the numerical values for the quark masses~\cite{mjamin}:
\begin{equation} \label{tab:masses}
\begin{tabular}{|ccc|}
\hline
\multicolumn{1}{|c|}{$m_{u}\left( 2\gev\right) $} & \multicolumn{1}{|c|}{$%
m_{d}\left( 2\gev\right) $} & \multicolumn{1}{|c|}{$m_s\left( 2\gev\right) 
$} \\ 
\multicolumn{1}{|c|}{$2.8\pm 0.6\mev$} & \multicolumn{1}{|c|}{$5.0\pm 1.0\mev
$} & \multicolumn{1}{|c|}{$95\pm 15\mev$} \\ \hline
\multicolumn{1}{|c|}{$m_{c}\left( m_{c}\right) $} & \multicolumn{1}{c|}{$%
m_{b}\left( m_{b}\right) $} & $m_{t}\left( m_{t}\right) $ \\ 
\multicolumn{1}{|c|}{$1.28\pm 0.05\gev$} & \multicolumn{1}{c|}{$4.22\pm
0.05\gev$} & $163\pm 3\gev$ \\ \hline
\multicolumn{1}{|c|}{$m_{u}\left( m_{Z}\right) $} & \multicolumn{1}{c|}{$%
m_{d}\left( m_{Z}\right) $} & $m_s\left( m_{Z}\right) $ \\ 
\multicolumn{1}{|c|}{$1.7\pm 0.4\mev$} & \multicolumn{1}{c|}{$3.0\pm 0.6\mev$%
} & $54\pm 8\mev$ \\ \hline
\multicolumn{1}{|c|}{$m_{c}\left( m_{Z}\right) $} & 
\multicolumn{1}{c|}{$m_{b}\left( m_{Z}\right) $} & $m_{t}\left( m_{Z}\right) 
$ \\ 
\multicolumn{1}{|c|}{$0.62\pm 0.03\gev$} & \multicolumn{1}{c|}{$2.87\pm
0.03\gev$} & $171\pm 3\gev$ \\ \hline
\end{tabular}
\end{equation}

\subsection{\boldmath{Form factors for $D$-term contributions to
$H^+$-plus-jet production}}

Here we present explicit results for
the form factors for $H^+$-plus-jet production induced 
by supersymmetric $D$ terms assuming MFV. For massless light quarks 
the form factors $\mathcal{F}_{4,5,6}^{ij,\sigma}$ vanish.
For massless quarks we also see $\mathcal{F}_{1 \ldots 6}^{ij,R}=0$, since
$D$~terms couple the $H^\pm$ only to the left-handed
squarks $(\sigma=L)$.
Following Eq.(\ref{eq:gaugeinvariance}) only two out of the remaining
form factors are independent.
Choosing $\mathcal{F}_{1,2}^{ij,L}$, 
the whole result can be expressed in terms of
\begin{alignat}{5}
\mathcal{F}_1^{ij,L}   = eg_s^3 \frac{m_W}{\pi^2} \;
                      \frac{V_{ij}^\ast \sin 2\beta }{12 \sqrt{2}\sin \theta_w}
           &   \Bigg[ - C_1(c_1) 
                      - C_1(c_2) 
                      + \frac{D_{00}(d_1)}{4}
                      + \frac{D_{00}(d_2)}{4}  \notag\\
           &          + \frac{9}{8} \Big[  
                                             D_0(d_3) m_{\tilde{g}}^2
                                           - D_1(d_3) m_{H^+}^2
                                           - D_2(d_3) u  
                                           - 2D_{00}(d_3) 
                                           - D_{11}(d_3) m_{H^+}^2
                                                        \notag\\
           &  \qquad                       
                                           - D_{12}(d_3) \left( m_{H^+}^2 +u \right)
                                           - D_{13}(d_3) \left( s+u \right) 
                                           - D_{22}(d_3) u
                                           - D_{23}(d_3) u
                                    \Big] 
               \Bigg],                       \notag\\
\mathcal{F}_2^{ij,L}   = eg_s^3 \frac{m_W}{\pi^2} \;
                      \frac{V_{ij}^\ast \sin 2\beta }{48 \sqrt{2}\sin \theta_w}
           &   \Bigg[  
                       D_{23}(d_1) 
                     - D_2(d_2) 
                     - D_{22}(d_2)
                     - D_{23}(d_2)   \notag\\
           &         - 9 \Big[   D_1(d_3) 
                               + D_2(d_3)
                               + D_{11}(d_3) 
                               + 2D_{12}(d_3) 
                               + D_{13}(d_3) 
                               + D_{22}(d_3)
                               + D_{23}(d_3)
                         \Big]
               \Bigg],
\label{eq:Dtermresult}
\end{alignat}
where the tensor coefficients $C_{i\dots}$, $D_{i\dots}$ are defined
as in Ref.~\cite{dittmaier}.
We use the following abbreviations to specify the arguments of
the three-point and four-point integrals:
\begin{alignat}{5}
c_1 &=\left(m_{H^+}^2,t,0,
            m_{\tilde{u}_j},m_{\tilde{d}_i},m_{\tilde{g}} 
      \right)    \; ,\notag\\
c_2 &=\left(m_{H^+}^2,u,0,
            m_{\tilde{d}_j},m_{\tilde{u}_i},m_{\tilde{g}}
      \right)    \; ,\notag\\
d_1 &=\left(0,0,m_{H^+}^2,0,s,t,
            m_{\tilde{g}},m_{\tilde{d}_j},m_{\tilde{d}_j},m_{\tilde{u}_i}
      \right)    \; ,\notag\\
d_2 &=\left(0,m_{H^+}^2,0,0,s,u,
            m_{\tilde{g}},m_{\tilde{d}_j},m_{\tilde{u}_i},m_{\tilde{u}_i}
      \right)    \; ,\notag\\
d_3 &=\left(m_{H^+}^2,0,0,0,t,u,
            m_{\tilde{u}_i},m_{\tilde{d}_j},m_{\tilde{g}},m_{\tilde{g}}
      \right),
\end{alignat}
They are connected to the 
ordering scheme for the arguments of the loop functions defined
in Ref.~\cite{dittmaier} as 
\begin{alignat}{5}
c &= \left(p_1^2,(p_1-p_2)^2,p_2^2,m_1,m_2,m_3\right) \equiv
\left(p_1,p_2,m_1,m_2,m_3\right),
\notag\\
d &= \left(
p_1^2,(p_1-p_2)^2,(p_2-p_3)^2,p_3^2,(p_1-p_3)^2,p_2^2,m_1,m_2,m_3,m_4
\right) \equiv \left(p_1,p_2,p_3,m_1,m_2,m_3,m_4\right).
\end{alignat}

\end{document}

%% file: udhg_sqcd.tex
\small
\unitlength=1.1bp%

\begin{feynartspicture}(432,100)(5,1)

\FADiagram{}
\FAProp(0.,15.)(10.,14.5)(0.,){/Straight}{1}
\FALabel(5.0774,15.8181)[b]{$u$}
\FAProp(0.,5.)(6.5,5.5)(0.,){/Straight}{-1}
\FALabel(3.36888,4.18457)[t]{$d$}
\FAProp(20.,15.)(13.5,5.5)(0.,){/ScalarDash}{1}
\FALabel(18.5763,11.3388)[tl]{$H$}
\FAProp(20.,5.)(10.,14.5)(0.,){/Cycles}{0}
\FALabel(13.2922,12.4512)[bl]{$g$}
\FAProp(10.,14.5)(10.,11.5)(0.,){/Straight}{1}
\FALabel(9.03,13.)[r]{$u$}
\FAProp(6.5,5.5)(13.5,5.5)(0.,){/ScalarDash}{-1}
\FALabel(10.,4.43)[t]{$\tilde d_L$}
\FAProp(6.5,5.5)(10.,11.5)(0.,){/Straight}{0}
\FAProp(7.0,5.5)(10.5,11.5)(0.,){/Cycles}{0}
\FALabel(7.60709,8.67503)[br]{$\tilde g$}
\FAProp(13.5,5.5)(10.,11.5)(0.,){/ScalarDash}{-1}
\FALabel(12.9252,7.84012)[bl]{$\tilde u_L$}
\FAVert(10.,14.5){0}
\FAVert(6.5,5.5){0}
\FAVert(13.5,5.5){0}
\FAVert(10.,11.5){0}

\FADiagram{}
\FAProp(0.,15.)(6.5,14.5)(0.,){/Straight}{1}
\FALabel(3.36888,15.8154)[b]{$u$}
\FAProp(0.,5.)(10.,5.5)(0.,){/Straight}{-1}
\FALabel(5.0774,4.18193)[t]{$d$}
\FAProp(20.,15.)(13.5,14.5)(0.,){/ScalarDash}{1}
\FALabel(16.6311,15.8154)[b]{$H$}
\FAProp(20.,5.)(10.,5.5)(0.,){/Cycles}{0}
\FALabel(14.8876,3.48281)[t]{$g$}
\FAProp(10.,5.5)(10.,8.5)(0.,){/Straight}{-1}
\FALabel(8.93,7.)[r]{$d$}
\FAProp(6.5,14.5)(13.5,14.5)(0.,){/ScalarDash}{1}
\FALabel(10.,15.57)[b]{$\tilde u_L$}
\FAProp(6.5,14.5)(10.,8.5)(0.,){/Straight}{0}
\FAProp(6.0,14.5)(9.5,8.5)(0.,){/Cycles}{0}
\FALabel(7.60709,11.325)[tr]{$\tilde g$}
\FAProp(13.5,14.5)(10.,8.5)(0.,){/ScalarDash}{1}
\FALabel(12.6089,11.199)[tl]{$\tilde d_L$}
\FAVert(6.5,14.5){0}
\FAVert(10.,5.5){0}
\FAVert(13.5,14.5){0}
\FAVert(10.,8.5){0}

\FADiagram{}
\FAProp(0.,15.)(6.5,13.5)(0.,){/Straight}{1}
\FALabel(3.59853,15.2803)[b]{$u$}
\FAProp(0.,5.)(6.5,6.5)(0.,){/Straight}{-1}
\FALabel(3.59853,4.71969)[t]{$d$}
\FAProp(20.,15.)(13.5,13.5)(0.,){/ScalarDash}{1}
\FALabel(16.4015,15.2803)[b]{$H$}
\FAProp(20.,5.)(13.5,6.5)(0.,){/Cycles}{0}
\FALabel(16.2441,4.03762)[t]{$g$}
\FAProp(6.5,13.5)(6.5,6.5)(0.,){/Straight}{0}
\FAProp(6.0,13.5)(6.0,6.5)(0.,){/Cycles}{0}
\FALabel(5.68,10.)[r]{$\tilde g$}
\FAProp(6.5,13.5)(13.5,13.5)(0.,){/ScalarDash}{1}
\FALabel(10.,14.57)[b]{$\tilde u_L$}
\FAProp(6.5,6.5)(13.5,6.5)(0.,){/ScalarDash}{-1}
\FALabel(10.,5.43)[t]{$\tilde d_L$}
\FAProp(13.5,13.5)(13.5,6.5)(0.,){/ScalarDash}{1}
\FALabel(14.57,10.)[l]{$\tilde d_L$}
\FAVert(6.5,13.5){0}
\FAVert(6.5,6.5){0}
\FAVert(13.5,13.5){0}
\FAVert(13.5,6.5){0}

\FADiagram{}
\FAProp(0.,15.)(6.5,13.5)(0.,){/Straight}{1}
\FALabel(3.59853,15.2803)[b]{$u$}
\FAProp(0.,5.)(6.5,6.5)(0.,){/Straight}{-1}
\FALabel(3.59853,4.71969)[t]{$d$}
\FAProp(20.,15.)(13.5,6.5)(0.,){/ScalarDash}{1}
\FALabel(17.9814,13.8219)[br]{$H$}
\FAProp(20.,5.)(13.5,13.5)(0.,){/Cycles}{0}
\FALabel(17.2754,6.00292)[tr]{$g$}
\FAProp(6.5,13.5)(6.5,6.5)(0.,){/Straight}{0}
\FAProp(6.0,13.5)(6.0,6.5)(0.,){/Cycles}{0}
\FALabel(5.68,10.)[r]{$\tilde g$}
\FAProp(6.5,13.5)(13.5,13.5)(0.,){/ScalarDash}{1}
\FALabel(10.,14.57)[b]{$\tilde u_L$}
\FAProp(6.5,6.5)(13.5,6.5)(0.,){/ScalarDash}{-1}
\FALabel(10.,5.43)[t]{$\tilde d_L$}
\FAProp(13.5,6.5)(13.5,13.5)(0.,){/ScalarDash}{-1}
\FALabel(12.43,10.)[r]{$\tilde u_L$}
\FAVert(6.5,13.5){0}
\FAVert(6.5,6.5){0}
\FAVert(13.5,6.5){0}
\FAVert(13.5,13.5){0}

\FADiagram{}
\FAProp(0.,15.)(13.5,13.)(0.,){/Straight}{1}
\FALabel(3.41023,15.5279)[b]{$u$}
\FAProp(0.,5.)(6.5,6.)(0.,){/Straight}{-1}
\FALabel(3.48569,4.44802)[t]{$d$}
\FAProp(20.,15.)(6.5,13.)(0.,){/ScalarDash}{1}
\FALabel(16.5898,15.5279)[b]{$H$}
\FAProp(20.,5.)(13.5,6.)(0.,){/Cycles}{0}
\FALabel(16.4079,3.75616)[t]{$g$}
\FAProp(13.5,13.)(6.5,13.)(0.,){/ScalarDash}{1}
\FALabel(10.,11.93)[t]{$\tilde u_L$}
\FAProp(13.5,13.)(13.5,6.)(0.,){/Straight}{0}
\FAProp(13.1,13)(13.1,6.)(0.,){/Cycles}{0}
\FALabel(14.32,9.5)[l]{$\tilde g$}
\FAProp(6.5,6.)(6.5,13.)(0.,){/ScalarDash}{-1}
\FALabel(5.43,9.5)[r]{$\tilde d_L$}
\FAProp(6.5,6.)(13.5,6.)(0.,){/Straight}{0}
\FAProp(6.5,5.6)(13.5,5.6)(0.,){/Cycles}{0}
\FALabel(10.,5.18)[t]{$\tilde g$}
\FAVert(13.5,13.){0}
\FAVert(6.5,6.){0}
\FAVert(6.5,13.){0}
\FAVert(13.5,6.){0}

\end{feynartspicture}

%% file: paper.bbl
\bigskip

\begin{thebibliography}{99}

\bibitem{hhg}
 See \eg:
 J.~F.~Gunion, H.~E.~Haber, G.~L.~Kane and S.~Dawson,
  %``THE HIGGS HUNTER'S GUIDE,''
  SCIPP-89/13;
  %\href{http://www.slac.stanford.edu/spires/find/hep/www?r=scipp-89\%2F13}{SPIRES entry}
 A.~Djouadi,
  %``The anatomy of electro-weak symmetry breaking. II: The Higgs bosons in the
  %minimal supersymmetric model,''
  arXiv:hep-ph/0503173.
  %%CITATION = HEP-PH 0503173;%%

\bibitem{lewwg}
 For the most recent results see: 
 \url{http://lepewwg.web.cern.ch}

\bibitem{wbf_susy}
 V.~B\"uscher and K.~Jakobs,
  %``Higgs boson searches at hadron colliders,''
  Int.\ J.\ Mod.\ Phys.\ A {\bf 20}, 2523 (2005);
  %[arXiv:hep-ph/0504099].
  %%CITATION = HEP-PH 0504099;%%
 T.~Plehn, D.~Rainwater and D.~Zeppenfeld,
  %``Probing the MSSM Higgs sector via weak boson fusion at the LHC,''
  Phys.\ Lett.\ B {\bf 454}, 297 (1999);
  %[arXiv:hep-ph/9902434].
  %%CITATION = HEP-PH 9902434;%%
 T.~Plehn, D.~Rainwater and D.~Zeppenfeld,
  %``A method for identifying H $\to$ tau tau $\to$ e+- mu-+ missing p(T)  at the
  %CERN LHC,''
  Phys.\ Rev.\ D {\bf 61}, 093005 (2000);
  %[arXiv:hep-ph/9911385].
  %%CITATION = HEP-PH 9911385;%%
 M.~Schumacher,
  %``Investigation of the discovery potential for Higgs bosons of the minimal
  %supersymmetric extension of the standard model (MSSM) with ATLAS,''
  arXiv:hep-ph/0410112.
  %%CITATION = HEP-PH 0410112;%%

\bibitem{duehrssen}
 M.~D\"uhrssen, S.~Heinemeyer, H.~Logan, D.~Rainwater, G.~Weiglein and D.~Zeppenfeld,
  %``Extracting Higgs boson couplings from LHC data,''
  Phys.\ Rev.\ D {\bf 70}, 113009 (2004)
  %[arXiv:hep-ph/0406323].
  %%CITATION = HEP-PH 0406323;%%

\bibitem{prod_top}
 R.~M.~Barnett, H.~E.~Haber and D.~E.~Soper,
  %``Ultraheavy Particle Production From Heavy Partons At Hadron Colliders,''
  Nucl.\ Phys.\ B {\bf 306}, 697 (1988);
  %%CITATION = NUPHA,B306,697;%%
 A.~C.~Bawa, C.~S.~Kim and A.~D.~Martin,
  %``Charged Higgs Production At Hadron Colliders,''
  Z.\ Phys.\ C {\bf 47}, 75 (1990);
  %%CITATION = ZEPYA,C47,75;%%
 V.~D.~Barger, R.~J.~Phillips and D.~P.~Roy,
  %``Heavy charged Higgs signals at the LHC,''
  Phys.\ Lett.\ B {\bf 324}, 236 (1994);
  %%CITATION = HEP-PH 9311372;%%
%\cite{DiazCruz:1992gg}
%\bibitem{DiazCruz:1992gg}
  J.~L.~Diaz-Cruz and O.~A.~Sampayo,
  %``Contribution of gluon fusion to the production of charged Higgs at hadron
  %colliders,''
  Phys.\ Rev.\  D {\bf 50}, 6820 (1994);
  %%CITATION = PHRVA,D50,6820;%%
 S.~Moretti and K.~Odagiri,
  %``Production of charged Higgs bosons of the minimal supersymmetric  standard
  %model in b-quark initiated processes at the Large Hadron  Collider,''
  Phys.\ Rev.\ D {\bf 55}, 5627 (1997).
  %[arXiv:hep-ph/9611374].
  %%CITATION = HEP-PH 9611374;%%

\bibitem{prod_top_nlo}
 S.~H.~Zhu,
  %``Complete next-to-leading order QCD corrections to charged Higgs boson
  %associated production with top quark at the CERN Large Hadron
  %Collider,''
  Phys.\ Rev.\ D {\bf 67}, 075006 (2003);
  %[arXiv:hep-ph/0112109].
  %%CITATION = HEP-PH 0112109;%%
 J.~Alwall and J.~Rathsman,
  %``Improved description of charged Higgs boson production at hadron colliders,''
  JHEP {\bf 0412}, 050 (2004);
  %[arXiv:hep-ph/0409094].
  %%CITATION = HEP-PH 0409094;%%
  N.~Kidonakis,
  %``Charged Higgs production via b g --> t H- at the LHC,''
  JHEP {\bf 0505}, 011 (2005).
  %[arXiv:hep-ph/0412422].

\bibitem{prod_tp1}
 T.~Plehn,
  %``Charged Higgs boson production in bottom gluon fusion,''
  Phys.\ Rev.\ D {\bf 67}, 014018 (2003).
  %[arXiv:hep-ph/0206121];
  %%CITATION = HEP-PH 0206121;%%

\bibitem{prod_tp2}
 E.~L.~Berger, T.~Han, J.~Jiang and T.~Plehn,
  %``Associated production of a top quark and a charged Higgs boson,''
  Phys.\ Rev.\ D {\bf 71}, 115012 (2005).
  %[arXiv:hep-ph/0312286].
  %%CITATION = HEP-PH 0312286;%%






\bibitem{bottom_pdf}
J.~C.~Collins and W.~K.~Tung,
  %``Calculating Heavy Quark Distributions,''
  Nucl.\ Phys.\ B {\bf 278}, 934 (1986);
  %%CITATION = NUPHA,B278,934;%%
 M.~A.~G.~Aivazis, J.~C.~Collins, F.~I.~Olness and W.~K.~Tung,
  %``Leptoproduction of heavy quarks. 2. A Unified QCD formulation of charged and neutral current processes from fixed target to collider energies,''
  Phys.\ Rev.\ D {\bf 50}, 3102 (1994);
  %%CITATION = HEP-PH 9312319;%%
 F.~I.~Olness and W.~K.~Tung,
  %``When Is A Heavy Quark Not A Parton? Charged Higgs Production And Heavy Quark Mass Effects In The QCD Based Parton Model,''
  Nucl.\ Phys.\ B {\bf 308}, 813 (1988);
  %%CITATION = NUPHA,B308,813;%%
 M.~Kr\"amer, F.~I.~Olness and D.~E.~Soper,
  %``Treatment of heavy quarks in deeply inelastic scattering,''
  Phys.\ Rev.\ D {\bf 62}, 096007 (2000);
  %%[arXiv:hep-ph/0003035].
  %%CITATION = HEP-PH 0003035;%%
 E.~Boos and T.~Plehn,
  %``Higgs-boson production induced by bottom quarks,''
  Phys.\ Rev.\ D {\bf 69}, 094005 (2004).
  %[arXiv:hep-ph/0304034].
  %%CITATION = HEP-PH 0304034;%%

\bibitem{dec_tau_ph}
 D.~P.~Roy,
  %``The hadronic tau decay signature of a heavy charged Higgs boson at LHC,''
  Phys.\ Lett.\ B {\bf 459}, 607 (1999).
  %[arXiv:hep-ph/9905542].
  %%CITATION = HEP-PH 9905542;%%

\bibitem{dec_tau_ex}
 K.~A.~Assamagan and Y.~Coadou,
  %``The Hadronic Tau Decay Of A Heavy H+- In Atlas,''
  Acta Phys.\ Polon.\ B {\bf 33}, 707 (2002);
  %%CITATION = APPOA,B33,707;%%
 Y.~Coadou,
  %``Searches for the charged Higgs at hadron colliders based on the tau lepton
  %signature,''
  FERMILAB-THESIS-2003-31;
  %\href{http://www.slac.stanford.edu/spires/find/hep/www?r=fermilab-thesis-2003-31}{SPIRES entry}
 R.~Kinnunen and A.~Nikitenko,
  report CMS note 2003/006.

\bibitem{dec_top_ph}
 J.~F.~Gunion,
  %``Detecting the t b decays of a charged Higgs boson at a hadron
  %supercollider,''
  Phys.\ Lett.\ B {\bf 322}, 125 (1994);
  %[arXiv:hep-ph/9312201].
  %%CITATION = HEP-PH 9312201;%%
 J.~A.~Coarasa, D.~Garcia, J.~Guasch, R.~A.~Jimenez and J.~Sola,
  %``Heavy charged Higgs boson decaying into top quark in the MSSM,''
  Phys.\ Lett.\ B {\bf 425}, 329 (1998);
  %[arXiv:hep-ph/9711472].
  %%CITATION = HEP-PH 9711472;%%
 S.~Moretti and D.~P.~Roy,
  %``Detecting heavy charged Higgs bosons at the LHC with triple b-tagging,''
  Phys.\ Lett.\ B {\bf 470}, 209 (1999).
  %[arXiv:hep-ph/9909435].
  %%CITATION = HEP-PH 9909435;%%

\bibitem{dec_top_ex}
 K.~A.~Assamagan, Y.~Coadou and A.~Deandrea,
  %``ATLAS discovery potential for a heavy charged Higgs boson,''
  report ATL-COM-PHYS-2002-002,
  arXiv:hep-ph/0203121;
  %%CITATION = HEP-PH 0203121;%%
 K.~A.~Assamagan and N.~Gollub,
  %``The ATLAS discovery potential for a heavy charged Higgs boson in g g  --> t
  %b H+- with H+- --> t b,''
  Eur.\ Phys.\ J.\ C {\bf 39S2}, 25 (2005);
  %[arXiv:hep-ph/0406013].
  %%CITATION = HEP-PH 0406013;%% 
 P.~Salmi, R.~Kinnunen and N.~Stepanov,
  %``Prospects of detecting massive charged Higgs from hadronic decay  H+- $\to$ t b in CMS,''
  arXiv:hep-ph/0301166.
  %%CITATION = HEP-PH 0301166;%%
 S.~Lowette, J.~D'Hondt and P.~Vanlaer,
  %``Charged MSSM Higgs boson observability in the H+- --> t b decay,''
  CERN-CMS-NOTE-2006-109;
  %\href{http://www.slac.stanford.edu/spires/find/hep/www?r=cern-cms-note-2006-109}{SPIRES entry}
 also see: S.~Lowette, Ph.D thesis
  \url{http://web.iihe.ac.be/~slowette} 

\bibitem{low_mass}
 F.~Borzumati, J.~L.~Kneur and N.~Polonsky,
  %``Higgs- and slepton-strahlung at hadron colliders,''
  Phys.\ Rev.\ D {\bf 60}, 115011 (1999).
  %%CITATION = HEP-PH 9905443;%%
 for the analysis results see: 
  CMS TDR, Volume~II, 
  CERN/LHCC 2006-021, p.386.

\bibitem{higgs_pairs}
 T.~Plehn, M.~Spira and P.~M.~Zerwas,
  %``Pair Production of Neutral Higgs Particles in Gluon-Gluon Collisions,''
  Nucl.\ Phys.\ B {\bf 479}, 46 (1996)
  [Erratum-ibid.\ B {\bf 531}, 655 (1998)];
  %[arXiv:hep-ph/9603205].
  %%CITATION = HEP-PH 9603205;%%
 S.~Dawson, S.~Dittmaier and M.~Spira,
  %``Neutral Higgs-boson pair production at hadron colliders:  {QCD}
  %corrections,''
  Phys.\ Rev.\ D {\bf 58}, 115012 (1998);
  %[arXiv:hep-ph/9805244].
  %%CITATION = HEP-PH 9805244;%%
 A.~Djouadi, W.~Kilian, M.~M\"uhlleitner and P.~M.~Zerwas,
  %``Production of neutral Higgs-boson pairs at LHC,''
  Eur.\ Phys.\ J.\ C {\bf 10}, 45 (1999);
  %[arXiv:hep-ph/9904287].
  %%CITATION = HEP-PH 9904287;%% 
 A.~A.~Barrientos Bendezu and B.~A.~Kniehl,
  %``Pair production of neutral Higgs bosons at the CERN Large Hadron
  %Collider,''
  Phys.\ Rev.\ D {\bf 64}, 035006 (2001);
  %[arXiv:hep-ph/0103018].
  %%CITATION = HEP-PH 0103018;%% 
 U.~Baur, T.~Plehn and D.~L.~Rainwater,
  %``Probing the Higgs self-coupling at hadron colliders using rare decays,''
  Phys.\ Rev.\ D {\bf 69}, 053004 (2004).
  %[arXiv:hep-ph/0310056].
  %%CITATION = HEP-PH 0310056;%%

\bibitem{prod_w}
 D.~A.~Dicus, J.~L.~Hewett, C.~Kao and T.~G.~Rizzo,
  %``W+- H-+ Production At Hadron Colliders,''
  Phys.\ Rev.\ D {\bf 40}, 787 (1989);
  %%CITATION = PHRVA,D40,787;%%
 A.~A.~Barrientos Bendezu and B.~A.~Kniehl,
  %``W+- H-+ associated production at the Large Hadron Collider,''
  Phys.\ Rev.\ D {\bf 59}, 015009 (1998);
  %%CITATION = HEP-PH 9807480;%%
 and
  %``Squark loop correction to W+- H-+ associated hadroproduction,''
  Phys.\ Rev.\ D {\bf 63}, 015009 (2001);
  %%CITATION = HEP-PH 0007336;%%
 O.~Brein, W.~Hollik and S.~Kanemura,
  %``The MSSM prediction for W+- H-+ production by gluon fusion,''
  Phys.\ Rev.\ D {\bf 63}, 095001 (2001);
  %%CITATION = HEP-PH 0008308;%%
 Z.~Fei, M.~Wen-Gan, J.~Yi, H.~Liang and W.~Lang-Hui,
  %``Improved calculation of W+- H=- associated production via gluon-gluon fusion
  %at the CERN LHC,''
  Phys.\ Rev.\ D {\bf 63}, 015002 (2000);
  %%CITATION = PHRVA,D63,015002;%%
 W.~Hollik and S.~H.~Zhu,
  %``O(alpha(s)) corrections to b anti-b $\to$ W+- H-+ at the CERN Large  Hadron Collider,''
  Phys.\ Rev.\ D {\bf 65}, 075015 (2002).
  %%CITATION = HEP-PH 0109103;%%

\bibitem{prod_pair}
 J.~F.~Gunion, H.~E.~Haber, F.~E.~Paige, W.~K.~Tung and S.~S.~Willenbrock,
  %``Neutral And Charged Higgs Detection: Heavy Quark Fusion, Top Quark Mass Dependence And Rare Decays,''
  Nucl.\ Phys.\ B {\bf 294}, 621 (1987);
  %%CITATION = NUPHA,B294,621;%%
 S.~S.~D.~Willenbrock,
  %``Pair Production Of Supersymmetric Charged Higgs Bosons,''
  Phys.\ Rev.\ D {\bf 35}, 173 (1987);
  %%CITATION = PHRVA,D35,173;%%
 A.~Krause, T.~Plehn, M.~Spira and P.~M.~Zerwas,
  %``Production of charged Higgs boson pairs in gluon gluon collisions,''
  Nucl.\ Phys.\ B {\bf 519}, 85 (1998);
  %%CITATION = HEP-PH 9707430;%%
 A.~A.~Barrientos Bendezu and B.~A.~Kniehl,
  %``H+ H- pair production at the Large Hadron Collider,''
  Nucl.\ Phys.\ B {\bf 568}, 305 (2000);
  %%CITATION = HEP-PH 9908385;%%
 O.~Brein and W.~Hollik,
  %``Pair production of charged MSSM Higgs bosons by gluon fusion,''
  Eur.\ Phys.\ J.\ C {\bf 13}, 175 (2000);
  %%CITATION = HEP-PH 9908529;%%
 E.~Eichten, I.~Hinchliffe, K.~D.~Lane and C.~Quigg,
  %``Super Collider Physics,''
  Rev.\ Mod.\ Phys.\  {\bf 56}, 579 (1984)
  [Addendum-ibid.\  {\bf 58}, 1065 (1986)];
  %%CITATION = RMPHA,56,579;%%
 N.~G.~Deshpande, X.~Tata and D.~A.~Dicus,
  %``Production Of Nonstandard Higgs Bosons In E+ E- Collisions,''
  Phys.\ Rev.\ D {\bf 29}, 1527 (1984);
  %%CITATION = PHRVA,D29,1527;%%
 for a recent update and overview, see \eg:
 A.~Alves and T.~Plehn,
  %``Charged Higgs boson pairs at the LHC,''
  Phys.\ Rev.\ D {\bf 71}, 115014 (2005).
  %[arXiv:hep-ph/0503135].
  %%CITATION = HEP-PH 0503135;%%

\bibitem{bottom_yuk}
 L.~J.~Hall, R.~Rattazzi and U.~Sarid,
  %``The Top quark mass in supersymmetric SO(10) unification,''
  Phys.\ Rev.\ D {\bf 50}, 7048 (1994);
  %%CITATION = HEP-PH 9306309;%%
 M.~Carena, M.~Olechowski, S.~Pokorski and C.~E.~Wagner,
  %``Electroweak symmetry breaking and bottom - top Yukawa unification,''
  Nucl.\ Phys.\ B {\bf 426}, 269 (1994);
  %%CITATION = HEP-PH 9402253;%%
 M.~Carena, D.~Garcia, U.~Nierste and C.~E.~Wagner,
  %``Effective Lagrangian for the anti-t b H+ interaction in the MSSM and  charged Higgs phenomenology,''
  Nucl.\ Phys.\ B {\bf 577}, 88 (2000);
  %%CITATION = HEP-PH 9912516;%%
 A.~Belyaev, D.~Garcia, J.~Guasch and J.~Sola,
  %``Prospects for supersymmetric charged Higgs boson discovery at the %Tevatron and the LHC,''
  Phys.\ Rev.\ D {\bf 65}, 031701(R) (2002);
  %[arXiv:hep-ph/0105053]
  %%CITATION = HEP-PH 0105053;%%
 J.~Guasch, P.~H\"afliger and M.~Spira,
  %``MSSM Higgs decays to bottom quark pairs revisited,''
  Phys.\ Rev.\ D {\bf 68}, 115001 (2003).
  %[arXiv:hep-ph/0305101].
  %%CITATION = HEP-PH 0305101;%%


\bibitem{mfv}
%\cite{Chivukula:1987py}
%\bibitem{Chivukula:1987py}
  R.~S.~Chivukula and H.~Georgi,
  %``Composite Technicolor Standard Model,''
  Phys.\ Lett.\  B {\bf 188}, 99 (1987); 
  %%CITATION = PHLTA,B188,99;%%
%\cite{D'Ambrosio:2002ex}
%\bibitem{D'Ambrosio:2002ex}
  G.~D'Ambrosio, G.~F.~Giudice, G.~Isidori and A.~Strumia,
  %``Minimal flavour violation: An effective field theory approach,''
  Nucl.\ Phys.\  B {\bf 645}, 155 (2002).
  %[arXiv:hep-ph/0207036].
  %%CITATION = NUPHA,B645,155;%%



%\cite{Hiller:2002um}
\bibitem{Hiller:2002um}
  G.~Hiller and M.~Schmaltz,
  %``Strong-weak CP hierarchy from non-renormalization theorems,''
  Phys.\ Rev.\  D {\bf 65}, 096009 (2002).
%  [arXiv:hep-ph/0201251].
  %%CITATION = PHRVA,D65,096009;%%

%\cite{Altmannshofer:2007cs}
\bibitem{Altmannshofer:2007cs}
  W.~Altmannshofer, A.~J.~Buras and D.~Guadagnoli,
  %``The MFV limit of the MSSM for low tan(beta): meson mixings revisited,''
  arXiv:hep-ph/0703200.
  %%CITATION = HEP-PH/0703200;%%


%\cite{Hall:1985dx}
\bibitem{Hall:1985dx}
  L.~J.~Hall, V.~A.~Kostelecky and S.~Raby,
  %``New Flavor Violations In Supergravity Models,''
  Nucl.\ Phys.\  B {\bf 267}, 415 (1986).
  %%CITATION = NUPHA,B267,415;%%

\bibitem{fcnc-susy}
%\cite{Hagelin:1992tc}
%\bibitem{Hagelin:1992tc}
  J.~S.~Hagelin, S.~Kelley and T.~Tanaka,
  %``Supersymmetric flavor changing neutral currents: Exact amplitudes and
  %phenomenological analysis,''
  Nucl.\ Phys.\  B {\bf 415}, 293 (1994);
  %%CITATION = NUPHA,B415,293;%%
%\cite{Gabbiani:1996hi}
%\bibitem{Gabbiani:1996hi}
  F.~Gabbiani, E.~Gabrielli, A.~Masiero and L.~Silvestrini,
  %``A complete analysis of FCNC and CP constraints in general SUSY extensions
  %of the standard model,''
  Nucl.\ Phys.\  B {\bf 477}, 321 (1996);
  %[arXiv:hep-ph/9604387];
  %%CITATION = NUPHA,B477,321;%%
%\cite{Misiak:1997ei}
%\bibitem{Misiak:1997ei}
  M.~Misiak, S.~Pokorski and J.~Rosiek,
  %``Supersymmetry and FCNC effects,''
  Adv.\ Ser.\ Direct.\ High Energy Phys.\  {\bf 15}, 795 (1998).
  %[arXiv:hep-ph/9703442].
  %%CITATION = 00319,15,795;%%


\bibitem{vacuum} 
 J.~A.~Casas and S.~Dimopoulos, 
  %``Stability bounds on flavor-violating trilinear soft terms in the MSSM,'' 
  Phys.\ Lett.\ B {\bf 387}, 107 (1996).
  %[arXiv:hep-ph/9606237]. 
  %%CITATION = HEP-PH 9606237;%%

%\cite{Nir:2007ac}
\bibitem{Nir:2007ac}
  See, \eg:
  Y.~Nir,
  %``Lessons from BaBar and Belle measurements of D-Dbar mixing parameters,''
  arXiv:hep-ph/0703235;
  %%CITATION = HEP-PH/0703235;%%
 {\it and references therein.}




\bibitem{b_s_gamma_ex}
 B.~Aubert {\it et al.}  [BABAR Collaboration],
  %``Measurements of the $B \to X_s \gamma$ branching fraction and photon
  %spectrum from a sum of exclusive final states,''
  Phys.\ Rev.\  D {\bf 72}, 052004 (2005);
  %%CITATION = PHRVA,D72,052004;%%
 P.~Koppenburg {\it et al.}  [Belle Collaboration],
  %``An inclusive measurement of the photon energy spectrum in b --> s gamma
  %decays,''
  Phys.\ Rev.\ Lett.\  {\bf 93}, 061803 (2004);
  %[arXiv:hep-ex/0403004].
  %%CITATION = PRLTA,93,061803;%%
 S.~Chen {\it et al.}  [CLEO Collaboration],
  %``Branching fraction and photon energy spectrum for b $\to$ s gamma,''
  Phys.\ Rev.\ Lett.\  {\bf 87}, 251807 (2001);
  %[arXiv:hep-ex/0108032].
  %%CITATION = HEP-EX 0108032;%%
 W.-M. Yao {\it et al.}, J.\ Phys.\ Lett.\ G {\bf 33}, 1 (2006);
 R.~Barate {\it et al.}  [ALEPH Collaboration],
  %``A measurement of the inclusive b $\to$ s gamma branching ratio,''
  Phys.\ Lett.\ B {\bf 429}, 169 (1998);
  %%CITATION = PHLTA,B429,169;%%
 H.~F.~A.~Group(HFAG),
  %``Averages of b-hadron properties as of winter 2005,''
  arXiv:hep-ex/0505100.
  %%CITATION = HEP-EX 0505100;%%



\bibitem{b_s_gamma_th}
 P.~L.~Cho, M.~Misiak and D.~Wyler,
  %``$K_L \to \pi~0 e~+ e~-$ and $B \to X_s \ell~+ \ell~-$ Decay in the MSSM,''
  Phys.\ Rev.\ D {\bf 54}, 3329 (1996);
  %%[arXiv:hep-ph/9601360].
  %%CITATION = HEP-PH 9601360;%%
 J.~L.~Hewett and J.~D.~Wells,
  %``Searching for supersymmetry in rare B decays,''
  Phys.\ Rev.\ D {\bf 55}, 5549 (1997);
  %%[arXiv:hep-ph/9610323].
  %%CITATION = HEP-PH 9610323;%%
 F.~M.~Borzumati and C.~Greub,
  %``2HDMs predictions for anti-B $\to$ X/s gamma in NLO {QCD},''
  Phys.\ Rev.\ D {\bf 58}, 074004 (1998)
  %%[arXiv:hep-ph/9802391].
  %%CITATION = HEP-PH 9802391;%%
 and
 %F.~M.~Borzumati and C.~Greub,
  %``Two Higgs doublet model predictions for anti-B $\to$ X/s gamma in  NLO {QCD}.
  %(Addendum),''
  Phys.\ Rev.\ D {\bf 59}, 057501 (1999);
  %%[arXiv:hep-ph/9809438].
  %%CITATION = HEP-PH 9809438;%%
 G.~Hiller and F.~Kr\"uger,
  %``More model-independent analysis of b $\to$ s processes,''
  Phys.\ Rev.\ D {\bf 69}, 074020 (2004).
  %%[arXiv:hep-ph/0310219].
  %%CITATION = HEP-PH 0310219;%%

\bibitem{rhogamma}
%\cite{Abe:2005rj}
%\bibitem{Abe:2005rj}
  K.~Abe {\it et al.},
  %``Observation of b --> d gamma and determination of |V(td)/V(ts)|,''
  Phys.\ Rev.\ Lett.\  {\bf 96}, 221601 (2006);
  %[arXiv:hep-ex/0506079];
  %%CITATION = PRLTA,96,221601;%%
%\cite{Aubert:2006pu}
%\bibitem{Aubert:2006pu}
  B.~Aubert {\it et al.}  [BABAR Collaboration],
  %``Branching fraction measurements of B+ --> rho+ gamma, B0 --> rho0 gamma,
  %and B0 --> omega gamma,''
  arXiv:hep-ex/0612017;
  %%CITATION = HEP-EX/0612017;%%
%%\cite{Bosch:2001gv}
%\bibitem{Bosch:2001gv}
%  S.~W.~Bosch and G.~Buchalla,
%  %``The radiative decays B --> V gamma at next-to-leading order in QCD,''
%  Nucl.\ Phys.\  B {\bf 621}, 459 (2002).
%  %[arXiv:hep-ph/0106081].
%  %%CITATION = NUPHA,B621,459;%%

\bibitem{bsll} 
%\cite{Aubert:2004it}
%\bibitem{Aubert:2004it}
  B.~Aubert {\it et al.}  [BABAR Collaboration],
  %``Measurement of the $B \to X_s \ell^+ \ell^-$ branching fraction with a sum
  %over exclusive modes,''
  Phys.\ Rev.\ Lett.\  {\bf 93}, 081802 (2004).
  %[arXiv:hep-ex/0404006];
  %%CITATION = PRLTA,93,081802;%%
%
%\cite{Abe:2004sg}
%\bibitem{Abe:2004sg}
  K.~Abe {\it et al.}  [Belle Collaboration],
  %``Improved measurement of the 
%electroweak penguin process B --> X/s l+  l-,''
  arXiv:hep-ex/0408119;
  %%CITATION = HEP-EX/0408119;%%
%\cite{Ali:2002jg}
%\bibitem{Ali:2002jg}
  A.~Ali, E.~Lunghi, C.~Greub and G.~Hiller,
  %``Improved model-independent analysis of semileptonic and radiative rare  B
  %decays,''
  Phys.\ Rev.\  D {\bf 66}, 034002 (2002);
  %[arXiv:hep-ph/0112300].
  %%CITATION = PHRVA,D66,034002;%%
%\cite{Grinstein:1988me}
%\bibitem{Grinstein:1988me}
  B.~Grinstein, M.~J.~Savage and M.~B.~Wise,
  %``B $\to$ X(s) e+ e- in the Six Quark Model,''
  Nucl.\ Phys.\  B {\bf 319}, 271 (1989);
  %%CITATION = NUPHA,B319,271;%%
%\cite{Guetta:1997fw}
%\bibitem{Guetta:1997fw}
  D.~Guetta and E.~Nardi,
  %``Searching for new physics in rare B --> tau decays,''
  Phys.\ Rev.\  D {\bf 58}, 012001 (1998).
%  [arXiv:hep-ph/9707371].
  %%CITATION = PHRVA,D58,012001;%%


\bibitem{susy-Zpenguin}
%\cite{Lunghi:1999uk}
%\bibitem{Lunghi:1999uk}
  E.~Lunghi, A.~Masiero, I.~Scimemi and L.~Silvestrini,
  %``B --> X/s l+ l- decays in supersymmetry,''
  Nucl.\ Phys.\  B {\bf 568}, 120 (2000);
  %[arXiv:hep-ph/9906286];
  %%CITATION = NUPHA,B568,120;%%
%\cite{Buchalla:2000sk}
%\bibitem{Buchalla:2000sk}
  G.~Buchalla, G.~Hiller and G.~Isidori,
  %``Phenomenology of non-standard Z couplings in exclusive semileptonic  b -->
  %s transitions,''
  Phys.\ Rev.\  D {\bf 63}, 014015 (2001);
  %[arXiv:hep-ph/0006136];
  %%CITATION = PHRVA,D63,014015;%%
%\cite{Hiller:2002ci}
%\bibitem{Hiller:2002ci}
  G.~Hiller,
  %``First hint of non-standard CP violation from B --> Phi K(S) decay,''
  Phys.\ Rev.\  D {\bf 66}, 071502 (2002).
  %[arXiv:hep-ph/0207356].
  %%CITATION = PHRVA,D66,071502;%%

\bibitem{B2pill}
%\cite{unknown:2007mm}
%\bibitem{unknown:2007mm}
    [BABAR Collaboration],
  %``Search for the rare decay B --> pi l+ l-,''
  arXiv:hep-ex/0703018;
  %%CITATION = HEP-EX/0703018;%%
%\cite{Ali:1999mm}
%\bibitem{Ali:1999mm}
  A.~Ali, P.~Ball, L.~T.~Handoko and G.~Hiller,
  %``A comparative study of the decays B --> (K,K*) l+ l- in standard  model and
  %supersymmetric theories,''
  Phys.\ Rev.\  D {\bf 61}, 074024 (2000);
  %[arXiv:hep-ph/9910221];
  %%CITATION = PHRVA,D61,074024;%%
%\cite{Ball:2004ye}
%\bibitem{Ball:2004ye}
  P.~Ball and R.~Zwicky,
  %``New results on B --> pi, K, eta decay formfactors from light-cone sum
  %rules,''
  Phys.\ Rev.\  D {\bf 71}, 014015 (2005).
  %[arXiv:hep-ph/0406232].
  %%CITATION = PHRVA,D71,014015;%%


%\cite{Yao:2006px}
%\bibitem{Yao:2006px}
\bibitem{data}
  W.~M.~Yao {\it et al.}  [Particle Data Group],
  %``Review of particle physics,''
  J.\ Phys.\ G {\bf 33} (2006) 1;
  %%CITATION = JPHGB,G33,1;%%
%
 E.~Barberio {\it et al.}  [Heavy Flavor Averaging Group (HFAG)],
  %``Averages of b-hadron properties at the end of 2005,''
  arXiv:hep-ex/0603003.
  %%CITATION = HEP-EX 0603003;%%
  \url{http://www.slac.stanford.edu/xorg/hfag}


\bibitem{bs_mix}
%\cite{Abazov:2006dm}
%\bibitem{Abazov:2006dm}
  V.~M.~Abazov {\it et al.}  [D0 Collaboration],
  %``First direct two-sided bound on the B/s0 oscillation frequency,''
  Phys.\ Rev.\ Lett.\  {\bf 97}, 021802 (2006);
  %[arXiv:hep-ex/0603029];
  %%CITATION = HEP-EX 0603029;%%
%\cite{Abulencia:2006ze}
%\bibitem{Abulencia:2006ze}
  A.~Abulencia {\it et al.}  [CDF Collaboration],
  %``Observation of $B^0_{s}$ - $\bar{B}^0$( $s^{)}$ Oscillations,''
  Phys.\ Rev.\ Lett.\  {\bf 97}, 242003 (2006).
  %[arXiv:hep-ex/0609040].
  %%CITATION = PRLTA,97,242003;%%

%\cite{Bertolini:1990if}
\bibitem{Bertolini:1990if}
  S.~Bertolini, F.~Borzumati, A.~Masiero and G.~Ridolfi,
  %``Effects of supergravity induced electroweak breaking on rare B decays and
  %mixings,''
  Nucl.\ Phys.\  B {\bf 353}, 591 (1991);
  %%CITATION = NUPHA,B353,591;%%
%
%\cite{Branco:1994eb}
%\bibitem{Branco:1994eb}
  G.~C.~Branco, G.~C.~Cho, Y.~Kizukuri and N.~Oshimo,
  %``Supersymmetric contributions to B0 - anti-B0 and K0 - anti-K0 mixings,''
  Phys.\ Lett.\  B {\bf 337}, 316 (1994).
  %[arXiv:hep-ph/9408229].
  %%CITATION = PHLTA,B337,316;%%

\bibitem{mixing-th}
% E.~Gabrielli and S.~Khalil,
%  % Constraining supersymmetric models from B/d - anti-B/d mixing and the  B/d -%-> J/psi K(S) asymmetry",
%   Phys. Rev. D {\bf 67}, 015008 (2003);
%  %hep-ph/0207288
%%
%
 %\cite{Becirevic:2001jj}
  D.~Becirevic {\it et al.},
  %``B/d anti-B/d mixing and the B/d --> J/psi K(S) asymmetry in general  SUSY
  %models,''
  Nucl.\ Phys.\  B {\bf 634}, 105 (2002);
  %[arXiv:hep-ph/0112303].
  %%CITATION = NUPHA,B634,105;%%
%%
%
  A.~Bartl, T.~Gajdosik, E.~Lunghi, A.~Masiero, W.~Porod, H.~Stremnitzer and O.~Vives,
  %``General flavor blind MSSM and CP violation,''
  Phys.\ Rev.\  D {\bf 64}, 076009 (2001);
  %[arXiv:hep-ph/0103324].
  %%CITATION = PHRVA,D64,076009;%%

P.~Ball, S.~Khalil and E.~Kou,
  %``B/s0-anti-B/s0 mixing and the B/s --> J/psi Phi asymmetry in
  %supersymmetric models,''
  Phys.\ Rev.\ D {\bf 69}, 115011 (2004);
  %[arXiv:hep-ph/0311361];
  %%CITATION = HEP-PH 0311361;%%
%\cite{Blanke:2006ig}
%\bibitem{Blanke:2006ig}
  M.~Blanke, A.~J.~Buras, D.~Guadagnoli and C.~Tarantino,
  %``Minimal Flavour Violation Waiting for Precise Measurements of Delta M_s,
  %S_{psi phi}, A^s_SL, |V_ub|, gamma and B^0_{s,d} -> mu+ mu-,''
  JHEP {\bf 0610}, 003 (2006);
  %[arXiv:hep-ph/0604057].
  %%CITATION = JHEPA,0610,003;%%
  T.~Goto, T.~Nihei and Y.~Okada,
  %``$B~0$--$\overline{B}~0$ mixing and $\epsilon_K$ parameter in the minimal
  %supergravity model,''
  Phys.\ Rev.\  D {\bf 53}, 5233 (1996)
  [Erratum-ibid.\  D {\bf 54}, 5904 (1996)].
  %[arXiv:hep-ph/9510286].
  %%CITATION = PHRVA,D53,5233;%%

\bibitem{single_higgs}
 J.~L.~Diaz-Cruz, H.~J.~He and C.~P.~Yuan,
  %``Soft SUSY breaking, stop-scharm mixing and Higgs signatures,''
  Phys.\ Lett.\ B {\bf 530}, 179 (2002);
  %[arXiv:hep-ph/0103178];
  %%CITATION = HEP-PH 0103178;%%
%
 H.~J.~He and C.~P.~Yuan,
  %``New method for detecting charged (pseudo-)scalars at colliders,''
  Phys.\ Rev.\ Lett.\  {\bf 83}, 28 (1999);
  %[arXiv:hep-ph/9810367];
  %%CITATION = HEP-PH 9810367;%%
%
C.~Balazs, H.~J.~He and C.~P.~Yuan,
  %``{QCD} corrections to scalar production via heavy quark fusion at hadron colliders,'' 
  Phys.\ Rev.\ D {\bf 60}, 114001 (1999);
  %[arXiv:hep-ph/9812263];
  %%CITATION = HEP-PH 9812263;%%
 S.~R.~Slabospitsky,
  %``Study of s-channel charged Higgs production in CMS,'' 
  arXiv:hep-ph/0203094. 
  %%CITATION = HEP-PH 0203094;%%

\bibitem{operators}
 C.~N.~Leung, S.~T.~Love and S.~Rao,
  %``Low-Energy Manifestations Of A New Interaction Scale: Operator Analysis,''
  Z.\ Phys.\ C {\bf 31}, 433 (1986);
  %%CITATION = ZEPYA,C31,433;%% 
 W.~Buchm\"uller and D.~Wyler,
  %``Effective Lagrangian Analysis Of New Interactions And Flavor
  %Conservation,''
  Nucl.\ Phys.\ B {\bf 268}, 621 (1986).
  %%CITATION = NUPHA,B268,621;%%



\bibitem{susyflavorRGE}
%\cite{Martin:1993zk}
%\bibitem{Martin:1993zk}
  S.~P.~Martin and M.~T.~Vaughn,
  %``Two Loop Renormalization Group Equations For Soft Supersymmetry Breaking
  %Couplings,''
  Phys.\ Rev.\  D {\bf 50}, 2282 (1994).
  %[arXiv:hep-ph/9311340].
  %%CITATION = PHRVA,D50,2282;%%


%\cite{Colangelo:1998pm}
\bibitem{Colangelo:1998pm}
  G.~Colangelo and G.~Isidori,
  %``Supersymmetric contributions to rare kaon decays: Beyond the single
  %mass-insertion approximation,''
  JHEP {\bf 9809}, 009 (1998).
  %[arXiv:hep-ph/9808487].
  %%CITATION = JHEPA,9809,009;%%


%\cite{Hahn:2000kx}
\bibitem{feynarts}
  T.~Hahn,
  %``Generating Feynman diagrams and amplitudes with FeynArts 3,''
  Comput.\ Phys.\ Commun.\  {\bf 140} (2001) 418;
%  [hep-ph/0012260];\\
  %%CITATION = HEP-PH 0012260;%%
%
%\cite{Hahn:2001rv}
%\bibitem{Hahn:2001rv}
  T.~Hahn and C.~Schappacher,
  % ``The implementation of the minimal supersymmetric standard model in
  %FeynArts and FormCalc,''
  Comput.\ Phys.\ Commun.\  {\bf 143} (2002) 54.
%  [hep-ph/0105349].
  %%CITATION = HEP-PH 0105349;%%

%\cite{Hahn:1998yk}
\bibitem{formcalclooptools}
  T.~Hahn and M.~Perez-Victoria,
  %``Automatized one-loop calculations in four and D dimensions,''
  Comput.\ Phys.\ Commun.\  {\bf 118} (1999) 153.
%  [hep-ph/9807565].
  %%CITATION = HEP-PH 9807565;%%

\bibitem{hadcalc}
 T.~Hahn and M.~Rauch,
  %``News from FormCalc and LoopTools,''
  Nucl.\ Phys.\ Proc.\ Suppl.\  {\bf 157}, 236 (2006).
  %%[arXiv:hep-ph/0601248].
  %%CITATION = HEP-PH 0601248;%%


\bibitem{Pumplin:2002vw}
J.~Pumplin, D.~R.~Stump, J.~Huston, H.~L.~Lai, P.~Nadolsky and W.~K.~Tung,
%%``New generation of parton distributions with uncertainties from global  QCD analysis,''
JHEP {\bf 0207} (2002) 012.
%[hep-ph/0201195].
%%CITATION = HEP-PH 0201195;%%



\bibitem{tev_limits}
 For the most recent limits see:
 \url{http://www-cdf.fnal.gov/physics/exotic/exotic.html}; 
 V.~M.~Abazov {\it et al.}  [D0 Collaboration],
  %``Search for squarks and gluinos in events with jets and missing  transverse
  %energy in p anti-p collisions at s**(1/2) = 1.96-TeV,''
  Phys.\ Lett.\ B {\bf 638}, 119 (2006).
  %[arXiv:hep-ex/0604029].
  %%CITATION = HEP-EX 0604029;%%


%\cite{Pierce:1996zz}
\bibitem{Pierce:1996zz}
  D.~M.~Pierce, J.~A.~Bagger, K.~T.~Matchev and R.~j.~Zhang,
  %``Precision corrections in the minimal supersymmetric standard model,''
  Nucl.\ Phys.\  B {\bf 491}, 3 (1997).
  %[arXiv:hep-ph/9606211].
  %%CITATION = NUPHA,B491,3;%%


\bibitem{higgs_mass} 
 H.~E.~Haber, R.~Hempfling and A.~H.~Hoang,
  %``Approximating the radiatively corrected Higgs mass in the minimal supersymmetric model,'' 
  Z.\ Phys.\ C {\bf 75}, 539 (1997);
  %[arXiv:hep-ph/9609331]. 
  %%CITATION = HEP-PH 9609331;%%
 M.~Frank, T.~Hahn, S.~Heinemeyer, W.~Hollik, H.~Rzehak and G.~Weiglein,
  %``The Higgs Boson Masses and Mixings of the Complex MSSM in the
  %Feynman-Diagrammatic Approach,''
  arXiv:hep-ph/0611326.
  %%CITATION = HEP-PH 0611326;%%


\bibitem{Btaunu}
%\cite{Ikado:2006un}
%\bibitem{Ikado:2006un}
  K.~Ikado {\it et al.},
  %``Evidence of the purely leptonic decay B- --> tau- anti-nu/tau,''
  Phys.\ Rev.\ Lett.\  {\bf 97}, 251802 (2006);
  %[arXiv:hep-ex/0604018];
  %%CITATION = PRLTA,97,251802;%%
%%\cite{Aubert:2006fk}
%\bibitem{Aubert:2006fk}
  B.~Aubert {\it et al.}  [BABAR Collaboration];
  %``A search for B+ --> tau+ nu recoiling against B- --> D0 l- anti-nu/l X,''
  arXiv:hep-ex/0608019;
  %%CITATION = HEP-EX/0608019;%%
 A.~G.~Akeroyd and S.~Recksiegel, 
  %``The effect of H+- on B+- --> tau+- nu/tau and B+- --> mu+- nu/mu,''
  J.\ Phys.\ G {\bf 29}, 2311 (2003).
  %[arXiv:hep-ph/0306037]. 
  %%CITATION = HEP-PH 0306037;%%


\bibitem{schwen} 
 S.~Heinemeyer, W.~Hollik, F.~Merz and S.~Penaranda, 
  %``Electroweak precision observables in the MSSM with non-minimal flavor violation,'' 
 Eur.\ Phys.\ J.\ C {\bf 37}, 481 (2004).
 %[arXiv:hep-ph/0403228]. 
 %%CITATION = HEP-PH 0403228;%%

\bibitem{top_fcnc} 
 J.~Cao, G.~Eilam, K.~i.~Hikasa and J.~M.~Yang,
 %``Experimental constraints on stop-scharm flavor mixing and implications in top-quark FCNC processes,'' 
 Phys.\ Rev.\ D {\bf 74}, 031701 (2006);
 %[arXiv:hep-ph/0604163].
 %%CITATION = HEP-PH 0604163;%%
J.~J.~Cao, G.~Eilam, M.~Frank, K.~Hikasa, G.~L.~Liu, I.~Turan and J.~M.~Yang,
  %``SUSY-induced FCNC top-quark processes at the Large Hadron Collider,''
  Phys.\ Rev.\  D {\bf 75}, 075021 (2007).
%  [arXiv:hep-ph/0702264].
  %%CITATION = PHRVA,D75,075021;%%


\bibitem{denner}
 A.~Denner,
  %``Techniques for calculation of electroweak radiative corrections at the one
  %loop level and results for W physics at LEP-200,''
  Fortsch.\ Phys.\  {\bf 41}, 307 (1993).
  %%CITATION = FPYKA,41,307;%%

%\cite{Smirnov:1996ng}
\bibitem{Smirnov:1996ng}
  V.~A.~Smirnov,
  %``Asymptotic expansions of Feynman diagrams on the mass shell in momenta  and
  %masses,''
  Phys.\ Lett.\ B {\bf 394} (1997) 205.
%  [arXiv:hep-th/9608151].
  %%CITATION = HEP-TH 9608151;%%

\bibitem{madevent}
 T.~Stelzer and W.~F.~Long,
  %``Automatic generation of tree level helicity amplitudes,''
  Comput.\ Phys.\ Commun.\  {\bf 81}, 357 (1994);
  %[arXiv:hep-ph/9401258].
  %%CITATION = HEP-PH 9401258;%%
 F.~Maltoni and T.~Stelzer,
  %``MadEvent: Automatic event generation with MadGraph,''
  JHEP {\bf 0302}, 027 (2003);
  %%[arXiv:hep-ph/0208156].
  %%CITATION = HEP-PH 0208156;%%
 \url{http://madgraph.phys.ucl.ac.be}

%\cite{Nir:1993mx}
\bibitem{Nir:1993mx}
  Y.~Nir and N.~Seiberg,
  %``Should squarks be degenerate?,''
  Phys.\ Lett.\  B {\bf 309}, 337 (1993).
  %[arXiv:hep-ph/9304307].
  %%CITATION = PHLTA,B309,337;%%


%\cite{Grossman:2007bd}
\bibitem{Grossman:2007bd}
  Y.~Grossman, Y.~Nir, J.~Thaler, T.~Volansky and J.~Zupan,
  %``Probing Minimal Flavor Violation at the LHC,''
  arXiv:0706.1845 [hep-ph].
  %%CITATION = ARXIV:0706.1845;%%


%\cite{Rosiek:1995kg}
\bibitem{Rosiek:1995kg}
  J.~Rosiek,
  %``Complete set of Feynman rules for the MSSM -- ERRATUM,''
  arXiv:hep-ph/9511250.
  %%CITATION = HEP-PH 9511250;%%

\bibitem{mjamin}
 See \eg:
 M.~Jamin,
  talk given at the University of Granada, March 2006.
  %''http://personal.ifae.es/jamin/my/talks/mq_granada06.pdf''

\bibitem{dittmaier}
 A.~Denner and S.~Dittmaier,
  %``Reduction schemes for one-loop tensor integrals,''
  Nucl.\ Phys.\ B {\bf 734}, 62 (2006).
  %[arXiv:hep-ph/0509141].
  %%CITATION = HEP-PH 0509141;%%


\end{thebibliography}
